\documentclass{JHEP3}
\pdfoutput=1
\usepackage{epsf}
\usepackage[utf8]{inputenc}
\usepackage{amstext}
\usepackage{graphicx}
\usepackage{amssymb}
\usepackage{esint}
\renewcommand{\vec}[1]{{\bf #1}}       %%%  vectors in bold
\def\beq{\begin{eqnarray}}    %%%  begequation/eqnarray
\def\eeq{\end{eqnarray}}      %%%  endequation/eqnarray
\newcommand{\Om}{\Omega_m}
\newcommand{\Omo}{\Omega_m^0}

\newcommand{\Oro}{\Omega_{r}^0}
\newcommand{\OL}{\Omega_{\Lambda}}

\newcommand{\OLo}{\Omega_{\Lambda}^0}

\newcommand{\rc}{\rho_c}
\newcommand{\rco}{\rho_{c}^0}
\newcommand{\rmo}{\rho_{m}^0}
\newcommand{\rro}{\rho_{r}^0}

\newcommand{\rmr}{\rho_m}
\newcommand{\pmr}{p_m}

\newcommand{\rR}{\rho_r}

\newcommand{\wm}{\omega_m}
\newcommand{\wCC}{\omega_\CC}

\newcommand{\rL}{\rho_{\CC}}

\newcommand{\rLo}{\rho_{\CC}^0}

\newcommand{\CC}{\Lambda}

\newcommand{\xiR}{\xi'}

\newcommand{\CH}{C_H}
\newcommand{\CHd}{C_{\dot{H}}}
\newcommand{\Hd}{\dot{H}}

\newcommand{\tetm}{\theta_{\rm m}}

\newcommand{\nueff}{\nu_{\rm eff}}
\newcommand{\zeff}{\zeta_{\rm eff}}

\newcommand{\rRo}{\rho_r^0}

\newcommand{\be}{\begin{equation}}
\newcommand{\ee}{\end{equation}}

\newcommand{\mysection}[1]{\section{#1}
\renewcommand{\theequation}{\thesection.\arabic{equation}}
\setcounter{equation}{0}}

\title{ Dynamical vacuum energy in the expanding Universe confronted with observations: a dedicated study}

\author{%
Adri\`a G\'omez-Valent, \, Joan Sol\`{a}\\
High Energy Physics Group, Dept.\ ECM, and Institut
de Ci\`encies del Cosmos,\\
Univ.\ de Barcelona, Av.\ Diagonal 647, E-08028 Barcelona,
Catalonia, Spain}

\author{Spyros Basilakos\\
Academy of Athens, Research Center for Astronomy and Applied
Mathematics,\\ Soranou Efesiou 4, 11527, Athens, Greece\\

\\E-mails:
\email{adriagova@ecm.ub.edu}, \email{sola@ecm.ub.edu}, \email{svasil@academyofathens.gr}}

\abstract{%
Despite the many
efforts, our theoretical understanding of the ultimate nature of the
dark energy component of the universe still lags well behind the
astounding experimental evidence achieved from the increasingly
sophisticated observational tools at our disposal.  While the
canonical possibility is a strict cosmological constant, or rigid
vacuum energy density $\rL=$const., the exceeding simplicity of this
possibility lies also at the root of its unconvincing theoretical
status, as there is no explanation for the existence of such
constant for the entire cosmic history. Herein we explore general
models of the vacuum energy density slowly evolving with the Hubble
function $H$ and/or its time derivative,
$\rho_\CC=\rho_\CC(H,\dot{H})$. Some of these models are actually
well-motivated from the theoretical point of view and may provide a
rich phenomenology that could be explored in future observations,
whereas some others have more limitations. In this work, we put them
to the test and elucidate which ones are still compatible with the
present observations and which ones are already ruled out. We
consider their implications on structure formation, in
combination with data on type Ia supernovae, the Cosmic
Microwave Background, the Baryonic Acoustic Oscillations, and the
predicted redshift distribution of cluster-size collapsed
structures. The relation of these vacuum models on possible evidence of dynamical dark energy recently pointed out in the literature is also briefly addressed.}

 \preprint{}

\begin{document}

\section{Introduction}

The existence of the dark energy (DE) component in our universe as
the purported physical cause for the accelerated expansion is
currently beyond dispute from the observational point of
view\,\cite{PlanckXVI2013}. The simplest explanation, namely in
the form of a cosmological constant (CC) in Einstein's field
equations, is apparently working quite well (it is the basis of the
standard $\CC$CDM cosmological model), but unfortunately it does not
provide a single clue about its origin in the context of fundamental
physics, i.e. according to quantum field theory (QFT), string theory
etc. In fact, the theoretical prediction within any sound framework
is well-known to overshoot by far the observational value, and this
dramatic discrepancy constitutes the famous CC
problem\,\cite{CCproblem1} -- see also the recent
review\,\cite{JSP-CCReview13}.

Dynamical models of the DE offer
better chances to mitigate the problem. Popular possibilities are,
among others, quintessence and phantom energy in its various
forms\,\cite{CCproblem2}. In the last few
years many proposals for modified gravity theories have flourished
with great impetus as an alternative to supersede the DE as a
physical substratum, see e.g.\,\cite{ModGrav1} and references
therein. In these models the DE appears usually as just a late-time
effect in the history of the universe (and hence they overlook the
impact of a huge vacuum energy density during the entire cosmic
evolution). However, in subsequent modified gravity models it is
possible to operate (at least technically) a dynamical adjustment of the DE value to its
current size, irrespective of its initial value in the early
universe, including a relaxation mechanism of the vacuum energy
density in astrophysical domains such as the Solar
System\,\cite{Relax}.

In another vein, there is the possibility that the DE is related to
the vacuum energy density of, say, QFT in curved spacetime\,\cite{BirrellDavis,ParkerToms}. In this
case, the DE need not be just a constant in an expanding universe.
In fact, one rather expects the vacuum energy density $\rL=\CC/(8\pi\,G_N)$ to be a
running quantity with the expansion rate. Such notion of dynamical vacuum energy can be formulated in more formal grounds inspired in the renormalization group\,\cite{SS-old1,BabicET02,SS05,Fossil07,SS09} (cf. \cite{JSP-CCReview13,MiniReview11} for reviews),
or from the more phenomenological point of view of time-evolving $\CC=\CC(t)$ or cosmological vacuum decay\,\cite{OldLambdaTime,WangMeng04,AlcanizLima05}.

It is intriguing to think of the measured $\rL$ as tracing the energy density difference with respect to the flat spacetime vacuum, and if so $\CC$ should be of order of the present curvature, namely
$R\sim H_0^2$. This interpretation
has been emphasized in \,\cite{JSP-CCReview13} where a parallelism
is made with the Casimir effect. Recall that the net Casimir force
is caused by the difference between the vibrational modes of the QED
vacuum in between the plates as compared to its absence. Although
the zero-point energy (ZPE) itself is not measurable (as it is
infinite and this infinity is shared by the original vacuum state
before we introduce the plates), ``changes in the ZPE'' are indeed
detectable when we modify the boundary conditions. Similarly, we may
view the evolution of the vacuum energy in an expanding background
with (dynamical) curvature $R\sim H^2(t)$ as the change that
remains of the disturbed vacuum energy density in the curved
background after we remove the flat spacetime result -- which is
also contained in the curved spacetime
calculation. This possibility could help to understand the CC problem from the renormalization framework in QFT\,\cite{JSP-CCReview13}.

Models of the aforementioned kind and related variants have been
used to describe a smooth time-evolving vacuum energy density around
its present value, $\rLo\sim 10^{-47}$ GeV$^4$, and successfully
compared with the modern cosmological observations on the background
cosmology, including in some cases the effect of cosmic perturbations, see e.g.
%\,\cite{BPS09,GSBP11,FritzschSola2012,BasPolarSola12,BasSola13a,BasSola14a,RunCCperturbations,LXCDMperturbations1,LXCDMperturbations2,Opher07}.
\cite{SS-old2}-\cite{BasSola14a} and
\cite{RunCCperturbations}-\cite{Opher07}. For recent additional studies of the ZPE in curved spacetime, cf. \cite{Maggiore,Bilic}. Furthermore, a generalized
class of the dynamical vacuum models has recently been applied to construct a
complete history of the cosmological evolution starting from
inflation up to the present days\,\cite{H2H4,Essay2014}.

Dynamical models of the vacuum energy may ultimately be necessary
not only as a new paradigm to improve the theoretical status of the
$\CC$-term in Einstein equations, but also phenomenologically so as to relax a number of
tensions between the standard $\CC$CDM predictions and the
observations that may be providing evidence of DE evolution
(cf. the recent analysis of \cite{SahniShafielooStarobinsky2014}
based on the Baryonic Acoustic Oscillations as a tool to determine
the expansion history of the universe). Last but not least, dynamical vacuum energy could also be linked to the frequent hints reported in the literature that the so-called fundamental constants of Nature might be slowly changing with the cosmic time\,\cite{Flambaum2001,FundamentalConstants}, see e.g. \cite{FritzschSola,Sola2014} for a direct application of these ideas.

In this paper we focus on a large class of dynamical models of the vacuum energy inspired in QFT in curved spacetime, namely on those based on the structure $\rL(H,\dot{H})$ which is well-motivated for a Friedmann-Lema\^\i tre-Robertson-Walker (FLRW)-like expanding universe characterized by the Hubble rate $H$. We solve not only their background cosmology but perform a detailed analysis of the corresponding cosmic perturbations and their implications for structure formation. For example, it is well-known that the so-called linear growth rate (the logarithmic derivative of the
linear growth factor $D=\delta\rmr/\rmr$ with respect to $\ln a$, i.e. $f=d\ln D/d\ln a$), can be in some cases a
good indicator of clustering, together with the growth rate index
$\gamma$ (used as the effective parameterization of the growth rate
through a power of the density parameter\,\cite{Peebles1993}). In general,
$\gamma$ is a function of the scale factor or
equivalently of the redshift, $z=(1-a)/a$, and the relation with
the growth rate reads: $f(z)=\Om(z)^{\gamma(z)}$. The asymptotic
value of the growth rate index parameter takes distinctive values for
different gravity models\,\cite{Athina2014}. In the case of the concordance $\CC$CDM model, $\gamma(0)\simeq 0.55-0.60$. In this work we examine these linear indicators of structure formation for the new
models of the vacuum energy, but we find also very convenient to study  nonlinear effects, such as the theoretically predicted
cluster-size halo redshift distributions. These ``number count'' observables can help to break degeneracies between the vacuum models and can be especially useful in the context of realistic and future X-ray and Sunyaev-Zeldovich cluster surveys such as eROSITA\,\cite{eROSITA} and SPT\,\cite{SPT1,SPT2}, as shown in previous work\,\cite{BPS09,GSBP11}. For other implications of dynamical models in the astrophysical domain, see e.g. \cite{BPS09b}.

We shall discuss the virtues and troubles associated to some of
these vacuum models according to the Hubble terms involved in the dynamical structure of the vacuum energy. Let us emphasize that not
all of the $\rho_\CC(H,\dot{H})$ models are equally favored from the
theoretical point of view.  Interestingly, those that are
theoretically more favored are in fact the ones that best fit the
structure formation data in combination with the other cosmic
observables such as type Ia supernovae, the Cosmic Microwave
Background and the Baryonic Acoustic Oscillations. At the same time
we find models that perform comparatively
not so good, and other that can be simply excluded by the current observations.

The plan of the paper is as follows. In Sec. 1 we define the
theoretical framework for the dynamical vacuum models
$\rho_\CC(H,\dot{H})$. In Sec. 2 we discuss the general dynamical
vacuum models that depend on powers of the Hubble function and its time
derivative. In Sect. 3 we single out the class of the running
vacuum models, more closely related to QFT. The corresponding
cosmological background solutions of these models is presented in
Sect. 4. In the next section we formulate the linear cosmic
perturbations for general dynamical vacuum models. The fitting of these models to the cosmic data is put forward in Sect. 6, where we also briefly address the implications that our dynamical vacuum models could have on possible evidence recently found on dynamical dark energy. In Sec. 7 we discuss how to distinguish the dynamical vacuum models by means of the redshift distribution of cluster-size halos. Finally, in Sect. 8 we
provide our discussion and conclusions. In two appendices we furnish
more technical details of our analysis, to wit: in Appendix A we discuss more closely why the cluster number counts method can crucially help
to distinguish the dynamical vacuum models under consideration, and
in Appendix B we summarize the calculation of the linear density threshold for collapse, $\delta_c$, for the models under consideration, a
quantity that is crucially needed for the determination of the cluster-size halo redshift distributions.

%%%%%%%%%%%%%%%%%%%%%%%%%%%%%%%%%%%%%%%%%%%%%%%%%%%%%%%%%%%%
%%%%%%%%%%%%%%%%%%%%%%%%%%%%%%%%%%%%%%%%%%%%%%%%%%%%%%%%%%%%

%%%%%%%%%%%%%%%%%%%%%%%%%%%%%%%%%%%%%%%%%%%%%%%%%%%%%%%%%%%%
%%%%%%%%%%%%%%%%%%%%%%%%%%%%%%%%%%%%%%%%%%%%%%%%%%%%%%%%%%%%
\section{Time-evolving vacuum energy in an expanding universe}
\label{sect:TimeEvolvingVacuum}

Let us consider the expanding universe as a perfect fluid with
matter-radiation density $\rho_{m}$ and vacuum energy density $\rL$.
The latter is usually associated to the value of the cosmological
term through $\CC=8\pi\,G\,\rL$, where $G$ is Newton's constant.
While we assume that $G$ remains strictly constant here, we do not
make the same assumption for $\rL$. The full energy-momentum tensor
of the cosmic fluid can be written as  $\tilde{T}_{\mu\nu}\equiv
T_{\mu\nu}^m+T_{\mu\nu}^{\CC} $, where $T_{\mu\nu}^m$ is the
ordinary matter energy-momentum tensor and $T_{\mu\nu}^{\CC}$
describes the vacuum part. The equation of state (EoS) for the
matter component reads $p_{m}=\wm\,\rmr$ (with $\wm=0$ for dust and
$\wm=1/3$ for radiation), and we introduce also the corresponding
EoS for the vacuum $p_{\CC}=\wCC\rL$, which we discuss below.

In the flat FLRW metric, on which we will hereafter exclusively
concentrate,  the two independent gravitational field equations
derived from Einstein's equations sourced by $\tilde{T}_{\mu\nu}$
are the following: \be
 3H^2= 8\pi G (\rmr+\rL)  \label{friedr1} \ee
and \be 2{\dot H}+3H^2=- 8\pi G (\wm\rmr+\wCC\rL)\;, \label{friedr2}
\ee where the overdot denotes derivative with respect to cosmic time
$t$. From these equations one can derive the rate of change of the
Hubble function,
\begin{equation}\label{rchangeH}
\dot{H}=-4\pi\,G\left[\rmr\,(1+\wm)+\rL\,(1+\wCC)\right]\,.
\end{equation}
A useful equation (actually a first integral) that follows from the
original system (\ref{friedr1})-(\ref{friedr2}) is the following:
\begin{equation}
\dot{\rho}_{m}+\dot{\rho}_{\Lambda}+3 H(1+\omega_{m})\rho_{m}+3
H(1+\wCC)\rL=0\,. \label{Bianchi1}
\end{equation}
All the above equations remain valid if we sum over all matter
components (relativistic and nonrelativistic). In the frequent
situation where there is a dominant matter component (e.g. cold matter or relativistic matter), it is possible
to obtain the evolution law for the Hubble function in terms of the
vacuum term and that matter component:
 \be
\label{EHmonocomp}
\dot{H}+\frac32\,(1+\wm)\,H^2=4\pi\,G\left(\wm-\wCC\right)\,\rL=\frac12\,\left(\wm-\wCC\right)\,\CC\,.
\ee
By integrating this equation we can obtain $H$ in the relevant epoch
of the cosmic evolution where that matter component dominates. This procedure will be frequently used in
our analysis.

Equation (\ref{Bianchi1}) constitutes the local conservation of the
full energy-momentum tensor $\tilde{T}_{\mu\nu}$ in the presence of
all contributions of matter and vacuum energy, namely it expresses
explicitly the covariant conservation law
$\nabla^{\mu}\tilde{T}_{\mu\nu}=0$ in the FLRW metric. Such law
holds for strictly constant $G$ since the left hand side of Einstein
equations  must have zero covariant derivative by virtue of the
Bianchi identity. Being the result of an identity,
Eq.\,(\ref{Bianchi1}) is not independent of the fundamental system
of Friedmann-Lema\^\i tre equations (\ref{friedr1})-(\ref{friedr2}),
but it is useful to provide a more physical interpretation of them.

Formally the above equations are valid whether the vacuum term $\rL$
is a rigid quantity or is represented by a dynamical variable
$\rL=\rL(t)$. In the last case the energy density $\rL(t)$ could
e.g. involve the dynamical behavior of various possible components
of the dark energy and/or the effects of additional terms in the
effective action, as for example in the modified gravity framework
of Ref.\,\cite{Relax} or in the $\CC$XCDM model of \cite{LXCDM},
where in both cases the overall EoS  parameter $\wCC$ may have a
nontrivial behavior.  However, here we will assume that $\rL(t)$ is
a true dynamical vacuum term whose time evolution is exclusively
associated to the quantum effects on $\CC$\,\,\cite{SS-old1}. From
this point of view, the corresponding EoS is still $\wCC=-1$. This
will be henceforth taken for granted.

In the present study the dynamical CC term is represented by a power
series of the Hubble function and its time derivative:
\begin{equation}\label{powerH}
\CC(H)=c_0+\sum_{k=1}\alpha_k H^k+\sum_{k=1}\beta_k\dot{H}^k+...
\end{equation}
where the leading term $c_0$ describes in good approximation the
current universe and the other terms introduce a mild dynamical
evolution. As we will see in the next section, the expression
(\ref{powerH}) generalizes previous forms that can  be motivated within the class of running vacuum
models\,\,\cite{SS-old1}-\cite{MiniReview11}, with certain restrictions, and hence they
are placed in the general context of QFT in curved spacetime\,\cite{BirrellDavis,ParkerToms,ShapiroReview08} --
see\,\cite{JSP-CCReview13} for a recent review focusing on these issues.

Lately these models have been successfully applied to describe the
complete history of the universe, as they involve the ingredients
capable of yielding a smooth transition from an early de Sitter
stage to a proper radiation and matter epochs\,\cite{H2H4}. In
practice, since structure formation is a relatively recent
phenomenon, we limit ourselves to consider the lowest powers of $H$.
Indeed, recall that the current vacuum energy density is of order
$M_P^2\,H_0^2$, where $H_0\sim 10^{-42}$ GeV is the current value of
the Hubble parameter and  $M_P=1/\sqrt{G}\sim 10^{19}$ GeV is the
Planck mass (in natural units). It follows that the power terms of
(\ref{powerH}) beyond $H^2$ and $\dot{H}$ are completely irrelevant
for the present universe. Thus, terms of the form $H^3$, $\dot{H}
H$, $H^4$, $\dot{H}^2$, $H^2 \dot{H}$, $\ddot{H}$ etc. will be
ignored for the present study, although they can be important for
the early universe\,\cite{H2H4,Essay2014,LimaTrodden96}.

For the above models of dynamical vacuum energy the corresponding
Eq.\,(\ref{Bianchi1}) simplifies as follows:
\begin{equation}
\dot{\rho}_{m}+3(1+\omega_{m})H\rho_{m}=-\dot{\rho_{\Lambda}}\,.
\label{Bianchi}
\end{equation}
Despite the obtained simplification, the nonvanishing
\textit{r.h.s.} of this equation signals a transfer of energy
between vacuum and matter. Needless to say, this transfer is absent
in the $\CC$CDM model for which $\rL=$const. The nonvanishing time
derivative of $\rL$ in the above conservation law involves the
relevant powers of the Hubble function in the Eq.\,(\ref{powerH}).

In this study we would like to check the effect of all terms that
can be phenomenologically significant for the recent universe.
Therefore, we will consider the linear term in $H$ as well as the
$H^2$ and $\dot{H}$ terms. The linear terms, however, have a
different status. They are not expected to have a fundamental origin
within in QFT in curved space-time, a fact that actually applies to
all the odd powers of the Hubble function\,\cite{Fossil07,SS09} as
they are, in principle, incompatible with the general covariance of
the effective action. However, the linear terms appear in various
dark energy models in the presence of phenomenological bulk
viscosity, see
e.g.\,\cite{BViscosity,RenMeng06,Komatsu2013}\,\footnote{The linear
terms also appear if nonperturbative infrared effects would be
possible in the cosmological context, in a manner similar to QCD
where lower dimensional terms squared in the gauge field, $A^2$, can
appear together with the usual $F^2\sim (\partial A+A^2)^2$ ones,
owing to the effect of IR renormalons. This issue is far from being
established at present, but it has been considered in different ways
in the literature, see e.g.
\,\cite{OddHpowers,ArielZhitnitsky,Carn08,Carn14}.}.

Therefore, following our aim to explore the various existing
possibilities from the phenomenological point of view, we will test
here the following list of six types or classes of dynamical vacuum models effectively leading to time-evolving $\CC$ scenarios:

\begin{eqnarray}\label{A1A2B1B2C}
A1: \phantom{XXX} \Lambda&=&a_0+a_2 H^2\nonumber\\
A2: \phantom{XXX}\Lambda&=&a_0+a_1 \dot{H}+a_2 H^2\nonumber\\
B1: \phantom{XXX}\Lambda&=&b_0+b_1 H\nonumber\\
B2:\phantom{XXX} \Lambda&=&b_0+b_1 H+b_2 H^2 \\
C1:\phantom{XXX} \Lambda&=&c_1 H+c_2 H^2\nonumber\\
C2:\phantom{XXX} \Lambda&=&c_1 \dot{H}+c_2 H^2\nonumber
\end{eqnarray}
All of them can be, in principle, relevant for the study of the current and recent
past cosmic history. Type A and B models, despite their differences,
share one important feature, to wit: they all have a well-defined
$\CC$CDM limit since they all tend to a constant value of $\Lambda$
when the coefficients of the powers of $H$ or $\dot{H}$ tend to
zero. In contrast, models C1 and C2 can never behave as a rigid
$\CC$ term. As we will see, this has important consequences for the  phenomenological consistency of these models, when faced against the expansion and structure formation data.

%%%%%%%%%%%%%%%%%%%%%%%%%%%%%%%%%%%%%%%%%%%%%%%%%%%%%%%%%%%%%%%%%
%%%%%%%%%%%%%%%%%%%%%%%%%%%%%%%%%%%%%%%%%%%%%%%%%%%%%%%%%%%%%%%%%

\section{Running vacuum versus time-evolving vacuum}
\label{sect:RunningVacuum}

In the current study we focus mainly on the class of a running vacuum
energy models. By this denomination we mean not only that $\rL$ is a
time-evolving quantity, but more specifically we assume that its
evolution is inherited indirectly from another dynamical variable
$\mu=\mu(t)$ on which $\rL$ is tied to, i.e. $\rL=\rL(\mu)$, rather
than from a direct phenomenological law of the type $\rL=\rL(t)$.
This point of view reminds of the renormalization group (RG) running
of the effective charges in gauge theories, and has been put forward
in the context of cosmology in the literature\,
\cite{SS-old1,BabicET02,SS05,Fossil07,SS09}. In the cosmological context
$\mu$ is a characteristic infrared cutoff scale typically associated
to the Hubble rate, $H(t)$, as this quantity is of the order of the
energy scale associated to the FLRW metric.  In general $\mu^2$ is
in correspondence with  $H^2$ and also with $\dot{H}\equiv dH/dt$
(which has the same dimension as $H^2$). As shown in the previous
references, one expects a RG equation for the vacuum energy density
of the general form
\begin{equation}\label{seriesLambda}
\frac{d\rL(\mu)}{d\ln\mu}=\frac{1}{(4\pi)^2}\left[\sum_{i}\,B_{i}M_{i}^{2}\,\mu^{2}
+\sum_{i}
\,C_{i}\,\mu^{4}+\sum_{i}\frac{\,D_{i}}{M_{i}^{2}}\,\mu^{6}\,\,+...\right]\,,
\end{equation}
where $M_{i}$ are the masses of the particles contributing in the
loops, and $B_{i},C_i,..$ are dimensionless parameters. The equation
(\ref{seriesLambda}) gives the rate of change of the quantum effects
on the CC as a function of the scale $\mu$. In practice we may cut
off the series at the quadratic contributions, i.e. only the
``soft-decoupling'' terms of the form $\sim M_i^2\,\mu^2$ will
remain  for the current universe. The $M_i^4$ ones would trigger a
too fast running of the CC term. Such effects are actually forbidden
by the RG condition that only the fields satisfying $\mu>M_i$ are to
be included as active degrees of freedom. Now, since $\mu$ is
associated to a scale of order $H$ it is  clear that such condition
cannot be satisfied by any known particle mass in the current or
recent universe. On the other hand the quartic contributions and
higher are, as indicated above, are suppressed at this
epoch\,\footnote{The main contribution to the running of $\rL$
clearly comes from the heaviest fields in a typical GUT near the
Planck scale, i.e. those with masses $M_i\sim M_X\lesssim M_P$. See
\cite{Fossil07} for a specific scenario of this sort connected to the effective action of QFT in curved spacetime,
and \,\cite{JSP-CCReview13,MiniReview11} for a review.}.

Let us recall that because of the general covariance of the
effective action one expects only powers of $H^2$ and $\dot{H}$,
whereas the linear terms in $H$ (and in general any term with an odd
number of derivatives of the scale factor) are not
expected\,\cite{SS-old1,Fossil07,SS09}. This explains the general
structure of the above RG equation.

Integrating (\ref{seriesLambda}) and relating $\mu^2$ to a linear
combination of $H^2$ and $\dot{H}$ as the characteristic physical
scale for the running\,\cite{BasPolarSola12}, we can express the
leading terms of the result as follows:
 \be
\rL(H,\dot{H}) =n_0+n_{\dot{1}}\,\dot{H}+n_2\,H^2= \frac{3}{8\pi
G}\left( C_0 + \CHd {\dot H}+C_H H^2\right)\,, \label{GRVE} \ee
where the second expression is convenient for the use of the new
dimensionless coefficients $C_H$ and $\CHd$ that control the
dynamical character of the vacuum energy density.

If we have a look to the list  of vacuum models under study,
Eq.\,(\ref{A1A2B1B2C}), we observe that model A2 comprises the
framework (\ref{GRVE}) that we have previously motivated from the RG
equation, and model A1 is a particular case of A2. On the other hand
models B1 and B2 involve a linear term in H which, as previously indicated, is not expected on
fundamental grounds but could appear as an effective contribution.

Let us first turn our attention to model A1 in the list. It is the
simplest model containing the expected ingredients. It is convenient
to normalize the additive term such that it coincides with the value
of the current CC density for $H=H_0$, and in addition we introduce
a (dimensionless) parameter $\nu$, which plays the role of
coefficient of the $\beta$-function for the running of the vacuum
energy\,\cite{SS-old1}. In this way the CC density for model A1 can
be cast as follows:
\begin{equation}\label{RGlaw2}
 \rL(H)=\rLo+ \frac{3\nu}{8\pi}\,M_P^2\,(H^{2}-H_0^2)\,.
\end{equation}
As desired, $\rL(H=H_0)=\rLo$. For $\nu=0$ the vacuum energy
remains strictly constant at all times, $\rL=\rLo$, whereas for
non-vanishing $\nu$  there is an obvious evolution of the vacuum
energy that departs as $H^2$ from a strictly constant value. This is
a mild evolution provided $\nu$ is small enough. Obviously this
model is a particular case of (\ref{GRVE}) with $C_H=\nu$ and
$C_{\dot{H}}=0$. Substituting (\ref{RGlaw2}) in the general
acceleration law for a FLRW-like universe in the presence of a
vacuum energy density $\rL$, we find
\begin{equation}\label{vacuuma}
\frac{\ddot{a}}{a}=-\frac{4\pi\,G}{3}\,(\rmr+3\pmr-2\rL)=-\frac{4\pi\,G}{3}\,\,(1+3\wm)\rmr+C_0+\nu\,H^2\,,
\end{equation}
where in this case
\begin{equation}\label{C0}
C_0=\frac{8\pi G}{3}\,\rLo-\nu\,H_0^2\,.
\end{equation}
%%%%%%%%%%%%%%%%%%%%%%%%%%%%%%%%%%%%%%%%%%%%%%%%%%%%%%%%%%%%
%%%%%%%%%%%%%%%%%%%%%%%%%%%%%%%%%%%%%%%%%%%%%%%%%%%%%%%%%%%%
%%%%%%%%%%%%%%%%%%%%%%%%%%%%%%%%%%%%%%%%%%%%%%%%%%%%%%%%%%%%
%%%%%%%%%%%%%%%%%%%%%%%%%%%%%%%%%%%%%%%%%%%%%%%%%%%%%%%%%%%%

Let us next consider model A2. It generalizes the previous one by
introducing the $\dot{H}$  contribution. Recall that the homogeneous terms $H^2$ and
$\dot{H}$ are in general independent variables. From the two
Friedmann's equations (\ref{friedr1}) and (\ref{friedr2}) it is easy
to show that
\begin{equation}\label{ratioH2dotH}
\frac{H^2}{\dot{H}}=-\frac{2}{3}\,\frac{1+r}{1+\wm}\,,
\end{equation}
where $r=\rL/\rmr$ is the ratio between the vacuum
and matter energy densities. Even for the $\CC$CDM model (where $\rL$
is strictly constant) $r$ is a dynamical variable. At present $r\sim
{\cal O}(1)$ ($r\sim 7/3$), whereas in the past $r\to 0$. In the
radiation epoch $H^2$ was just minus half the value of $\dot{H}$,
whereas at present  $H_0^2$ is roughly minus twice the value of
$\dot{H}_0$.

The acceleration equation for the scale factor of the model class A2 reads
\begin{equation}\label{vacuumgeneral}
\frac{\ddot{a}}{a}=-\frac{4\pi\,G}{3}\,(1+3\wm)\rmr+C_0+\nu
H^2+\CHd\,\Hd\,.
\end{equation}
We still denote $\CH\equiv\nu$ since this parameter is closely
related to the simplest running vacuum model
(\ref{RGlaw2})-(\ref{vacuuma}) first introduced in \cite{SS-old1}.
Equation (\ref{vacuumgeneral}) can be rewritten in terms of the
deceleration parameter $q$ and the usual cosmological parameters
$\Om=\rmr/\rc$ and $\OL=\rL/\rc$, where $\rc$ stands for the
critical density $\rc=3H^2/8\pi G$. We find:
\begin{equation}\label{decelparam}
q=-\frac{\ddot{a}}{aH^2}=-1-\frac{\dot{H}}{H^2}=\frac12\,(1+3\wm)\Omega_m-\Omega_{\CC}\,.
\end{equation}
In the current epoch (where radiation can be safely neglected) we
obtain the following relation,
$\CHd\,\dot{H}_0=-\CHd\,(q_0+1)\,H_0^2=-(3/2)\,\CHd\,\Omo\,H_0^2$, which
is now helpful to determine $C_0$ in (\ref{vacuumgeneral}) after we
impose the boundary condition $\rL(H_0)=\rLo$ in (\ref{GRVE}):
\begin{equation}\label{C0CHd}
C_0=H_0^{2}\left(\OLo-\nu+\frac32\,\,\Omo\CHd\right)\,.
\end{equation}
This relation clearly generalizes Eq.\,(\ref{C0}) for nonvanishing
$\CHd$.

While models A (whether version A1 or the extended A2) are directly
related to the RG approach based on Eq.\,(\ref{seriesLambda}),
models B and C are more phenomenological by the reasons explained
before. We will solve the cosmological equations for all these
models in the next sections.

Before solving these models, let us stress that in the case of the
running vacuum models deriving from the general RG equation
(\ref{seriesLambda}), the solution compatible with the
general covariance of the effective action is of the form
\begin{equation}\label{ GeneralRG}
\rL(t)=c_0+\sum_{k=1} \alpha_{k} H^{2k}(t)+\sum_{k=1}
\beta_{k}\dot{H}^{k}(t)\,,
\end{equation}
with $c_0\neq 0$. This is the particular form that
Eq.\,(\ref{powerH}) takes for the running vacuum
case\,\cite{JSP-CCReview13}. That is to say, one obtains in general
an ``affine'' function constructed out of  powers of
$H^2=\left(\dot{a}/a\right)^2$ and $\dot{H}=\ddot{a}/a-H^2$, hence
with an even number of time derivatives of the scale factor $a$. The
higher order powers once more are irrelevant for the current
universe and for this reason the model types A,B,C singled out above
have been cut off at the lowest significant powers. As previously
emphasized, the higher powers of $H$ can play a very significant
role in the early universe. This role has been studied in detail in
Refs. \,\cite{H2H4,Essay2014}  where it is shown in particular that
they can lead to an inflationary scenario with graceful exit of the
vacuum phase (de Sitter regime) into the radiation phase. It means
that in this kind of dynamical vacuum fraweworks  one can formulate
a unified model of the cosmological evolution covering both the
early, the recent and the present universe. For the rest of the
paper we focus on the last two stretches of the cosmic history.

\mysection{Background solution of the cosmological equations}
\label{sect:solving}

Whenever possible we will solve the background equations for the
models (\ref{A1A2B1B2C}) by providing the matter and vacuum energy
densities, as well as the Hubble function, in terms of the
scale factor $a$  or the cosmic time. In general the most useful
form is in terms of the scale factor since this is the variable
which can be more easily related with the cosmological redshift
$z=(1-a)/a$. However this will not always be possible and in some
cases the analytic solution can be given only in terms of the cosmic
time. In this section we provide the analytical solution of the
background cosmologies corresponding to these models, obtained by extending the
analysis of Refs.\,\cite{BPS09}-\cite{BasSola14a} to which we refer
the reader for further details. The perturbations equations will be
analyzed in subsequent sections.

\subsection{Models A1 and A2} \label{sect:solvingA1A2}

We start from the local covariant conservation law (\ref{Bianchi})
in the matter dominated epoch and insert Eq.\,(\ref{GRVE}) on
its \textit{r.h.s.} A straightforward calculation making use of
(\ref{rchangeH}) and its time derivative yields: \be {\dot \rho}_m
+3 H \frac{ 1 - C_H}{ 1 - \frac{3}{2} C_{\dot H} }~\rho_m=0\,.
\label{generalizedConserv} \ee Trading the cosmic time variable
for the scale factor through $d/dt=aH d/da$ we can solve for the
energy densities as a function of the scale factor as follows:
\begin{equation}\label{MatterdensityCCtCDM}
\rho_m(a) =  \rmo~a^{-3 \xi}
\end{equation}
and
\begin{equation}\label{CCdensityCCtCDM}
\rL(a)=\rLo+{\rmo}\,\,(\xi^{-1} - 1) \left(a^{-3\xi} -1  \right)\,,
\end{equation}
with
\begin{equation}\label{defxiM}
\xi= \frac{ 1 - \nu }{ 1 - \alpha }\,,
%\simeq 1+\alpha-\nu\,,
\end{equation}
where $\nu=C_H$ as before, and we have introduced  $\alpha=3\CHd/2$.
Obviously, for $\alpha=0$ model A2 becomes model A1 (for which
$\xi=1-\nu$).

The corresponding Hubble function can now be obtained from the above
energy densities using (\ref{friedr1}), resulting in the following
expression:
\begin{equation}\label{HubbleA}
H^2(a) =H_0^2\,\left[1+\frac{\Omo}{\xi}\,\left(~a^{-3
\xi}-1\right)\right] \,.
\end{equation}
It satisfies the correct normalization $H^2(a=1)=H_0^2$.

Let us also compute the corresponding inflection point where
there is a transition of the Hubble expansion from the decelerating
to the accelerating regime. From the definition
(\ref{decelparam}) of deceleration parameter it is easy to show
that it can be rewritten as follows:
\begin{equation}\label{defqa}
q=-1-\frac{a}{2 H^2(a)}\,\frac{dH^2(a)}{da}\,.
\end{equation}
From this expression we can comfortably compute the point where
$q=0$ from Eq.\,(\ref{HubbleA}). Let us deliver the final result for
model A2 in terms of the redshift value at the transition point:
\begin{equation}\label{zIA}
z_I=\left[\frac{2(\xi-\Omo)}{(3\xi-2) \Omo}\right]^{1/3\xi}-1\,.
\end{equation}
The result for model A1 is just obtained by setting $\alpha=0$
(hence $\xi=1-\nu$). The standard $\CC$CDM result $z_I^{\CC{\rm
CDM}}$ is, as always, recovered for $\xi=1$:
\begin{equation}\label{zILCDM}
z_I^{\CC{\rm CDM}}=\left[\frac{2\OLo}{\Omo}\right]^{1/3}-1\,.
\end{equation}
The numerical value is $z_I^{\CC{\rm CDM}}\simeq 0.726 (0.645)$ for
$\Omo= 0.28 (0.31)$. As we can see the result is quite sensitive to the precision of $\Omo$ from
observation\,\footnote{$\Omo h^2 = 0.1426 \pm 0.0025$, with
$h=0.673\pm 0.012$
%$h=0.706\pm 0.032$
for the standard $\Lambda$CDM model from
Planck+WP\,\cite{PlanckXVI2013}.}.  Computing the departure of the
new transition point (\ref{zIA}) from the standard result for
small $\nu$ and $\alpha$, we obtain:

\begin{eqnarray}\label{aIRG2}
z_I-z_I^{\CC{\rm
CDM}}=(\nu-\alpha)\left[\frac{2\OLo}{\Omo}\right]^{1/3}\,
\left[1+\frac13\left(\ln\frac{2\OLo}{\Omo}-\frac{1}{\OLo}\right)\right]\,.
\end{eqnarray}
The numerical difference will be small to the extend that $\nu$ and
$\alpha$ are small. The corresponding fit to these parameters will
be made in Sect.\ref{sect:fitting}.

Although the above equations provide the exact background solution of type-A models
for arbitrary values of $\xi$, and hence of the original parameters
$\nu$ and $\alpha$, the natural range for these parameters is
\begin{equation}\label{naturalness}
|\nu|\ll 1\,,\ \ \ \ \ |\alpha|\ll 1\,.
\end{equation}
In this range the overall parameter $\xi$ can be expressed, in
linear approximation in terms of an effective $\nueff$ parameter:
\begin{equation}\label{smalllimit}
\xi=\frac{1-\nu}{1-\alpha}\simeq 1-(\nu-\alpha)\equiv 1-\nueff\,.
\end{equation}

\subsection{Models B1 and B2} \label{sect:solvingB1B2}

The vacuum energy density for model B2 can be parameterized
as follows:

\begin{equation}\label{B1B2}
\rho_{\Lambda}(H)=\frac{3}{8\pi G}(C_0+C_{1}H+C_{2}H^2)
\end{equation}
%
%\newline
The solution of model B1 obviously
ensues by simply setting $C_2=0$ in the background solution of model B2. However, the technical
difficulty in solving these models resides already in the simplest
``affine'' model B1 since the linear term in $H$ is harder to handle
than the quadratic one. On the other hand, phenomenologically the
reason to single out the case B1 is because this model is able to
reasonable fit the data provided we maintain the additive term
$C_0\neq 0$. For $C_0=0$, in contrast, the pure lineal model $\rL\propto
H$ is unable to accommodate the data on structure
formation\,\cite{BPS09,BasSola14a}. This feature has an important
impact on some theoretical and phenomenological proposals in the
literature\,\cite{OddHpowers,ArielZhitnitsky,Carn08,Carn14}.

In order to solve model (\ref{B1B2}) analytically we proceed as
follows. First of all it is convenient to re-express $\rL(H)$ in
terms of dimensionless coefficients. We set $C_1\equiv\epsilon H_0$
(where $\epsilon$ is dimensionless) and we continue identifying
$C_2$ with $\nu$, that is $C_2\equiv\nu$ as the basic parameter of
the simplest viable model (\ref{RGlaw2}) compatible with the RG
formulation. With this notation the expression that relates $C_0$
with $\nu$ and $\epsilon$ is:

\begin{equation}\label{C0TypeB}
C_0=\frac{8\pi
G}{3}\rLo-H_0^2(\epsilon+\nu)=H_0^2(\OLo-\epsilon-\nu)
\end{equation}
If we substitute (\ref{B1B2}) in the basic differential equation
(\ref{EHmonocomp}) for $H(t)$ in the matter dominated epoch, we
find
\begin{equation}\label{diffEq.TypeB}
\frac{2}{3}\dot{H}+\zeta\,H^2-\epsilon H_0 H=C_0
\end{equation}
where we have defined $\zeta=1-\nu$, not to be confused with
Eq.\,(\ref{defxiM}).
Upon direct integration we determine the Hubble function for this
model explicitly in terms of the cosmic time:
\begin{equation} \label{Hubblef}
H(t)=\frac{H_0}{2\,\zeta}\left[\mathcal{F}\,\coth\left(\frac{3}{4}H_0\mathcal{F}\,t\right)+\epsilon\right]\,,
\end{equation}
with
\begin{equation}\label{defmathF}
\mathcal{F}(\OLo,\epsilon,\nu)\equiv\sqrt{\epsilon^2+4\,\zeta(\OLo-\epsilon-\nu)}\,.
\end{equation}
Notice the presence of an additive constant term  in
(\ref{Hubblef}) (proportional to $\epsilon$) apart from the
hyperbolic one. This feature is precisely what makes the treatment
of the type-B models for $\rho_\Lambda(H)$ (the models with the
linear term in $H$) more complicated from the technical point of
view.

The time evolution of the  pressureless matter density can be
obtained from (\ref{rchangeH}) and the explicit form of the Hubble
function (\ref{Hubblef}), with the result:
\begin{equation}\label{rhom}
\rho_m(t)=-\frac{\dot{H}(t)}{4\pi G}=\frac{3H_0^2}{32\pi
G\,\zeta}\mathcal{F}^2\,{\rm
csch}^2\left(\frac{3}{4}H_0\mathcal{F}t\right)
\end{equation}
Similarly, from the previous equation and with the help of
(\ref{Hubblef}) and (\ref{friedr1}), we infer the expression of
the vacuum energy density:
\begin{equation}\label{rhoL}
\rho_\Lambda(t)=\frac{3H_0^2}{32\pi
G\zeta^2}\left[\mathcal{F}^2+\epsilon^2+2\epsilon\mathcal{F}\,\coth\left(\frac{3}{4}H_0\mathcal{F}t\right)+\nu
\mathcal{F}^2\,{\rm
csch}^2\left(\frac{3}{4}H_0\mathcal{F}t\right)\right]\,.
\end{equation}
Let us note that in the far past ($t\to 0$) we have
$\rL(t)/\rmr(t)\simeq \nu/(1-\nu)$ and therefore since $|\nu|\ll 1$
the vacuum energy is suppressed in this period, as expected. The
same conclusion applies in the radiation period, if we would include
the corresponding radiation term. On the other hand, from (\ref{rhom}) we have $\rmr(t)\to 0$ for $t\to\infty$  , as expected.

\noindent For this model it is impossible to obtain analytically the
matter and vacuum energy densities in terms of the scale factor,
except for $C_0=0$. We can however obtain $a(t)$
by direct integration of the Hubble function (\ref{Hubblef}):
\begin{equation}\label{atBtype}
a(t)=\left(\frac{[(2\zeta-\epsilon)^2-\mathcal{F}^2]^{1+\frac{\epsilon}{\mathcal{F}}}}{\mathcal{F}^2(\mathcal{F}+2\zeta-\epsilon)^{\frac{2\epsilon}
{\mathcal{F}}}}\right)^{\frac{1}{3\zeta}}e^{\frac{\epsilon
H_0}{2\,\zeta}t}\,\sinh^{\frac{2}{3\zeta}}\left(\frac{3}{4}H_0\mathcal{F}t\right)\,,
\end{equation}
where the complicated normalization factor (referred to as $A$ later on) is fixed by using
$H(t_0)=H_0$ and $a(t_0)=1$. We can check that for $\epsilon=\nu=0$
it reduces to the standard $\CC$CDM one.

We remark that the two basic parameters $(\epsilon,\nu)$,
or equivalently $(\epsilon,\zeta$), of type-B models are completely independent and cannot be mimicked by a single effective parameter in a given matter-dominated or radiation-dominated epoch. This is different from type-A
models, which can effectively be described by the unique parameter
$\xi$, defined in (\ref{defxiM}), in the matter-dominated epoch.

Finally, one can show that type-B models have also an inflection point very close to that of the standard model for small $|\epsilon|$ and $|\nu|$. In particular, for type-B2 with $\epsilon\ll\nu$ the transition point is essentially given by Eq.(\ref{zIA}) with $\xi=1-\nu$.

\subsection{Models C1 and C2} \label{sect:solvingC1C2}

Type C1 and C2 models are actually quite different from the previous
ones and at the same time different from each other. They have in
common the fact that do not have a well defined $\CC$CDM limit for
any value of the parameters, and therefore can never behave
sufficiently close to a pure $\CC$ model. This raises some doubts
about their possible viability, but they have nevertheless been
discussed in the literature for different theoretical and
phenomenological reasons.

For example, recently they have been discussed from the point of
view of their possible relation with the ``entropic-force''
scenario\,\cite{Verlinde10} and its implications for the dark
energy\,\cite{Easson10,Easson10b}. Sometimes type-C models are
presented in some drastically simplified forms as e.g. when only one
of the two terms is present (say, models of the form $\CC\propto H$
or $\CC\propto H^2$), see e.g.
\cite{OddHpowers,ArielZhitnitsky,Carn08,Carn14} and
\cite{ArcuriWaga94}. Here we will summarize very shortly the situation for completeness and also to highlight the important differences of type C with respect to type A and type B models.

Let us briefly comment on the model subclass C2 first. This is the canonical version of the mentioned class of models that acquired some
relevance recently in connection to the entropic-force scenario and
its possible cosmological implications\,\cite{Easson10,Easson10b}.
Many authors have analyzed recently this scenario for dark energy and
generalizations thereof, cf. e.g.
\,\cite{JapaneseHolog1,JapaneseHolog2,GrandaOliveros2008,KoivistoVisser,Entropic-others}. We mention in particular the studies in Ref.
\cite{BasPolarSola12,BasSola14a}, where it is shown in detail that model C2 is excluded. Basically, the acceleration parameter of this model remains constant,
i.e., it does not change with the expansion. Thus this kind of model
cannot have an inflection point from deceleration to acceleration.

In particular, models of the form $\rL\propto H^2$ or
$\rL\propto\dot{H}$ are definitely ruled out. Recently it has been
shown that even if the constant acceleration regime of model C2
would just be a partial description of a more complete model of
expansion where a transition point would exist, the corresponding
linear growth of cosmic perturbations is also strongly disfavored by
the current data\,\cite{BasSola14a}. This result puts this specific kind of entropic-force cosmologies against the wall.

Let us now move to model C1. Let us summarize briefly the situation -- (details are given in \cite{BasSola14a}. It is a particular case of B2 in the
limit $C_0\to 0$ of Eq.\,(\ref{B1B2}), where in this case $C_0$ is
defined in Eq.\,(\ref{C0TypeB}). Thus for this model we have
$\OLo=\epsilon+\nu$, which implies that $\epsilon$ and $\nu$ cannot
be both arbitrarily small. The Hubble function can be
expressed in this case in terms of the scale factor: \be\label{E2aC1} E(a)=1+
\frac{\Omo}{\zeta} \left(a^{-3\zeta/2}-1\right)\,, \ee where we have
defined the normalized Hubble rate $E(a)\equiv {H(a)}/{H_{0}}$. The matter and vacuum energy densities read:
\begin{equation}\label{rhomaC1}
\rmr(a)=\rco\left[\zeta\,E^2(a)-(\zeta-\Omo) E(a)\right]
\end{equation}
and
\begin{equation}\label{rhoLaC1}
\rL(a)=\rco\left[(1-\zeta)\,E^2(a)+(\zeta-\Omo) E(a)\right]\,.
\end{equation}
In the past
$\rL/\rmr\propto(1-\zeta)/\zeta=\nu/(1-\nu)$, and as a result we
must require $|\nu|$ to be sufficiently small to avoid domination of
vacuum energy. Since $\nu$ and $\epsilon$ cannot be
simultaneously small for C1, we must have $|\nu|\ll\epsilon$ for
this specific model class.

In contraposition to model C2, model C1 has a
well-defined inflection point in the expansion regime. It is located at redshift
\begin{eqnarray}
      \label{inflectionC1}
z_{I}^{\rm
(C1)}=\left[\frac{2(\zeta-\Omo)}{(3\zeta-2)\Omo}\right]^{2/3\zeta}-1\,.
\end{eqnarray}
Because of $|\nu|\ll\epsilon$, all C1 models have an inflection point near the case $\nu=0$. For this situation we find e.g. $z_I\simeq 1.979
(1.706)$ for $\Omo= 0.28 (0.31)$. These predictions are
substantially different from the $\CC$CDM ones, see
Eq.\,(\ref{zILCDM}). From
(\ref{rhomaC1}) and (\ref{E2aC1}) we see that, in the past, the
behavior of the matter density for the C1 models is abnormal:
$\rmr(a)\sim\rco \zeta\,E^2(a)\sim
\rco\,\left(\Omo\right)^2\,\zeta^{-1}\,a^{-3\zeta}$. Even for
$\nu=0$ (equivalently, $\zeta=1$, corresponding to the pure linear
model $\rL\propto H$) there is an extra factor of $\Omo$ as compared
to the $\CC$CDM. Because of this anomaly, when the model is
confronted with the cosmological data the preferred value of $\Omo$
is significantly larger than in the $\CC$CDM, see
\cite{BasSola14a} and \cite{BPS09}.

Needless to say, for a full assessment of the possibilities of the
various models we have to consider the analysis of cosmic
perturbations. We do this in Sect.\,\ref{sect:perturbations}.

\subsection{Including the effect of radiation} \label{sect:radiation}

Up to now we have not considered the effect of relativistic matter
in the solution of the background cosmological equations of the dynamical vacuum models since we
have focused on the form of the solution near our time. For the
study of cosmological perturbations  in Sect.\,\ref{sect:perturbations} the effect of radiation is not
necessary. However, for the fitting  of
the current models to the Baryonic Acoustic Oscillations (BAO)
and the Cosmic Microwave Background (CMB) in Sect.\ref{sect:fitting}  it is convenient to test this effect.

%\subsection{Radiation component for type-A models}
%\label{sect:radiationA}

Including the radiation component in Eq.\,(\ref{Bianchi1}) leads to
\begin{equation}\label{RadA}
{\dot \rho}_m + {\dot \rho}_r + {\dot \rho}_{\Lambda} + 3 H \rho_m +
4 H \rho_r=0\,.
\end{equation}
For type-A vacuum models, where $\rL$ is given by (\ref{GRVE}), we
obtain the following generalized form of
Eq.\,(\ref{generalizedConserv}):
\be\label{BianchitypeA2} {\dot \rho}_m +
3\,H\,\xi\,\rmr+\dot{\rho}_r+4\,H\,\xiR\,\rR=0\,, \ee where we used the
familiar parameter $\xi$, defined in (\ref{defxiM}), and we have now
introduced a new one that is related with the radiation component:
\begin{equation}\label{defxiR}
\xiR= \frac{ 1 - \nu }{ 1 - 4\alpha/3 }\,.
\end{equation}
We do not want to describe in detail the exchange dynamics of matter and radiation here (despite it was certainly relevant in some epoch of the universe), but only the situation when one of the components dominates.  We can therefore solve separately for the matter and radiation parts of (\ref{BianchitypeA2}) -- the same kind of assumption as in the standard model case. In this way we obtain:
\begin{eqnarray}
\rho_m &=& \rho_m^0 ~a^{-3 \xi} \label{splitAsolutionM} \\
\rho_r &=& \rho_r^0 ~a^{-4 \xiR}\,.\label{splitAsolutionR}
\end{eqnarray}
The presence of $\xi$ and $\xiR$, when different from $1$, denotes the anomaly in the corresponding conservation laws.  Of course such anomaly must be
small (i.e. $\xi\simeq \xiR\simeq 1$) in the natural physical region
(\ref{naturalness}), but the small deviation from $1$ is
exactly what permits the vacuum energy to evolve with the expansion.
By repeating the integration procedure, the vacuum energy density in the presence of radiation evolves as follows:
\begin{equation}\label{rLArad}
\rL(a)=\rLo+{\rmo}\,\,(\xi^{-1} - 1) \left( a^{-3\xi} -1  \right) +
{\rRo}\,\,(\xiR^{-1} - 1) \left( a^{-4\xiR} -1\right)\,.
\end{equation}
Similarly, the corresponding normalized
Hubble rate in the presence of radiation reads
\begin{equation}\label{HArad}
 E^2(a) =1+\frac{\Omo}{\xi}\left(a^{-3\xi}-1\right)+\frac{\Oro}{\xiR}\left(a^{-4\xiR}-1\right)\,,
\end{equation}
where the various normalized cosmological parameters satisfy now the
constraint
\begin{equation}\label{sumruleRad}
 \Omo + \Oro+\OLo = 1\,.\end{equation}
Obviously we have found a generalizion of the equations of
Sect.\,\ref{sect:solvingA1A2} for when the radiation component is
taken into account.

%\subsection{Radiation component for type-B models}
%\label{sect:radiationB}

Consider now type-B models.  As already mentioned, in this case we cannot avoid using the cosmic time variable for the analytical solution. Under similar assumptions as before, we obtain the corresponding matter densisties as follows:
\begin{equation}\label{rhomNR}
\rho_m(t,a)=\rmo\,e^{\frac{3}{2}\epsilon H_0(t-t_0)}\,a^{-3\zeta}
\end{equation}
\begin{equation}\label{rhorNR}
\rho_r(t,a)=\rRo\,e^{2\epsilon H_0(t-t_0)}\,a^{-4\zeta},
\end{equation}
where $\zeta=1-\nu$ and $t_0$ is the value of the cosmic time at present.

We assume that the effects of radiation are relatively small in the relevant periods of our interest and hence cannot modify dramatically the
matter-dominated solution. Such is the case around the time when the CMB was released. To generate a consistent solution
within this period, we initially assume that the scale factor evolves with time as in the nonrelativistic epoch and solve for it. Introducing the result in  (\ref{rhomNR}) and (\ref{rhorNR}) we obtain:
\begin{equation}\label{rhomNR2}
\rho_m(t)=\rmo\,B\,e^{-\frac{3}{2}\epsilon H_0 t_0}\,{\rm
csch}^2\left(\frac{3}{4}H_0\mathcal{F}t\right)\,,
\end{equation}
\begin{equation}\label{rhorNR2}
\rho_r(t)=\rro\,B^{4/3}\,e^{-2\epsilon H_0 t_0}\,{\rm
csch}^{8/3}\left(\frac{3}{4}H_0\mathcal{F}t\right)\,,
\end{equation}
where the normalization constant is $B=A^{-3\zeta}$, with $A$ the one in Eq.\,(\ref{atBtype}).

The Hubble function including the effects of radiation in the
matter-dominated epoch can now be computed from
\begin{equation}\label{HtypeBRad}
H^2(t)=\frac{8\pi G}{3}\left[\rmr(t)+\rR(t)+\rL(t)\right]=\frac{8\pi
G}{3}\left[\rmr(t)+\rR(t)\right]+C_0+C_1\,H(t)+C_2\,H^2(t)\,,
\end{equation}
where we have used (\ref{B1B2}) and we understand that $\rmr(t)$ and
$\rR(t)$ are given by (\ref{rhomNR2}) and (\ref{rhorNR2})
respectively. After some calculations we arrive at the following result:
\begin{equation}\label{hubbleimpro}
H(t)=\frac{H_0}{2\zeta}\,\left[\epsilon +\sqrt{\epsilon^2+4\zeta
s(t)}\right]=\frac{H_0}{2\zeta}\,\left[\epsilon
+\mathcal{F}\,\coth{\left(\frac{3}{4}H_0\mathcal{F}t\right)\,\sqrt{1+\Delta(t)}}\right]\,,
\end{equation}
where
\begin{equation}\label{Delta}
\Delta(t)=\frac{\Oro}{\Omo}\,\left(\frac{\mathcal{F}^2}{4\zeta\Omo}\right)^{1/3}\,{\rm
csch}^{2/3}\left(\frac{3}{4}H_0\mathcal{F}t\right){\rm
sch}^2\left(\frac{3}{4}H_0\mathcal{F}t\right)\,.
\end{equation}
The form of the result (\ref{hubbleimpro}) enables us to better compare with the original form\,(\ref{Hubblef}). The correction from the radiation is
clearly represented by the function ${\Delta}(t)$. Obviously this
effect is negligible at the present time since the prefactor in it reads $\Oro/\Omo=\left(1+0.227\,N_{\nu}\right)\,\Omega_{\gamma}^0/\Omo=4.15\times
10^{-5} h^{-2}/\Omo\simeq 3\times 10^{-4}$ (including photons and
$N_{\nu}=3$ neutrino species, and assuming $\Omo\,h^2\simeq 0.14$).
However, at decoupling (i.e. at the time $t_{*}$ of last scattering
of radiation with matter) the hyperbolic function in
(\ref{Delta}) rockets into a numerical value of order $\sim 10^3$ and $\Delta(t_{*})$ can become quite
sizeable. In fact, the fraction of radiation at decoupling can be
around $\sim 23\%$ in the $\CC$CDM case. This is of course also the
case for the generalized type A and B vacuum models deviating mildly
with respect to the $\CC$CDM (i.e. for small values of the
parameters).

From the numerical integration of (\ref{hubbleimpro}) we can obtain the corresponding improved version of the scale factor, and with it
the function $H(a)$ and the energy densities $\rmr(a)$ and $\rL(a)$ numerically.

\section{Linear perturbations for dynamical vacuum  models}
\label{sect:perturbations} After discussing the most relevant
aspects of the background cosmological solution, the next essential
step in our study is to analyze the linear perturbations equations.
The structure formation properties obviously play an essential role
to discriminate between the three kinds A,B and C of dynamical
vacuum models considered in this paper. In this section we discuss
the perturbations in the presence of a variable vacuum energy. While
we are not going to introduce perturbations for the vacuum energy
itself, only for matter, we incorporate the dynamical character of
$\rL$ in the matter perturbation equations. In other words,
$\rL=\rL(t)$ is time evolving, but homogeneous in first
approximation. This approach will suffice to clarify the fingerprint
differences between the A,B and C model types as far as structure
formation is concerned.

\subsection{Perturbation equations for dynamical vacuum models
$\rL=\rL(t)$} \label{sect:Generalperturbations}

We consider the linear perturbation equations for a system composed
of the dynamical vacuum fluid $\rL=\rL(t)$ and the matter fluid
$\rmr=\rmr(t)$, assuming that there are matter perturbations
$\delta\rmr$ but no perturbation in $\rL$. Although the inclusion of
perturbations of the $\rL$ component is
possible\,\cite{RunCCperturbations,LXCDMperturbations1,LXCDMperturbations2},
it is in general model-dependent and is not necessary for the
present study. In the synchronous gauge the relevant system of first
order differential equations in the epoch of structure formation can
be extracted from the general framework of \,\cite{LXCDMperturbations1} as follows\footnote{In previous studies\,\cite{RunCCperturbations} it
was found that the perturbation equations in two different gauges
(synchronous and conformal Newtonian) leads to perfectly consistent
results for this kind of models.}:
\begin{eqnarray}\label{diffsystem}
&&\ddot{h}+2\,H\dot{h}= - 8\pi G\delta\rmr\nonumber\\
&& \delta\dot{\rho}_m+\rmr \left(\tetm+\frac{\dot{h}}{2}\right)+3\,H\,\delta\rmr= 0 \\
&&\rmr\dot{\theta}_m+\left(\dot{\rho}_m+5H\rmr\right)\,\tetm =0\,,
\nonumber
\end{eqnarray}
where ${h}$ is the trace of the metric perturbation $\delta
g_{\mu\nu}$, and $\tetm=\nabla_{\mu}\delta U^{\mu}$ is the
divergence of the perturbed matter velocity.  The last equation of
the system (\ref{diffsystem}) can be cast as
\begin{equation}\label{difftheta}
\dot{\theta}_m+\left(2H+\Psi\right)\tetm=0\,,
\end{equation}
where we have used Eq.\,(\ref{Bianchi}), and defined
\begin{equation}\label{defQ}
\Psi\equiv
-\frac{\dot{\rho}_{\CC}}{\rmr}=3H+\frac{\dot{\rho}_m}{\rho_m}\,.
\end{equation}
Introducing the linear growth factor $D\equiv\delta\rmr/\rmr$, the first
two equations of the system (\ref{diffsystem}) can be combined with
(\ref{difftheta}) to yield the desired second order differential
equation for $D$ in the presence of a dynamical vacuum term:
\begin{equation}\label{diffeqD}
\ddot{D}+\left(2H+\Psi\right)\,\dot{D}-\left(4\pi
G\rmr-2H\Psi-\dot{\Psi}\right)\,D=\Psi\,\tetm\,,
\end{equation}
In the $\CC$CDM model we have $\rL=$const. and hence $\Psi=0$, so that
the above equation correctly shrinks to the standard
one\,\cite{Peebles1993}:
\begin{equation}\label{diffeqDCC}
\ddot{D}+2H\,\dot{D}-4\pi G\rmr\,D=0\,.
\end{equation}
We will henceforth set the \textit{r.h.s.} of (\ref{diffeqD}) to
zero.  For the vacuum models under consideration the parameters
$p_i=\nu,\alpha,\epsilon$ of the various models under consideration
are small, $|p_i|\ll 1$. Taking into account that $\theta_m$ is a perturbation term and that $\dot{\rho}_{\CC}$ is proportional to at least one  $p_i$ (as otherwise $\rL$ would remain constant and $\Psi=0$),  the product ${|\Psi \tetm|}\sim |{\cal O}(p_i)\tetm|$ is of second order and can be neglected.

In this approximation, we can rewrite the homogeneous form of
(\ref{diffeqD}) in terms of the scale factor as independent variable
as follows:
\begin{equation}\label{diffeqDa}
{D}''(a)+\left[\frac{3}{a}+\frac{H'(a)}{H(a)}+\frac{\Psi(a)}{aH(a)}\right]\,{D}'(a)-\left[\frac{4\pi
G\rmr(a)}{H^2(a)}-\frac{2\Psi(a)}{H(a)}-a\frac{\Psi'(a)}{H(a)}\right]\,\frac{D(a)}{a^2}=0\,,
\end{equation}
where primes here indicate $d/da$ differentiation, and we have traded the cosmic time variable for the scale factor through $d/dt=aH(a) d/da$. In particular, notice
that when $p_i=0$ we have $\Psi=0$ and (\ref{diffeqDa}) reduces to
the standard perturbation equation for the
$\CC$CDM\,\cite{Dodelson2003}:
\begin{equation}\label{diffeqDaLCDM}
{D}''(a)+\left[\frac{3}{a}+\frac{H'(a)}{H(a)}\right]\,{D}'(a)-\frac32\,\Omo\,\frac{H_0^2}{H^2(a)}\,\frac{D(a)}{a^5}=0\,,
\end{equation}
where in the last term of the \textit{l.h.s.} we have used the fact
that matter is covariantly conserved in the $\CC$CDM and hence
$\rmr=\rmo a^{-3}$. The decaying mode solution can be shown to be $D\propto H$, but this is not the one we want. The growing mode solution of
(\ref{diffeqDaLCDM}), which is the relevant one, is well-known and reads as follows:
\begin{equation}\label{soldiffeqDaLCDM}
D(a)=\frac52\,\Omo\,E(a)\,\int_0^a\,\frac{da'}{\left(a'\,E(a')\right)^3}\,,
\end{equation}
where as before $E(a)\equiv H(a)/H_0$. Early on in the matter
dominated epoch, when $E(a)=\sqrt{\Omo}a^{-3/2}$,
eq.(\ref{soldiffeqDaLCDM}) yields the standard result for the
linear growth factor: $D(a)\propto a$. The effect of a
nonvanishing $\rL>0$ is to suppress this linear growing rate, and in our case this effect is dynamical since $\rL=\rL(t)$.

\subsection{Perturbations for type-A models}
\label{sect:perturbationsTypeA}

In the following we solve the perturbation equation (\ref{diffeqDa})
for model A2 and then derive the solution of A1 as a particular
case. Recall that for these models the background solution can be
fully expressed in terms of the combined parameter $\xi$, which
depends on $\nu$ and $\alpha$ as indicated in (\ref{defxiM}). It
is convenient to introduce the following change of independent
variable, which can be operated on the cosmic time or the scale
factor as follows:
\begin{equation}\label{changet}
x=\coth\left[\frac{3}{2}\,H_{0}\sqrt{\xi\,(\xi-\Omo)}\;t\right]
\end{equation}
and
\begin{equation}\label{changea}
x^2=\frac{\xi}{\xi-\Omo}E^2(a)=1+\frac{\Omo}{\xi-\Omo}a^{-3\xi}\,.
\end{equation}
These changes of variable are associated respectively to the
perturbation equations (\ref{diffeqD}) and (\ref{diffeqDa}).
Starting from any of these equations and applying the corresponding
change of variable (\ref{changet}) or (\ref{changea}) we arrive,
after some lengthy algebra, at the following result:
\begin{eqnarray}\label{eqdiff}
3\xi^2 (x^{2}-1)^{2}\,\frac{d^2D(x)}{dx^2}&+&2\,\xi\,(6\xi-5)\,
x(x^{2}-1)\,\frac{dD(x)}{dx}\\ &-& 2\,[(2-\xi)\,(3\xi-2)\,x^{2}-\xi
(4-3\xi)]\,D(x)=0\,. \nonumber\end{eqnarray}
Notice that for $\alpha=0$ (hence $\xi=\zeta=1-\nu$) the previous
equation reduces to the one for the type-A1 model, first analyzed in
Ref.\,\cite{BPS09}. Thus, we have extended the perturbations
analysis of that reference so as to include the more general class
of models A2, which had not been considered before. We find that the
basic parameter of models A1 and A2 are $\zeta$ and $\xi$
respectively, and this holds both at the background and perturbation
levels. Remarkably, the basic perturbation equation turns out to be
formally the same in each model after exchanging the parameters
$\zeta ({\rm A1})\leftrightarrow\xi ({\rm A2})$. Not only so, this
means that type A2 model effectively behaves as a single parameter
model $\xi$ for all purposes.

The solution of \ref{eqdiff} can be expressed as follows:
\begin{equation}\label{solperturb}
D(x)=(x^2-1)^{\frac{5-3\xi}{6\xi}}Q_n^m(x),
\end{equation}
where $Q_n^m(x)$ is the associated Legendre's function of the second
kind,  and

\begin{equation}
m=\frac{1}{3\xi}-1\qquad n=\frac{1}{3\xi}.
\end{equation}
Using standard properties of the Legendre
functions\,\cite{Gradshteyn} (see also Appendix B of
Ref.\,\cite{BPS09}) and restoring the scale factor variable through
Eq.\,(\ref{changea}), we can finally express the solution within the
natural parameter domain (\ref{naturalness}) in the following way:
\begin{equation}
\label{solchangextoa} D(a)=A_1 a^{\frac{9\xi-4}{2}}E(a)\
F\left(\frac{1}{3\xi}+\frac{1}{2},\, \frac{3}{2};
\,\frac{1}{3\xi}+\frac{3}{2};\,-\frac{\xi-\Omo}{\Omo}\,\,a^{3\xi}
\right)\,,
\end{equation}
where $A_1$ is a constant to be adjusted by an initial condition,
and $F$ is the conventional hypergeometric series\,\,\cite{Gradshteyn}.

Recall that the natural range of the parameters for type-A models is given by Eq.\,(\ref{naturalness}), where $\nueff$ is the single effective parameter. Thus, in practice we
can replace $\xi\to\zeff$ in the equations (\ref{eqdiff}) and
(\ref{solchangextoa}), where
$\zeff\equiv1-\nueff$\,.

Let us note that, in the particular case $\nu=\alpha=0$ (hence
$\zeff=1$), the formula (\ref{solchangextoa}) leads to the
$\CC$CDM solution (\ref{soldiffeqDaLCDM}) after using a standard
integral representation of the hypergeometric
series\,\cite{Gradshteyn}. In the early epochs of matter domination,
i.e. for sufficiently large values of $a$ when the cosmological term
can be neglected and $E(a)\simeq \sqrt{\Omo}a^{-3/2}$, the solution
(\ref{solchangextoa}) takes on the simple form $D(a)\sim a$. This
behavior was also obtained previously from
(\ref{soldiffeqDaLCDM}) and it can be used as initial condition
to determine $A_1$.

Finally, let us mention that in Sect.\ref{sec:Number counts} (and
Appendix B), we will extend the structure formation analysis to the
non-linear regime (specifically to the formation of collapsed
structures).

\subsection{Perturbations for type-B and C models}
\label{sect:perturbationsTypeB}
The corresponding perturbations analysis starts in this case directly from Eq.\,(\ref{diffeqD}). It proves convenient the
following change of independent variable
\begin{equation}\label{defx}
y(t)=\coth\left(\frac{3}{4}H_0\mathcal{F}t\right)\,,
\end{equation}
The matter density (\ref{rhom}) in terms of the new variable reads
\begin{equation}\label{rhoy}
\nonumber \rho_m(y)=\frac{3H_0^2}{32\pi
G\zeta}\mathcal{F}^2(y^2-1)\,.
\end{equation}
Consider next the function $\Psi$ defined in (\ref{defQ}). With the
help of (\ref{rhoL}) and after straightforward algebra we
can write it also in terms of $y$ and express the result in a rather
compact form:
\begin{equation}\label{Qdot2}
\Psi(y)=-\frac{\dot{\rho}_{\CC}(t)}{\rmr(t)}=\frac{3
H_0}{2\zeta}\,\left[\epsilon+(1-\zeta)\mathcal{F}\coth\left(\frac{3}{4}H_0\mathcal{F}t\right)\right]=
\frac{3H_0}{2\zeta}\,\left[y(1-\zeta)\,\mathcal{F}+\epsilon\right]\,.
\end{equation}
With the above formulae and performing the substitution
(\ref{defx}) in the original Eq.\,(\ref{diffeqD}) we arrive at
the differential equation for the growth factor  $D(y)$ in the
new variable:
\begin{eqnarray}\label{perturbTypeB2}
&&(y^2-1)\frac{d^2D(y)}{dy^2}+\left[\frac{2y}{3\zeta}(6\zeta-5)-\frac{10\epsilon}{3\zeta\mathcal{F}}\right]\frac{dD(y)}{dy}\phantom{XXXXXXXXXXXXXXXXXX}\\
&&\phantom{XXXXXXXx}+\frac{2}{3\zeta}\left[\frac{4}{\zeta(y^2-1)}\left(\frac{\epsilon^2}{\mathcal{F}^2}+(1-\zeta)y^2+\frac{\epsilon}{\mathcal{F}}(2-\zeta)y\right)+3\zeta-4\right]D(y)=0\nonumber\,.
\end{eqnarray}
This equation cannot be solved analytically, not even in terms of
standard special functions. The case of the simpler type-B1
model, for which $\nu=0$, is no exception.
Thus, for both types B1 and B2 we are forced to use
the numerical techniques, for instance the standard method of finite
differences.

During the matter-dominated epoch $D(a)\propto a$, so we take as
initial conditions the value of the growth factor and its
derivative at very high redshifts $y_i=y(z_i\gg z^*)$ (recall that
at redshifts of order $z^*={\cal O}(1)$ the vacuum energy density
begins to dominate the Universe's dynamics for the $\CC$CDM and the
other models considered here, see Sect. \ref{sect:solving}). The
scale factor can be expressed in terms of $y$ with the help of
(\ref{atBtype}):
\begin{equation}
a(y)=A\,(y^2-1)^{-\frac{1}{3\zeta}}\left(\frac{y+1}{y-1}\right)^{\frac{\epsilon}{3\zeta\mathcal{F}}}\,.
\end{equation}
We normalize the growth factor with the value $D(z=0)$, i.e.
$D(a=1)$, and we take $D(a)=a$ at very high redshifts.

Concerning type-C1 models the perturbation equations have been studied in detail in Ref.\,\cite{BasSola14a}. We have explicitly checked that the results of that reference are correctly retrieved from the general Eq.\,(\ref{perturbTypeB2}) in the limit $C_0\to 0$ in Eqs.\,(\ref{B1B2})
and (\ref{C0TypeB}). Finally, we recall that in Ref.\,\cite{BasSola14a} the perturbations analysis of type-C2 model was also considered and it was shown that it does not lead to a growing mode solution for any reasonable value of the cosmological parameters, which is a fatal blow for the C1 class.

\section{Fitting the models to the observational data}
\label{sect:fitting}

\begin{figure}[!t]
%\begin{center}
%\includegraphics[scale=0.85]{BAO_d.ps}
\includegraphics[scale=0.65]{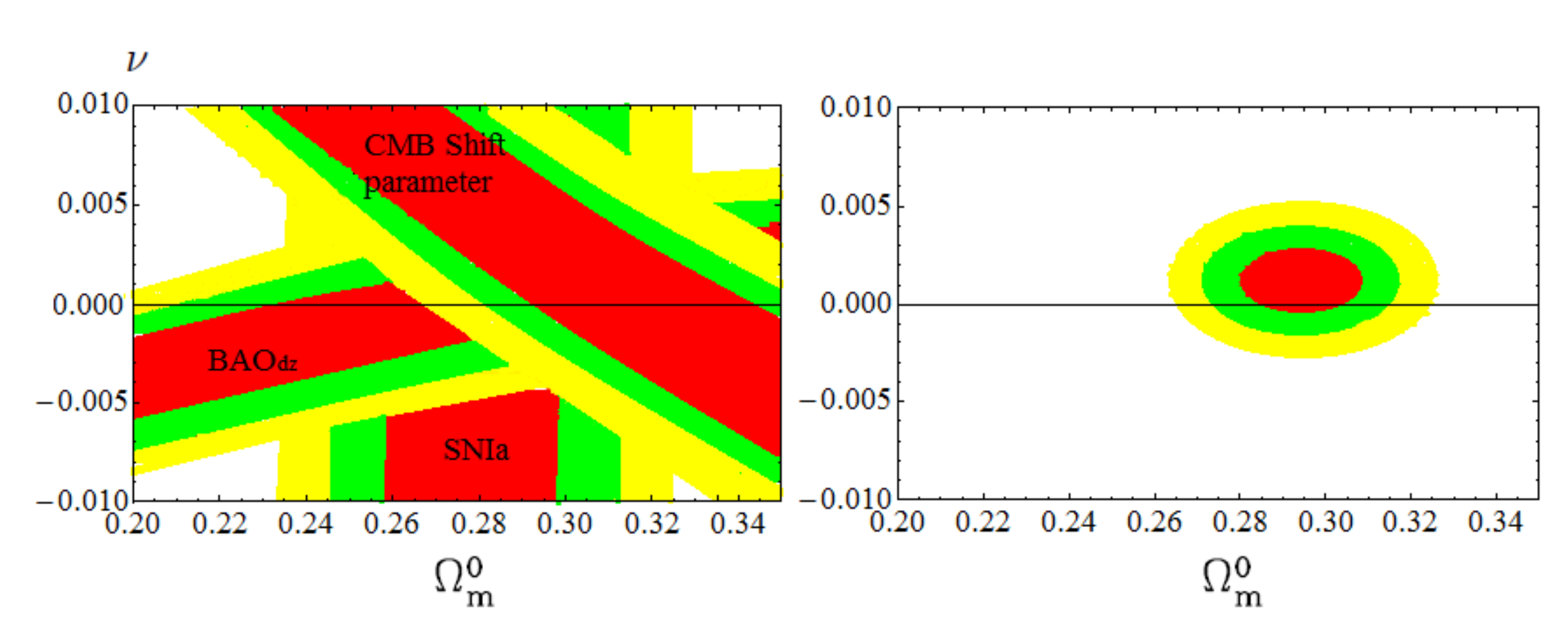}
\caption{\footnotesize{Likelihood contours (for $-2{\rm ln}{\cal
L}/{\cal L}_{\rm max}$ equal to 2.30, 6.16 and 11.81, corresponding
to 1$\sigma$, 2$\sigma$ and $3\sigma$ confidence levels) in the
$(\Omega_{m}^{0},\nu)$ plane for the A1
vacuum model ($\alpha\equiv 0$).
%The corresponding
%cosmological functions in the effective $\nueff$-approximation are
%given in eq.\,(\ref{expansionsnualpha}). For the CMB analysis we
%include also radiation as indicated in (\ref{mixturev}).
The left panel shows the contours based on the SNIa data (represented by approximate vertical bands), BAO$_{dz}$ (diagonal bands) and CMB shift parameter (antidiagonal bands). On the right panel we show the corresponding
contours based on the joint statistical analysis (SNIa+CMB+BAO$_{dz}$
data. From the inner to the outer regions (successively in red, green and yellow) we find the aforementioned 1$\sigma$, 2$\sigma$ and $3\sigma$ confidence levels, respectively.}\label{fig:contoursA1 BAOdz}}
%\end{center}
\end{figure}

\begin{figure}[!t]
\begin{center}
\includegraphics[scale=0.45]{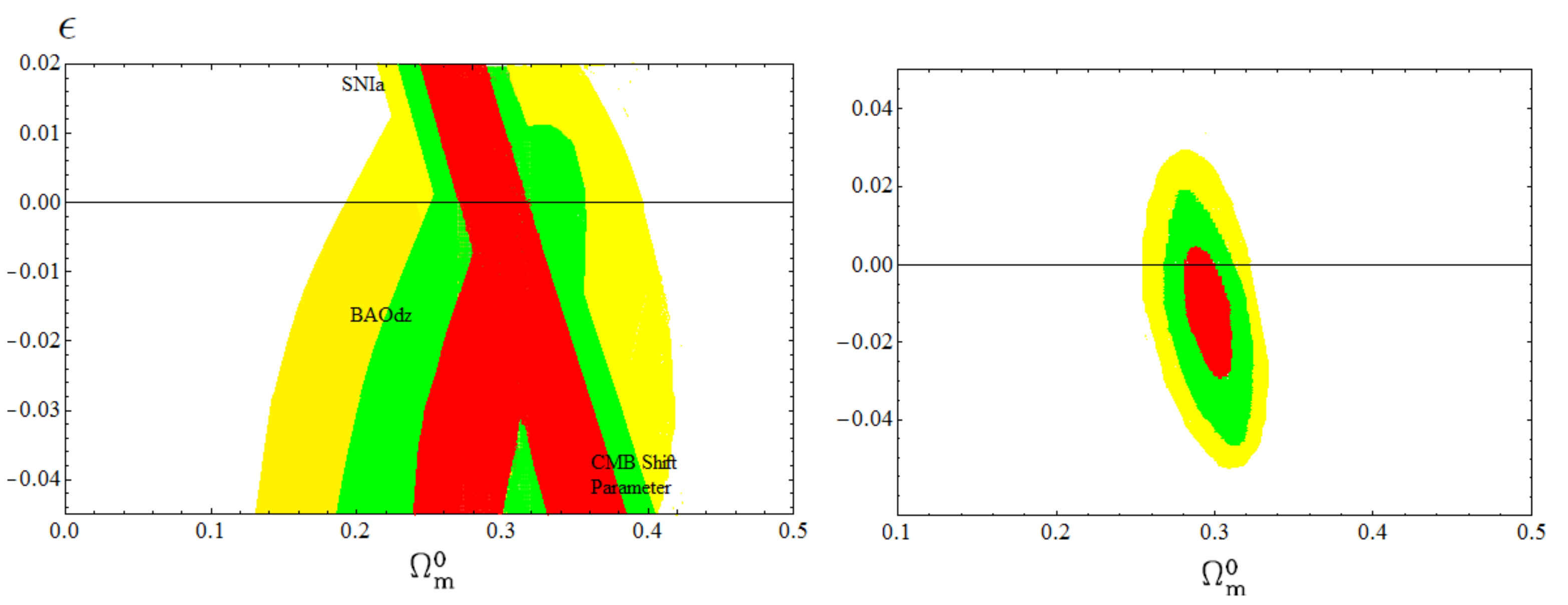}
\caption[]{\footnotesize{Likelihood contours for the B1
vacuum model, with BAO$_{dz}$ data.
The left panel shows the contours based on the SNIa data, BAO$_{dz}$ and CMB shift parameter. The meaning of the shaded regions is as in Fig.\,\ref{fig:contoursA1 BAOdz}. The various bands follow a similar pattern as in the previous figure, but here the overlapping of the SNIa and BAO regions is larger. The right panel shows
the joint contours of SNIa+CMB+BAO$_{dz}$.}\label{fig:contoursB1 BAOdz}}
\end{center}
\end{figure}

\begin{figure}[!t]
\begin{center}
\includegraphics[scale=0.42]{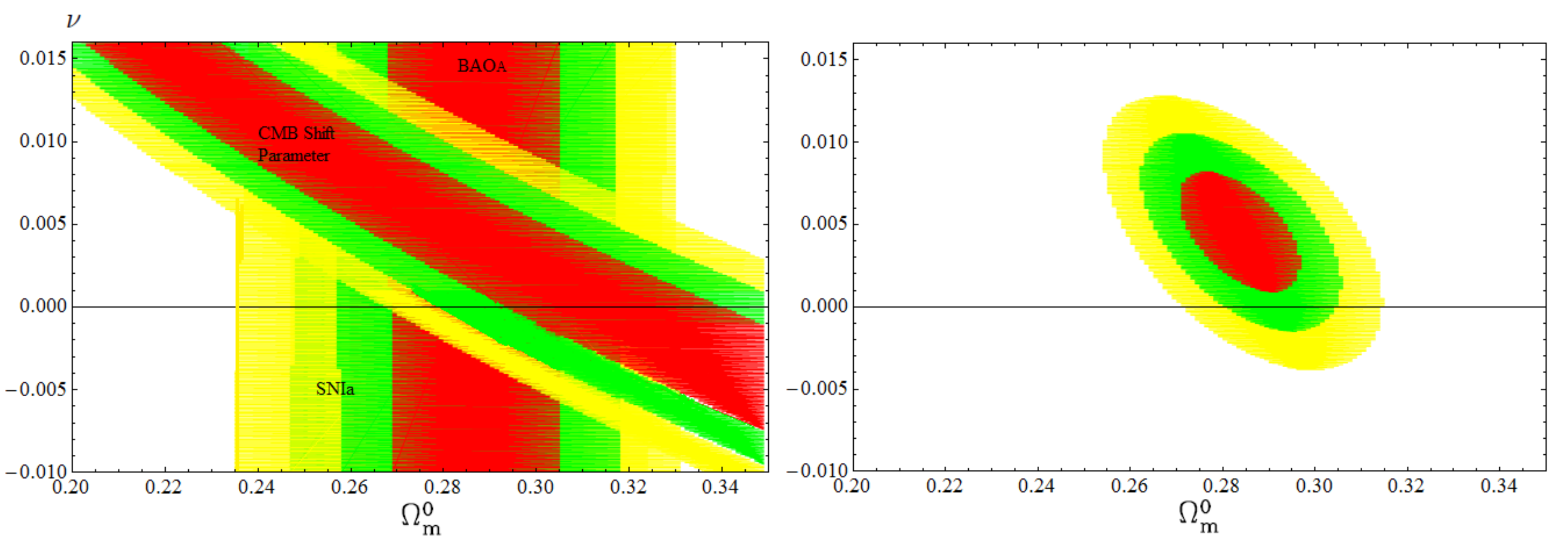}
\caption[]{\footnotesize{Likelihood contours for the A1
vacuum model, this time with BAO$_A$ data.
The left panel shows the contours based on the SNIa data, BAO$_{A}$ and CMB shift parameter indicated in a similar way as in Fig.\,\ref{fig:contoursA1 BAOdz}. The SNIa data and BAO$_{A}$ bands appear as almost vertical and with significant overlap. The right panel shows
the joint contours of SNIa+CMB+BAO$_{A}$. A well defined final region is projected with $\nu>0$} at the $1\sigma$ level.}
\label{fig:contoursA1 BAOA}
\end{center}
\end{figure}

\begin{figure}[!t]
\begin{center}
\includegraphics[scale=0.60]{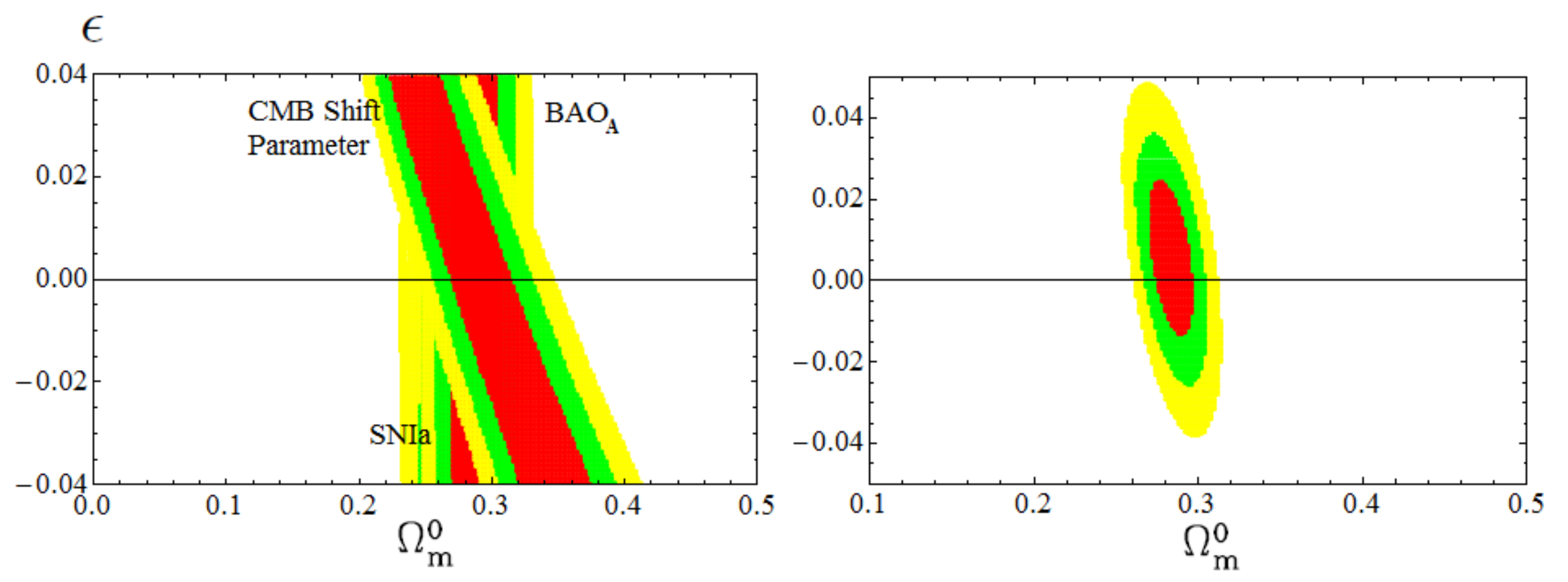}
\caption{\footnotesize{Likelihood contours for the B1
vacuum model, in this case with BAO$_A$ data.
The meaning of the shaded regions in the two panels is as in the previous figures.}\label{fig:contoursB1 BAOA}}
\end{center}
\end{figure}

In the following, we describe the statistical method, the
observational samples and data statistical analysis that will be
adopted to constrain the parameters of the dynamical vacuum models
presented in the previous sections. We extract our fit from the
combined data on type Ia supernovae (SNIa), the shift parameter of
the Cosmic Microwave Background (CNB), and the data on the Baryonic
Acoustic Oscillations (BAOs). The basic fitting results to the dynamical models under consideration are presented in a nutshell in Figs.\,\ref{fig:contoursA1 BAOdz}-\ref{fig:contoursB1 BAOA} and Tables 1 and 2. We devote the rest of this section to explain these results and also to analyze the implications for linear structure formation.

\subsection{The global fit to SNIa, CMB and BAOs}

First of all, we use the {\em Union 2.1} set of 580 type Ia supernovae
of Suzuki et al.~\cite{Suzuki:2011hu}.
%Amanullah et al. ~\cite{Ama11}\,\footnote{Note that the data can
%be found in: http://supernova.lbl.gov/Union/.}.
The corresponding
$\chi^{2}_{\rm SNIa}$ function, to be minimized, is:
\begin{equation}\label{xi2SNIa}
\chi^{2}_{\rm SNIa}({\bf p})=\sum_{i=1}^{580} \left[ \frac{ {\cal
\mu}_{\rm th} (z_{i},{\bf p})-{\cal \mu}_{\rm obs}(z_{i}) }
{\sigma_{i}} \right]^{2} \;,
\end{equation}
where $z_{i}$ is the observed redshift for each data point. The
fitted quantity ${\cal \mu}$ is the distance modulus, defined as
${\cal \mu}=m-M=5\log{d_{L}}+25$, in which $d_{L}(z,{\bf p})$ is the
luminosity distance:
\begin{equation}\label{LumDist}
d_{L}(z,{\bf p})={c}{(1+z)} \int_{0}^{z} \frac{{\rm d}z'}{H(z')}\;,
%d_{L}(z,{\bf p})=\frac{c}(1+z) \int_{0}^{z} \frac{{\rm
%d}x}{H(x)} \;,
\end{equation}
with $c$ the speed of light (now included explicitly in some of these formula for better clarity) and ${\bf p}$ a vector containing the
cosmological parameters of the models that we wish to fit for. In
equation (\ref{xi2SNIa}), the theoretically calculated distance
modulus $\mu_{\rm th}$ for each point follows from using
(\ref{LumDist}), in which the Hubble function is the one
corresponding to each model, see Sect.\,\ref{sect:solving}. Finally,
$\mu_{\rm obs}(z_{i})$ and $\sigma_i$ stand for the measured
distance modulus and the corresponding $1\sigma$ uncertainty for
each SNIa data point, respectively. The previous formula
(\ref{LumDist}) for the luminosity distance applies only for
spatially flat universes, which we are assuming throughout.

%%%%%%%%%%%%%%%%%%%%%%%%%%%%%%%%%%%%%%%%%%%%%%%%%%%%%%%%%%%%%%%%%

%%%%%%%%%%%%%%%%%%%%%%%%%%%%%%%%%%%%%%%%%%%%%%%%%%%%%%%%%%%%%%%%%%%%%%%%%%%%%

Furthermore, a very accurate and deep geometrical probe of dark
energy is the angular scale of the sound horizon at the last
scattering surface (i.e. at the time of decoupling of radiation from
matter). The probe is described by the  CMB ``shift
parameter''~\cite{Bond:1997wr,Nesseris:2006er} and is encoded in the
location $l_1^{TT}$ of the first peak of the Cosmic Microwave
Background (CMB) temperature perturbation spectrum. It provides the reduced distance to the last scattering surface. For spatially
flat cosmologies it is given by
\begin{equation}\label{shiftparameter}
R=\sqrt{\Omega_{m}^0}\int_{0}^{z_{*}} \frac{dz}{E(z)}\,.
\end{equation}
The measured shift parameter according to the Planck
data~\cite{PlanckXVI2013,Shaf13} is $R=1.7499\pm 0.0088$ at the redshift of decoupling (viz. at the last
scattering surface), $z_{*}$. Its precise value depends weakly on
the parameters, and it is obtained from the fitting
formula\,\cite{HuSugiyama}:
\begin{equation}\label{zlastscatt}
z_{*}=1048\,\left[1+0.00124\,\left(\Omega_b^0
h^2\right)^{-0.738}\right]\left[1+g_1 \left(\Omo
h^2\right)^{g_2}\right]\,,
\end{equation}
with
\begin{equation}
g_1=0.0783\,\frac{\left(\Omega_b^0
h^2\right)^{-0.238}}{1+39.5\,\left(\Omega_b^0
h^2\right)^{0.763}}\,,\ \   \qquad
g_2=\frac{0.560}{1+21.1\left(\Omega_b^0 h^2\right)^{1.81}}\,.
\end{equation}

In this case, the $\chi^{2}_{\rm CMB}$ function is given by:
\begin{equation}
\chi^{2}_{\rm CMB}({\bf p})=\frac{[R({\bf
p})-1.7499]^{2}}{0.0088^{2}}\;.
\end{equation}
As emphasized in the previous section, when dealing with the CMB
shift parameter we have to include both the matter and radiation
terms in the total normalized matter density entering the $E(z)$
function in (\ref{shiftparameter}) since the total radiation
contribution at the last scattering amounts to some $\sim 23\%$ of
the total energy density associated to matter and is therefore not entirely negligible. It means that when we
compute the CMB shift parameter we have to use the modified formulas
for the Hubble function that we  have found in Sect. \ref{sect:radiation}
for type A and B
models respectively.

Finally, we also consider the BAO scale produced in the last
scattering surface by the competition between the pressure of the
coupled baryon-photon fluid and gravity. The resulting acoustic
waves leave (in the course of the evolution) an overdensity
signature at certain length scales of the matter distribution.
They appear as regular, periodic fluctuations of visible matter density in large-scale structure (LSS) resulting from sound waves propagating
in the early Universe.
Evidence of this excess has been found in the clustering properties
of the SDSS galaxies (see~\cite{Eis05,Perc10,Blake11}). In recent years, measurements of BAO have proven useful as a ``standard ruler'' or geometric probe that we can employ to constrain dark energy models.

In this work we use the results of Blake et al. \cite{Blake11} (cf. Table 3 of this reference)
which are given in terms of the
parameter $d_z(z_{i})=r_{s}(z_{d})/D_{\rm V}(z_{i})$, where
$D_{V}(z_{i})$ is the effective distance measure
\cite{Eis05} and $z_{i}$ is a reference redshift for
observations. Moreover $r_{s}(z_d)$ is the comoving sound horizon
size at the baryon drag epoch\,\cite{Eis98} (i.e. the epoch at which
baryons are released from the Compton drag of photons), and
$z_{d}\sim {\cal O}(10^{3})$ is the corresponding redshift of that
epoch, closely related to that of last scattering-- the precise
expression is given below, see Eq.\,(\ref{zdrag}).

%%%%%%%%%%%%%%%%%%%%%%%%%%%%%%%%%%%%%%%%%%%%%%%%%%%%%%%%%%%%%%%%%%%%%%%%%%%%%%%%

%%%%%%%%%%%%%%%%%%%%%%%%%%%%%%%%%%%%%%%%%%%%%%%%%%%%%%%%%%%%%%%%%%%%%%%%%%%%%

Since $r_{s}(z_d)$ is the comoving distance that light can
travel prior to redshift $z_d$, it can be computed as follows:
\begin{equation}
\label{drag}
r_{s}(z_{d})=\int_{0}^{t(z_d)}\,\frac{c_s\,dt}{a}=\int_{0}^{a_{d}}
\frac{c_s(a)\,da}{a^{2} H(a)}=\int_{z_d}^{\infty}
\frac{c_s(z)\,dz}{H(z)}\;,
\end{equation}
where $a_{d}=(1+z_{d})^{-1}$, and
\begin{equation}\label{cs2}
c_s(a)=c\,\left(\frac{\delta {p}_{\gamma}}{\delta
{\rho}_{\gamma}+\delta {\rho}_b}\right)^{1/2}=
\frac{c}{\sqrt{3\,\left(1+{\cal R}(a)\right)}}
\end{equation}
is the sound speed in the baryon-photon plasma. Here we assume
adiabatic perturbations and we have used $\delta {p}_b=0$ and
$\delta {p}_{\gamma}=(1/3)\,\delta{\rho}_{\gamma}$, and defined
${\cal R}(a)=\delta{\rho}_b/\delta{\rho}_{\gamma}$. If the scaling
laws for non-relativistic matter and radiation were those of the
standard model, we would have ${\cal
R}(a)=3\rho_b/4\,\rho_{\gamma}$, which can be finally cast as ${\cal
R}^{\CC
CDM}(a)=\left({3\Omega_{b}^{0}}/{4\Omega_{\gamma}^0}\right)\,a$,
where $\Omega_{b}^{0}h^{2}\simeq 0.02205$ and
$\Omega_{\gamma}^0\,h^2\simeq 2.46\times 10^{-5}$ are the current
values of the normalized baryon and photon
densities\footnote{We use
$\Omega_{r}^{0}=4.153\times 10^{-5}h^{-2}$ \cite{Shaf13},
$\Omega_{\gamma}^{0}=\frac{\Omega_{r}^{0}}{1+0.2271N_{\nu}}$
(with $N_{\nu}\simeq 3.04$ and $h=0.673$ \cite{PlanckXVI2013}).}.
However, when we consider cosmologies beyond the $\CC$CDM a modification of these
formula for ${\cal R}(a)$ has to be implemented.  We explain the
details in the next section.

%%%%%%%%%%%%%%%%%%%%%%%%%%%%%%%%%%%%%%%%%%%%%%%%%%%%%%%%%%%%%%%%%%%%%%%%%%%%%

%%%%%%%%%%%%%%%%%%%%%%%%%%%%%%%%%%%%%%%%%%%%%%%%%%%%%%%%%%%%%%%%%%%%%%%%%%%%%

The remaining ingredients of the BAO analysis are as in the standard
case, in particular the effective distance is (see \cite{Eis05}):
\begin{equation}
D_{\rm V}(z)\equiv \left[ (1+z)^{2} D_{A}^{2}(z)
\frac{cz}{H(z)}\right]^{1/3}\;,
\end{equation}
where $D_{A}(z)=(1+z)^{-2} d_{L}(z,{\bf p}) $ is the angular
diameter distance. It follows from the foregoing that the $d_z$ estimator for BAO analysis explicitly reads:
\begin{equation}\label{fromula dz}
d_z(z_i)=\frac{r_s(z_d)}{\displaystyle{\left[\left(\int_0^{z_i}\frac{cd{z}}{H({z})}\right)^2\,\left(\frac{cz_i}{H(z_i)}\right)\right]^{1/3}}}\,.
\end{equation}
The fitted formula for the baryon drag redshift, $z_d$, is
given by\,\cite{Eis98}:
\begin{equation}\label{zdrag}
z_d=1291\,\frac{(\Omo\,h^2)^{0.251}}{1+0.659(\Omo\,h^2)^{0.828}}[1+\beta_1(\Omega_b^{0}\,h^2)^{\beta_2}]\,,
\end{equation}
with
\begin{equation}
\beta_1=0.313(\Omo\,h^2)^{-0.419}[1+0.607(\Omo\,h^2)^{0.674}]\,,\ \
\qquad \beta_2=0.238(\Omo\,h^2)^{0.223}\,.
\end{equation}
As we see the drag epoch ends at a redshift which is somewhat more
strongly dependent on the parameters than the decoupling
redshift $z_{*}$. Numerically, $z_d$  is not very different from
$z_{*}$, both being of order $10^3$ (with $z_{*}>z_d$). Typically $z_{*}\simeq 1090$ and $z_d\simeq 1060$ for the Planck results\,\cite{PlanckXVI2013}.

At this point, we would like to stress that
Blake et al. \cite{Blake11} also provide
BAO measurements in terms of the acoustic parameter $A(z)$,
first introduced by Eisenstein et al. \cite{Eis05}. Acoustic oscillations in the photon-baryon plasma prior to recombination give rise to a peak in the correlation function of galaxies, whose value is given by the mentioned $A(z)$-estimator for BAO analysis:
\begin{equation}\label{defBAOA}
A({z_i,\bf p})=\frac{\sqrt{\Omo}}{[z_i^{2} E(a_{i})]^{1/3}}
\left[\int_{a_{i}}^{1} \frac{da}{a^{2}E(a)} \right]^{2/3}=\frac{\sqrt{\Omo}}{E^{1/3}(z_{i})}
\left[\frac{1}{z_i}\int_{0}^{z_i} \frac{dz}{E(z)} \right]^{2/3}\,,
\end{equation}
with $a_{i}=(1+z_{i})^{-1}$, and $z_i$ is the redshift at which the acoustic scale has been measured.
%
%%%%%%%%%%%%%%%%%%%%%%%%%%%%%%%%%%%%%%%%%%%%%%%%%%%%%%%%%%%%%%%%%%%%%%%%%%%%%

%%%%%%%%%%%%%%%%%%%%%%%%%%%%%%%%%%%%%%%%%%%%%%%%%%%%%%%%%%%%%%%%%%%%%%%%%%%%%%
%
According to \cite{Blake11} the $A(z)$ measurements are
approximately uncorrelated with respect to $\Omega_{m}^{0}h^{2}$, while
this is not the case for the $d_{z}$ measurements.
Therefore, it is natural to use both BAO estimators, $d_{z}$ and $A(z)$,
in our statistical analysis (confer section 6.3)
in order to check the
range of validity of the free parameters included in the various
vacuum models. Specifically, here and henceforth
we consider the following
two notations: BAO$_{dz}$ for the $d_{z}$ measurements, and
BAO$_A$ for those based on the $A(z)$ estimator.
Therefore, the corresponding $\chi^{2}$-functions for BAO analysis are defined as:
\begin{equation}\label{BAOd}
\chi^{2}_{\rm BAO_{dz}}({\bf p})=\sum_{i=1}^{6} \left[ \frac{ d_{z,\rm th} (z_{i},{\bf p})-d_{z,\rm obs}(z_{i})}
{\sigma_{z,i}} \right]^{2}
\end{equation}
and
\begin{equation}\label{BAOA}
\chi^{2}_{\rm BAO_{A}}({\bf p})=\sum_{i=1}^{6} \left[ \frac{ A_{\rm th} (z_{i},{\bf p})-A_{\rm obs}(z_{i})}
{\sigma_{A,i}} \right]^{2} \;,
\end{equation}
where $z_{i}$, $d_{z,\rm obs}$, $\sigma_{z,i}$, $A_{\rm obs}$ and $\sigma_{A,i}$
can be found in Table 3 of \cite{Blake11}.

\subsection{Adapting the BAO analysis for dynamical vacuum models}

In this section we describe the necessary modifications to ${\cal
R}(a)$ for the BAO$_{dz}$  analysis when the cosmological vacuum is dynamical. The modifications are
necessary since the scaling laws for non-relativistic matter and
radiation are slightly different as compared to the $\CC$CDM. For
example, for type-A models the generalization is simple
\,\cite{BasPolarSola12}. From the anomalous scaling laws
(\ref{splitAsolutionM}) and (\ref{splitAsolutionR}) we easily find
the following modification of the ${\cal R}^{\CC CDM}(a)$
function:

\begin{equation}\label{Ra2}
{\cal R}^{({\rm type
A})}(a)=\frac34\,\frac{\xi}{\xiR}\,\frac{{\rho}_b(a)}{\,{\rho}_{\gamma}(a)}
=\frac34\,\frac{1-4\alpha/3}{1-\alpha}\frac{{\Omega}_{b}^{0}}{{\Omega}_{\gamma}^0}\,
a^{4\xiR-3\xi}\,.
\end{equation}
Of course for $\nu=0$ and $\alpha=0$ ($\xi=\xiR=1$) we recover ${\cal R}^{\CC CDM}(a)$ as given in the previous
section.

For type-B models the corresponding ${\cal R}^{\CC CDM}(a)$ also
differs from the standard one, but is more complicated. For this model we have to use the cosmic
time rather than the scale factor, i.e.
$R(t)={\delta\rho_b(t)}/{\delta\rho_{\gamma}(t)}$. Thus, upon differentiating
(\ref{rhomNR2}) and (\ref{rhorNR2}) with respect to $t$, and after some calculations we obtain:
\begin{equation}\label{eq:RtypeB}
{\cal R}^{({\rm type
B})}(t)=\frac{3\Omega_b^{0}}{4\Omega_{\gamma}^{0}}\left[\frac{\sinh\left(\frac{3}{4}\,H_0\,\mathcal{F}\,t\right)}{\sinh\left(\frac{3}{4}\,H_0\,\mathcal{F}\,t_0\right)}\right]^{2/3}.
\end{equation}
One can easily check that for $\nu=\epsilon=0$, we have
$\mathcal{F}\to 2\sqrt{\OLo}$ and we recover the corresponding
$\CC$CDM result:
\begin{equation}
\left.{\cal R}^{({\rm type B})}(t)\right|_{\epsilon=\nu=0}
=\frac{3\Omega_b^{0}}{4\Omega_{\gamma}^{0}}\left[\frac{\sinh\left(\frac{3}{2}\,H_0\,\OLo\,t\right)}{\sinh\left(\frac{3}{4}\,H_0\,\OLo\,t_0\right)}\right]^{2/3}=\frac{3\Omega_b^{0}}{4\Omega_{\gamma}^{0}}\,a(t)={\cal
R}^{\CC CDM}(a)\,.
\end{equation}
Let us mention the following specification for fitting the BAO$_{dz}$ observable (\ref{BAOd}) for type-B models. In this case the computation of the parameter $r_{s}(z_{d})$ -- the comoving distance traveled by light to the drag epoch, Eq.\, (\ref{drag}) --
is performed in two steps: in the first step we integrate from $a=0$ up to the decoupling (or last scattering) point $a_{*}$ by neglecting the vacuum energy corrections since dark energy effects are negligible for $a<a_{*}$; in the second step we integrate from $a=a_{*}$ to the drag epoch $a=a_{d}$ using the correction from the radiation component discussed above.

\subsection{Combined likelihood function}
\label{subsect:combined likelihood}

In order to place tighter constraints on the corresponding parameter
space of our model, the probes described above must be combined
through a joint likelihood analysis\footnote{Likelihoods are
normalized to their maximum values. In the present analysis we
always report $1\sigma$ uncertainties on the fitted parameters. Note
also that if the total number of data points is
$N_{\rm tot}$ the associated degrees of freedom is:
$dof = N_{\rm tot}-n_{\rm fit}$, where $n_{\rm fit}$ is
the model-dependent number of fitted parameters.}, given by the
product of the individual likelihoods according to:
\begin{equation}\label{overalllikelihood} {\cal L}_{\rm tot}({\bf p})=
{\cal L}_{\rm SNIa} \times {\cal L}_{\rm
CMB}\times {\cal L}_{\rm BAO_{X}}\,. \end{equation}
This translates into an addition of the joint
$\chi^2$ function:
\begin{equation}\label{overalllikelihoo}
\chi^{2}_{\rm tot}({\bf p})=\chi^{2}_{\rm SNIa}+\chi^{2}_{\rm CMB}+\chi^{2}_{\rm
BAO_{X}}\;,
\end{equation}
where $X$ denotes the kind of BAO measurements used in the statistical
analysis, namely $X=d_z$ or $X=A(z)$.
In our $\chi^2$ minimization procedure, for the vacuum models
(running and concordance $\Lambda$CDM) we use the following range
and steps of the fitted parameters: $\Omega_{m}^{0} \in [0.1,1]$ in
steps of 0.001 and $\nu \in [-0.02,0.02]$  in steps of
$10^{-4}$.

Since the current vacuum models contain different number of free
parameters, as a further statistical test we use the
({\em corrected}) Akaike Information Criterion (AIC)
relevant to our case ($N_{\rm tot}/n_{\rm fit}>40$)\,\cite{Akaike1974}, which is defined,
for the case of Gaussian errors, as follows:
\begin{equation}
{\rm AIC}=\chi^2_{\rm tot}+2n_{fit} \;.
\end{equation}
%where $n_{fit}$ is the number of free parameters.
It is well known that
a smaller value of AIC points to a better model-data fit. In this context,
we have to mention that small
differences in AIC are not necessarily significant and therefore, in order
to test the effectiveness of the models themselves, it is
important to calculate the model pair difference
$\Delta$AIC$ = {\rm AIC}_{y} - {\rm AIC}_{x}$. The larger the
value of $|\Delta{\rm AIC}|$, the higher the evidence against the
model with larger value of ${\rm AIC}$, with a
difference $|\Delta$AIC$| \ge 2$ indicating a positive
such evidence and $|\Delta$AIC$| \ge 6$
indicating a strong such evidence, while a value $\le 2$ indicates
consistency among the two comparison models.

\begin{table}[!t]
\begin{center}
%\begin{scriptsize}
\begin{tabular}{| c |  c |c | c | c | c |}
\multicolumn{1}{c}{Model} & \multicolumn{1}{c}{$\Omo$} & \multicolumn{1}{c}{$\nu$} & \multicolumn{1}{c}{$\epsilon$} & \multicolumn{1}{c}{$\chi^2/dof$} &\multicolumn{1}{c}{AIC} \\\hline
$\CC$CDM & $0.293\pm 0.013$ & - & - & $567.8/586$ & $569.8$\\\hline
$A1$ & $0.292\pm 0.014$ & $+0.0013\pm 0.0018 $ & - & $566.3/585$ & $570.3$ \\\hline
$A2$ & $0.290 \pm 0.014$ & $+0.0024\pm  0.0024 $ & - & $565.6/585$ & $569.6$ \\\hline
$B1$ & $ 0.297^{+0.015}_{-0.014}$ & -  & $-0.014^{+0.016}_{-0.013}$ & $587.2/585$ & $591.2$ \\\hline
$B2$ & $ 0.300^{+0.017}_{-0.003}$ &$ - 0.0039^{+0.0020}_{-0.0021}$  & $- 0.0039^{+0.0020}_{-0.0021}$ & $583.1/585$ & $587.1$ \\\hline
 \end{tabular}
% \end{scriptsize}
\caption{The fit values for the various models using SNIa+CMB+BAO$_{dz}$ data, together with their statistical significance according to $\chi^2$ and AIC statistical tests. Notice that for type-A2 models the quoted value of $\nu$ stands actually for  $\nueff$ under the conditions explained in the text, and for B2 we have set $\nu=\epsilon$ (see the text as well). \label{tableFitBAOdz}}
\end{center}
\end{table}

Let us next present the basic results of the overall statistical analysis.
\begin{enumerate}
\item {\it We first consider SNIa+CMB+BAO$_{dz}$.}

\begin{itemize}

\item In the case of the concordance $\Lambda$CDM cosmology
we find $\Omega_{m}^{0}=0.293\pm 0.013$ with statistic significance $\chi_{\rm
tot}^{2}(\Omega_{m}^{0})/dof \simeq 567.8/586$
(AIC$_{\Lambda}$$\simeq 569.8$). For comparison, the determination by Planck+WP\,\cite{PlanckXVI2013} reads: $\Omo\,h^2=0.1426\pm0.0025$. With $h=0.673\pm 0.012$ (also from the same source) this yields $\Omo\simeq0.315\pm 0.016$, which agrees with our value within slightly more than $1\sigma$. We shall comment further on the possible implications of the $\Omo$ determination in the next section.
Another important cosmological parameter that will enter later
our analysis is the rms mass fluctuation
on $R_{8}=8 h^{-1}$ Mpc scales at $z=0$. In this work we use as a prior for such parameter the value $\sigma_{8,\Lambda}=0.829$ in the $\CC$CDM, i.e. the Planck+WP result\,\cite{PlanckXVI2013}.
Then, with the aid of this value,
we can calculate the corresponding $\sigma_{8}$ values for the vacuum models
through Eq.(\ref{s88general}) below [see discussion in section \ref{sec:PSformalism} and Tables 3 and 4].

\item In Fig.\,\ref{fig:contoursA1 BAOdz} we present the results of our analysis for the type-A1 running
vacuum model, which is characterized by the $\nu$-parameter. We have sampled this parameter in the interval $[-0.02,0.02]$  in steps of
$10^{-4}$.  The left panel in that figure shows the fitted regions at 1$\sigma$,
2$\sigma$ and $3\sigma$ confidence levels in the
$(\Omega_{m}^{0},\nu)$ plane, from the  SNIa,  BAO$_{dz}$ and CMB shift parameter data. The right panel in the figure shows  the fit contours when the three types of data are intersected. Using the SNIa data alone it is evident (from the left panel)  that although the $\Omega_{m}^{0}$ parameter is tightly  constrained ($\sim 0.29$), the $\nu$ parameter remains completely unconstrained (in the shown interval). However, as it is manifest from the  right panel of that figure, the above degeneracy is broken when we use the joint likelihood analysis with all the
cosmological data. Indeed the overall likelihood function peaks at
$\Omega_{m}^{0}=0.292\pm 0.014$ and
$\nu=+0.0013\pm 0.0018$ with $\chi_{\rm
tot}^{2}(\Omega_{m}^{0},\nu) \simeq 566.3$ (AIC$_{A1}$$\simeq 570.3$)
for $585$ degrees
of freedom\footnote{Note that in \cite{GSBP11} we used the earlier
BAO results of Percival \cite{Perc10} and the {\em Constitution} set
of 397 SNIa \cite{Hic09}. We would like to mention here that those
results are in agreement with the current results within $1\sigma$
uncertainties.}.
The $\CC$CDM value of AIC ($\simeq 569.8$) is smaller
with respect to that of the A1 model. This indicates that the
fit of the concordance model to the combined data is slightly better than that of the A1 vacuum model.
However, the differential AIC value $|\Delta {\rm AIC}|$=$|{\rm AIC}_{\Lambda}-{\rm AIC}_{A1}|=1.5$ is actually $\le 2$. This tells us that
the cosmological data are perfectly consistent with the A1 model in a way comparable to the $\CC$CDM.
Furthermore, we find that $\sigma_{8}=0.813$.

\item Let us now address the more general case of the type-A2 model,whose statistical vector ${\bf p}$ contains 3 free parameters, namely
${\bf p}=(\Omega_{m}^{0},\nu,\alpha)$.
Our minimization analysis provides strongly degenerate results between
$\nu$  and  $\alpha$, rendering impossible to put any
significant constraints on their values\,\footnote{This is similar to the situation with the CPL parameterization\,\cite{CPL} of the dark energy, where in the statistical vector ${\bf p}=(\Omega_{m}^{0},w_0, w_1)$ one actually has to fix a parameter to fit the other two in an efficient way}.  Note, however that
the value of $\Omega_{m}^{0}$ is well constrained ($\simeq 0.29$).
Therefore we adopt -- as in Ref.\,\cite{BasPolarSola12}-- the
additional setting $\xiR \equiv 1$, which is tantamount to assume that there is no modification in the scaling law of radiation and hence radiation behaves in the strict standard way, in contraposition to dust. This occurs  when
$\alpha= \frac{3}{4} \nu$ (see equation \ref{defxiR}), implying
from Eq.(\ref{smalllimit})
that $\xi\simeq 1-\nueff$, in which $\nueff\simeq (\nu-\alpha)=\nu/4$.
The statistical vector
reduces in this case to ${\bf p}=(\Omega_{m}^{0},\nueff)$, and we sample $\nueff \in [-0.02,0.02]$  in steps of
$10^{-4}$.
The joint minimization provides now $\Omega_{m}^{0}=0.29\pm 0.014$,
$\nueff=0.0024 \pm 0.0024$ ($\nu\simeq 0.0096$ and $\alpha \simeq 0.0072$)
with $\chi_{\rm tot}^{2}(\Omega_{m}^{0},\nueff)/dof \simeq 565.6/585$
and AIC$_{A2}$$\simeq 569.6$.
%In Figure 2 we plot the overall likelihood
%contours in the $(\Omega_{m}^{0},\nu)$ and
%$(\alpha,\nu)$ and planes by
%marginalizing the former over $\alpha$ and the
%latter over $\Omega_{m}^{0}$.
%Notice, that we sample $\alpha \in [-0.02,0.02]$  in steps of
%$10^{-4}$.
Notice that in the present case, opposite to the A1 model, the fit to the combined data from the A2 model is slightly better than in the $\CC$CDM model (cf. Table \ref{tableFitBAOdz}). However, utilizing the AIC information criterium, and because
AIC$_{\Lambda}\simeq$AIC$_{A2}$ we have that
the A2 vacuum model is statistically equivalent ($|\Delta {\rm AIC}|\leq 2$) with the $\Lambda$CDM model.
In this case we obtain $\sigma_{8}=0.797$.

\item   We next face model B1 with BAO$_{dz}$ data. As in the A1 model, here we have only one characteristic parameter, in this case $\epsilon$  (apart from the generic one $\Omo$). We have dealt with the fitting procedure of $\epsilon$ in a similar way as $\nu$ for the A1 model. The fitting regions in the $(\Omo,\epsilon)$-plane are shown in Fig.\,\ref{fig:contoursB1 BAOdz}, left panel, whereas the  1$\sigma$, 2$\sigma$ and $3\sigma$ contour lines for the combined SNIa+CMB+BAO$_{dz}$ data are displayed on the right panel of that figure. As we can see, the determination of $\Omo$ is rather sharp around $0.3$, to be precise: $\Omo= 0.297^{+0.015}_{-0.014}$. However the parameter $\epsilon$ is not so well bounded by the data as it was the case of $\nu$, specifically we find $\epsilon=-0.014^{+0.016}_{-0.013}$. The central value and the errors are roughly one order of magnitude bigger than before, and the fit quality is significantly poorer. This is clear from the $\chi^2/dof$ and AIC statistical diagnostics in Table \ref{tableFitBAOdz}, which give substantially larger values than those in the $\CC$CDM model. If we attend strictly the AIC statistical criterion we should conclude that the type-B1 model does not fit at all the combined data in a way comparable to type-A1 or A2 models and the $\CC$CDM. Concerning the rms mass fluctuation for this model, we find $\sigma_{8}=0.859$.

\item As for the B2 model, we have once more a situation with three fit parameters $(\Omo,\nu,\epsilon)$. But to avoid similar difficulties as those mentioned with the A2 model, we fix a correlation between the two model parameters $\nu$ and $\epsilon$. Of course, both are small in absolute value, but let us note that for $\nu\ll\epsilon$ model B2 must reduce to B1, whereas for $\epsilon\ll\nu$ it essentially behaves as A1. Therefore, the parameter region which is left unexplored is when $\epsilon\simeq\nu$, and for definiteness we will fix $\epsilon=\nu$ and shall perform a two-parameter fit in $(\epsilon,\Omo)$. Under these conditions we find  $\epsilon\simeq -0.0039$ and  $\Omo\simeq 0.30$ (cf. Table \ref{tableFitBAOdz} for more precise values and errors). As with the B1 case, the quality of the fit is worse than for the $\CC$CDM or any of the type-A models.  Moreover, for the B2 model, we find $\sigma_{8}=0.896$.

%With
%these numerical values we see from (\ref{an1}) that $C_1$ is of
%order $H_0$, as could be expected.
%we find that the entropic-force model is unable to fit the
%combined observational data (SNIa+BAO+CMB). In particular, using
%either SNIa or BAO alone we can place tight constraints on $\xi$.
%Indeed in the case of SNIa we find $\xi=0.38\pm 0.015$ (or
%$n_{eff}=0.62$) with $\chi_{\rm SNIa}^{2}(\xi)/dof \simeq
%552.82/556$ while using BAO the likelihood fuction peaks at
%$\xi=0.266\pm 0.01$ (or $n_{eff}=0.73$) with $\chi_{\rm
%BAO}^{2}(\xi)/dof \simeq 2.26/5$. In this context, the combined
%SNIa/BAO data provides $\xi=0.31\pm 0.01$ (or $n_{eff}=0.69$) with
%$\chi_{\rm SNIa/BAO}^{2}(\xi)/dof \simeq 589.33/562$. Despite the
%fact that the entropic-force model fits well the cosmological data
%at relatively low redshifts ($z<2$) it fails catastrophically to fit
%the CMB shift parameter. Interestingly, the addition of one more
%point (the CMB shift parameter) increases the overall likelihood
%function by a factor of $\sim 4$ with $\xi\simeq 0.99$, which
%implies that the Hubble flow of the entropic-force model (see
%eq.\ref{Hentropic}) tends to that of the Einstein de-Sitter model
%namely, $E(z)=(1+z)^{3/2}$ with $D_{+}(a)=a$.

\end{itemize}

\item {\it In the following we are based on SNIa+CMB+BAO$_{A}$ data.}

\begin{itemize}

\item We describe now the situation concerning the various models for when we use the alternative BAO option. The corresponding results are clearly displayed in Table\,\ref{tableFitBAOA}.

For the $\Lambda$CDM model
we obtain $\Omega_{m}^{0}=0.292\pm 0.011$
with $\chi_{\rm tot}^{2}(\Omega_{m}^{0})/dof \simeq 567.5/586$
(AIC$_{\Lambda}$$\simeq 569.5$). As it could be expected, we obtain almost the same
results with those of the previous fit with SNIa+CMB+$BAO_{dz}$ data.

\item A1 vacuum model: Here the overall likelihood function peaks at a lower value of the mass parameter
$\Omega_{m}^{0}=0.282\pm 0.012$ (see the concluding comments of this section) and higher running parameter
$\nu=0.0048^{+0.0032}_{-0.0031}$, with $\chi_{\rm
tot}^{2}(\Omega_{m}^{0},\nu) \simeq 563.8$ (AIC$_{A1}$$\simeq 567.8$)
for $585$ degrees of freedom.
The shape of the BAO$_A$-contours in Fig.\,\ref{fig:contoursA1 BAOA}
are quite different from those of BAO$_{dz}$ (Fig.\,\ref{fig:contoursA1 BAOdz}). This is somehow related
with the fact that unlike for the case of $A(z)$, the
$d_{z}$ BAO measurements are correlated with the
$\Omega_{m}^{0}h^{2}$ \cite{Blake11} as well as with the
necessary modifications to ${\cal R}(a)$
introduced in the BAO$_{dz}$ analysis (see section 7.2).
The rms mass fluctuation is found to be $\sigma_{8}=0.758$. Let us note the remarkable fact that using the BAO$_A$ observable, instead of BAO$_{dz}$, the value of $\nu$ is not compatible with zero at $1\sigma$, showing a slight tendency to favor nonvanishing values of $\nu$ rather than the $\CC$CDM result.

\item A2 vacuum model:
the overall minimization provides $\Omega_{m}^{0}=0.283\pm 0.012$,
$\nueff=+0.0048 \pm 0.0031$ ($\nu\simeq 0.019$ and $\alpha \simeq 0.014$)
with $\chi_{\rm tot}^{2}(\Omega_{m}^{0},\nueff)/dof \simeq 563.8/585$
and AIC$_{A2}$$\simeq 567.8$.
The rms mass fluctuation is $\sigma_{8}=0.757$.

At this point we would like to make some
comments for the A1-A2 model.
Generally, the $\Omega_{m}^{0}$ values are in agreement (with 1$\sigma$ errors)
with those of SNIa+CMB+ BAO$_{dz}$. However, as far as $\nueff$
(or $\nu$) is concerned we find differences among the parameters which
could reach up to a factor of $\sim 3.7$.
In this context, the SNIa+CMB+BAO$_{A}$ data
analysis highlights the fact that
the values of AIC$_{A1-A2}$($\simeq 567.8$) are actually smaller
with respect to those of the concordance $\CC$CDM cosmology. In other words, it turns out that the
type-A1 and A2 vacuum models appear now to
fit slightly better than the $\Lambda$CDM the
observational data.
Still, the $|\Delta {\rm AIC}|$=$|{\rm AIC}_{A1-A2}-{\rm AIC}_{\Lambda}|$
values (ie., $\le 2$) indicate that
the cosmological data are simultaneously consistent with the A1,A2 and the
$\Lambda$CDM models.

\begin{table}[!t]
\begin{center}
%\begin{scriptsize}
\begin{tabular}{| c |  c |c | c | c | c |}
\multicolumn{1}{c}{Model} & \multicolumn{1}{c}{$\Omo$} & \multicolumn{1}{c}{$\nu$} & \multicolumn{1}{c}{$\epsilon$} & \multicolumn{1}{c}{$\chi^2/dof$} &\multicolumn{1}{c}{AIC} \\\hline
$\CC$CDM & $0.292\pm 0.011$ & - & - & $567.5/586$ & $569.5$\\\hline
$A1$ & $0.282\pm 0.012$ & $+0.0048^{+0.0032}_{-0.0031} $ & - & $563.8./585$ & $567.8$ \\\hline
$A2$ & $0.283\pm 0.012$ & $+0.0048\pm 0.0031$ & - & $563.8/585$ & $567.8$ \\\hline
$B1$ & $ 0.283^{+0.012}_{-0.011}$ & -  & $+0.005^{+0.018}_{-0.015}$ & $563.7/585$ & $567.7$ \\\hline
$B2$ & $ 0.283^{+0.011}_{-0.012}$ & $+0.0015^{+0.0025}_{-0.0030}$ & $ +0.0015^{+0.0025}_{-0.0030}$ & $563.8/585$ & $567.8$ \\\hline
$C1$ & $ 0.296\pm 0.017$ & $-0.189\pm 0.008$ & - & $568.3/585$ & $570.3$ \\\hline
 \end{tabular}
% \end{scriptsize}
\caption{The fit values for the various models using the same data and statistical tests as before except for the BAO observable, which now is BAO$_{A}$, i.e., overall we use SNIa+CMB+BAO$_{A}$ data. In this case we include also the C1 model (see text). Same notation as the previous table. \label{tableFitBAOA}}
\end{center}
\end{table}

\item B1 model with BAO$_A$ data. We find  $\Omo=0.283^{+0.012}_{-0.011}$, so it remains similar to the type-A models, with $\epsilon=+0.005^{+0.018}_{-0.015}$. Interestingly, the fit quality is also in this case slightly better than in the $\CC$CDM model, but still with $|\Delta {\rm AIC}|\leq 2$, and hence statistically comparable.
The rms mass fluctuation is $\sigma_{8}=0.820$.

\item For the B2 model we proceed here with a similar strategy as with the BAO$_{dz}$ case, and we find $\Omo=0.283$ and $\epsilon\simeq \nu\simeq +0.0015$. The quality of the fit is once more comparable, but better, than for the concordance $\CC$CDM model.
The rms mass fluctuation is $\sigma_{8}=0.791$.

\item Finally, the observational viability of the C1 model has been
tested previously in Basilakos \& Sol{\`a}
\cite{BasSola14a}, and thus for the rest of the paper we use their
SNIa+CMB+BAO$_{A}$ analysis of the
$(\Omega_{m}^{0},\zeta)$ pair.  Specifically,
we remind the reader that  the following results were found\, \cite{BasSola14a}:
$\Omega_{m}^{0}=0.296\pm 0.017$, $\zeta=1.189
\pm 0.008$ with $\chi_{\rm tot}^{2}(\Omo,\zeta)/dof\simeq 568.3/585$. Recall that for C1 models $\zeta=1-\nu$, and the fitted value of $\nu$ is the one indicated for this model in Table\,\ref{tableFitBAOA}. Because of the absence of the constant additive term for the C1 model (cf. Sect. \ref{sect:solvingC1C2}), the value of $\nu$ is forced to be much larger than in the other models.
Notwithstanding, the corresponding AIC value is $570.3$, which is statistically comparable to that of the $\CC$CDM model, and therefore at least from the point of view of the Hubble expansion data and the shift parameter, the C1 model seems to present a respectable status. But it is only an ostensible good status. The situation for this model will undergo a radical change when we test the structure formation data at low redshifts, as we shall see in the next section. The trouble is related to the aforementioned absence of the additive term. A first hint of decline of this model appears when we compute the corresponding rms mass fluctuation, namely $\sigma_{8}=1.365$, which is clearly anomalously large.

\end{itemize}

\end{enumerate}

We conclude this section with the following observation. The fitting functions (\ref{zlastscatt}) and (\ref{zdrag}) for the redshifts of decoupling and baryon drag epochs respectively, have been obtained\,\cite{HuSugiyama,Eis98} under the assumption that the precise scaling laws for matter are as in the $\CC$CDM concordance model. This is not the case for our dynamical vacuum models, see e.g. (\ref{splitAsolutionM}) and (\ref{splitAsolutionR}). Therefore it is natural to assess if this can have a significant effect in our analysis.
In regard to Eq.\,(\ref{zlastscatt}) the small deviations around the standard value of $z_{*}$ caused by the anomalous scaling laws in the natural region (\ref{naturalness}) do not affect significantly the value of the integral (\ref{shiftparameter}) (the shift parameter). As for the baryon drag redshift $z_d$ we have roughly estimated the possible effects by considering the method of Ref.\,\cite{Marteens09} in combination with the framework of\,\cite{HuSugiyama}. In this way we can test the sensibility of our fits to the vacuum models under study, with and without the scaling correction on the $R$-functions of the BAO analysis, i.e. equations (\ref{Ra2}) and (\ref{eq:RtypeB}), which enter the determination of $z_d$ with the method of\,\cite{Marteens09} (cf. Appendix A of this reference). While we have obtained some differences, in all cases they are compatible with the parameter errors indicated in Tables 1 and 2. Let us also notice that these corrections can only affect the BAO data associated to the $d_z$ observable (\ref{fromula dz}), but not the BAO analysis based on the $A(z)$ estimator (\ref{defBAOA}). If in the future more accurate cosmological data becomes available, one may have to consider the effect of these corrections for a better determination of the parameters potentially responsible for the vacuum dynamics.

\subsection{Discussion of the fitting results and implications for dynamical DE}

From the previous analysis one could tentatively say that type-A models are preferred to type-B ones from the point of view of the quality fits to the combined data. This is indeed suggested by the results involving BAO$_{dz}$. In contrast,  BAO$_{A}$ data does not seem to point so strongly to this conclusion. This is somehow understandable if we take into account that the BAO$_{A}$ data are exclusively based on the imprints of baryonic acoustic oscillations left at low redshifts during the early epochs of galaxy clusters formation, which means at relatively recent times, whereas the BAO$_{dz}$ data is also sensitive to the model behavior of these oscillations at earlier epochs in between the decoupling and baryon drag epochs. More observational work will be necessary to decide about the best vacuum models.

We come now to a point mentioned in passing in the previous section concerning the fitting values of $\Omo$. For the $\CC$CDM we have found $\Omo\simeq 0.293$ (virtually independent of the type of BAO data used). This is smaller than the Planck+WP value $\Omo\simeq0.315$. Similarly, for the vacuum models A and B we have found $\Omo$ smaller than the Planck+WP value. This holds not only for BAO$_{dz}$ data (cf. Table 1) but even more pronounced when the fit is done using  BAO$_{A}$ data, where the value of $\Omo$ lessens significantly for all vacuum models at around $\Omo\simeq0.282-0.283$ (cf. Table 2). At the same time one obtains, in the last case, a slightly improved fit quality with respect to the $\CC$CDM for all the dynamical vacuum models A and B. The difference with respect to the Planck+WP value of $\Omo$ is now larger and, as we will see in Sect.\,\ref{sec:Number counts}, it does matter as far as the possible implications on the predicted cluster number counts for the dynamical models. At this stage of precision cosmology it is difficult to make a final selection between the two types of BAO  data, and therefore we have decided to present the results separately for each BAO set.

The importance of the BAOs measurements cannot be underemphasized. They are sensitive to the physics of large scales and hinge primarily on the well-known principles of the linear regime of gravitational instability. Recall the recent hint of dynamical dark energy based on BAOs data mentioned in the introduction -- cf. Ref.\,\cite{SahniShafielooStarobinsky2014}.  These authors utilize the ``$Om(z)$-diagnostic''\,\cite{OmzSahni08} for dark energy from the recent measurement of $H(z)$ made, in particular, on the basis of BAOs in the Ly$\alpha$ forest of BOSS DR11 quasars\,\cite{Delubac2014}. Such DE diagnostic is defined as
\begin{equation}\label{Omz}
Om(z)\equiv\frac{E^2(z)-1}{(1+z)^3-1}.
\end{equation}
For the $\CC$CDM one finds the constant value $Om(z)=\Omo$. According to the test, a departure from this result points to dynamical vacuum energy. A related test is the two-point diagnostic \cite{SahniShafielooStarobinsky2014} defined as follows. Introducing  $h(z)=E(z)\,h$ (where $h$ is the reduced Hubble parameter), the test is based on the quantity
\begin{equation}\label{Omz2}
Omh^2(z_2,z_1)\equiv\frac{h^2(z_2)-h^2(z_2)}{(1+z_2)^3-(1+z_1)^3}.
\end{equation}
It has the following properties. To start with, $Omh^2(z,0)=Om(z)\,h^2$. In addition, for the  $\CC$CDM we have $Omh^2(z_2,z_1)=\Omo\,h^2$, which is constant for any pair of points $z_2$ and $z_1$.
Using the previous diagnostic and the known observational information on $H(z)$ at the three redshift values $z=0,0.57,2.34$ (the last one being from \,\cite{Delubac2014}) the authors of \cite{SahniShafielooStarobinsky2014} observe that the average result is: $Omh^2=0.122\pm0.01$, with very little variation from any pair of points taken. The obtained result is significantly smaller than the  corresponding  Planck+WP  value of the two-point diagnostic, which is obviously constant and given by $Omh^2=\Omo\,h^2=0.1426\pm0.0025$. Since a departure of $Omh^2$ from this result
should, according to \cite{SahniShafielooStarobinsky2014}, signal that the DE is not the $\CC$-cosmology, the obvious tension found between the  BAO observations and the CMB measurements  (assuming the concordance $\CC$CDM model) cannot be explained within the $\CC$CDM model.

The above mentioned authors make a case for this result and conclude that the spotted difference provides model-independent evidence for dynamical DE. Although it is probably too early to draw definite conclusions from these results before we get more statistics on $H(z)$ at high redshifts (cf. e.g. \,\cite{HuMiaoZhang14}), we can at least say that this kind of scenario is roughly consistent with the results of the current analysis. We have indeed found that our vacuum dynamical framework, when confronted with the presently available SNIa+CMB+BAO data, tends to emphasize significantly smaller values of $\Omo$. Therefore, in case that the  claims of dynamical DE would be confirmed at some point, the vacuum models presented here could provide an explanation.

We can understand analytically the possible origin of these results in our theoretical framework. Let us take e.g. a general type-A model. From the formulae of Sect.\ref{sect:solvingA1A2} we can easily compute the corresponding $Om(z)$-diagnostic. The result is:
\begin{equation}\label{OmzA}
Omh^2(z)=\frac{\Omo}{\xi}\,\frac{\left(1+z\right)^{3\xi}-1}{(1+z)^3-1}\,,
\end{equation}
with $\xi$ given in Eq.\,(\ref{defxiM}). It is pretty obvious that for $\xi=1$ we recover the $\CC$CDM result, which remains pegged to $Om(z)=\Omo\ (\forall z)$. However, as soon we allow a small dynamical running of vacuum (meaning $\nu$ and or $\alpha$ different from zero) we obtain a small departure of $\xi$ from $1$ and therefore the DE diagnostic $Om(z)$ deviates from $\Omo$. Actually, in this case (\ref{OmzA}) evolves with time (or redshift). According to the $Om(z)$ diagnostic this implies that the vacuum energy is dynamical. By the same token, the two-point diagnostic for type-A models can be computed:
\begin{equation}\label{Omh2zA}
Omh^2(z_2,z_1^2)=\frac{\Omo}{\xi}\,\frac{\left(1+z_2\right)^{3\xi}-\left(1+z_1\right)^{3\xi}}{(1+z_2)^3-1+z_1)^3}\,.
\end{equation}
Clearly, the result depends on $z_i$ for $\xi\neq 1$. Only for the $\CC$CDM case ($\xi=1$) it remains anchored to $\Omo\,h^2$ for any $z_i$.
Similar considerations hold for type-B models. The upshot is that the detailed numerical analysis of both model types of dynamical vacuum models, in the light of the available observations, confirms that such vacuum dynamics leads to smaller $\Omo$ -- cf. Sect. \ref{subsect:combined likelihood}.

%%%%%%%%%%%%%%%%%%%%%%%%%%%%%%%%%%%%%%%%%%%%%%%%%%%%%%%%%%%%%%%%%%%%%%%%%%%%%
\begin{figure}[!t]
\begin{center}
\includegraphics[scale=0.42]{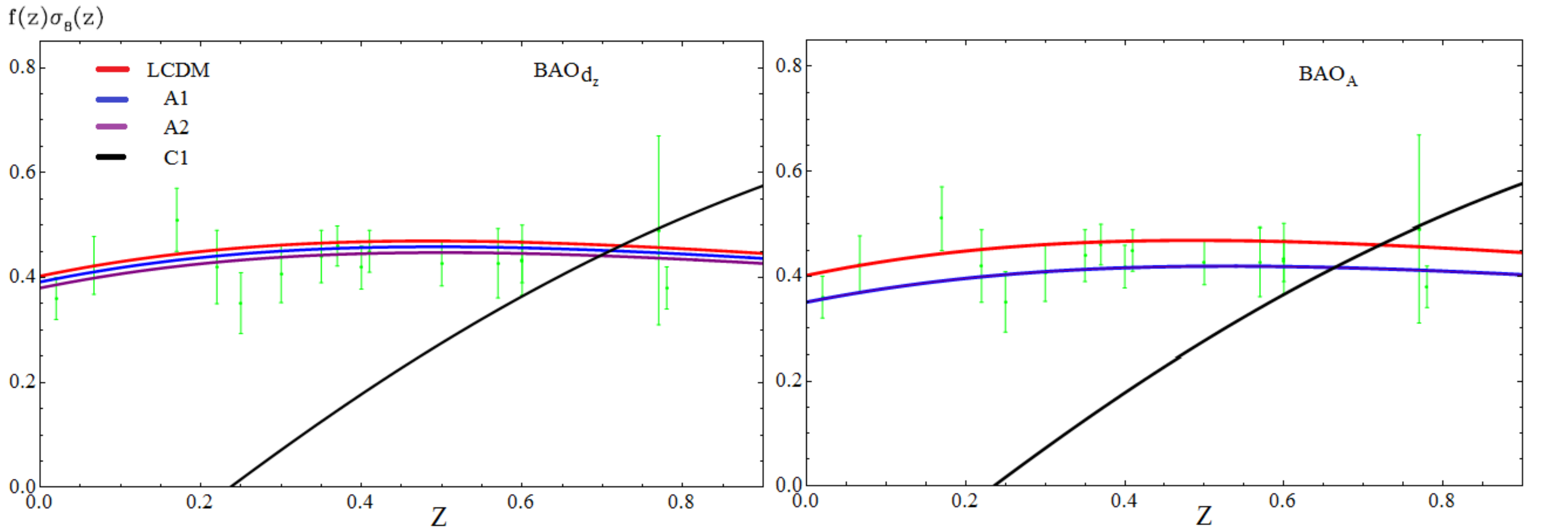}
\caption[]{\footnotesize{
Comparison of the observed (solid points with vertical error bars) and
theoretical evolution of the weighted growth
rate $f(z)\sigma_{8}(z)$ for the models A1,A2 and C1. The uppermost (red) line corresponds to the $\CC$CDM model, used as a reference. The subsequent ones (from top to bottom) correspond to the vacuum models A1 (blue line) and A2 (purple line). On the right panel, the lines for A1 and A2 are one essentially overlapping. The curve that deviates significantly from the others in the two panels and loses power quickly near our time corresponds to model
C1 (black line). The curves have been obtained for the best
fit values of the cosmological parameters as discussed in Sec.\,\ref{subsect:combined likelihood} (for a summary, cf. Tables \ref{tableFitBAOdz} and \ref{tableFitBAOA}).
The left panel shows the results based on the
SNIa+CMB+BAO$_{dz}$ fitting while the right panel those of the
SNIa+CMB+BAO$_{A}$ analysis. The C1 curve is obtained only for SNIa+CMB+BAO$_{A}$ data (and therefore is the same in both panels).
 \label{sigma8TypeA}}
}
\end{center}
\end{figure}
%%%%%%%%%%%%%%%%%%%%%%%%%%%%%%%%%%%%%%%%%%%%%%%%%%%%%%%%%%%%%%%%%%%%%%%%%%%%

\subsection{The linear growth rate of clustering and the
$\gamma$ index}\label{sect:growthrate}

In this section we analyze the linear perturbation growth
regime for the various models.
Although one could do it by means of the
power spectrum, we follow the approach of\,\cite{BPS09} and will
test herein the implications of the various models on structure
formation through the study of the linear growth rate of clustering\,\cite{Peebles1993}.
This important (dimensionless)
indicator is defined as the logarithmic derivative of the linear growth factor $D(a)$ with respect to the variable $\ln a$. Therefore,
\begin{equation}\label{growingfactor}
f(a)\equiv \frac{1}{D}\frac{dD}{d\ln a}=\frac{d{\rm ln}D}{d{\rm ln}a}=-(1+z) \frac{d{\rm
ln}D}{dz}\,,
\end{equation}
where $D(a)$ is obtained from solving
the differential equation (\ref{diffeqDa}) for each model. The physical significance of $f(a)$ is that it determines the amplitude of redshift distortions, and also of the peculiar velocity flows. The latter can be seen by witting $f(a)=(\dot{D}/D)/(\dot{a}/a)=\dot{D}/(D\,H)$, which is the ratio of the peculiar flow rate to the Hubble rate.

In order to investigate the performance of our vacuum models, we
compare the theoretical growth prediction with the latest growth data
(as collected e.g. by \cite{BasilakosNes2013} and references therein),
which are based on the combined observable $f(z)\sigma_{8}(z)$, viz. the ordinary growth rate weighted by the rms mass fluctuation field.
It has been found that this estimator is almost a model-independent
way of expressing the observed growth history
of the universe, in particular it is found to be independent of the galaxy density bias (see \cite{Song09}).
The theoretical functional form
of $\sigma_{8}(z)$ will be studied in Sect.\ref{sec:PSformalism} -- see Eq.(\ref{ss88}).

%%%%%%%%%%%%%%%%%%%%%%%%%%%%%%%%%%%%%%%%%%%%%%%%%%%%%%%%%%%%%%%%%%%%%%%%%%%%%%%%%%%%
\begin{figure}[!t]
\begin{center}
\includegraphics[scale=0.45]{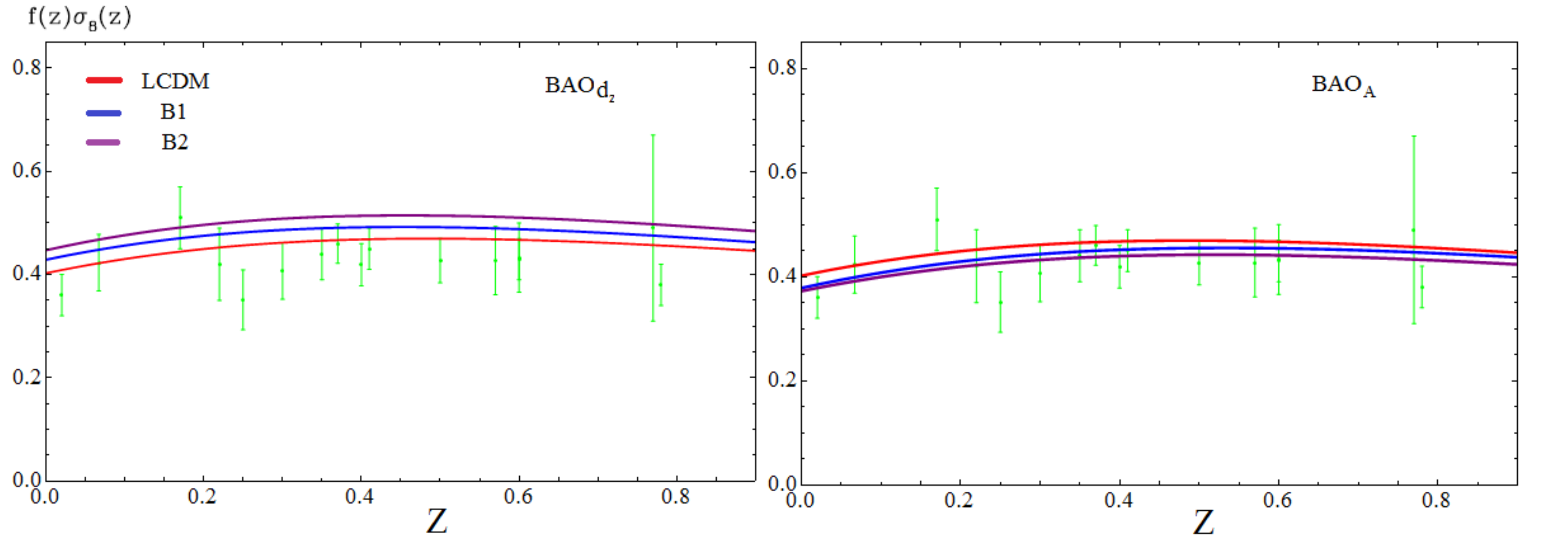}
\caption[]{\footnotesize{
Comparison of the observed and
theoretical evolution of the weighted growth
rate $f(z)\sigma_{8}(z)$ for the models B1 and B2. As in Fig.\,\ref{sigma8TypeA}, the left and right panels show the results based on the SNIa+CMB+BAO$_{dz}$ and
SNIa+CMB+BAO$_{A}$ best fit values, respectively.  On the left panel, the lowermost (red) line corresponds to the $\CC$CDM model, used as a reference. The closest one on top of it at $z=0$ corresponds to model B1 (blue line) and the next to closest at this point is model B2 (purple line). On the right panel, the B1 and B2 lines essentially overlap below the $\CC$CDM one, the B1 being in the middle.
\label{sigma8TypeB}}
}
\end{center}
\end{figure}

In Figs.\,\ref{sigma8TypeA} and \ref{sigma8TypeB} we display the predicted $f(z)\sigma_{8}(z)$
together with the observed linear growth data, for the various vacuum models A, B and C1. No information is provided on C2 which we already discarded.
Notice, that the theoretical curves on the left and the right panels correspond
to fitted values of the cosmological parameters
derived from SNIa+CMB+BAO$_{dz}$ and SNIa+CMB+BAO$_{A}$ respectively.
Obviously, despite the fact that the C1 model fits well the expansion history (cf. Sect. \ref{subsect:combined likelihood}),
it is finally ruled out by the growth data\,\cite{BasSola14a}.
%The lack of structure formation near our time
%is exceedingly evident (confer Fig.\,\ref{sigma8TypeA}) as compared to the other vacuum models
%-- for more details, see \cite{BasSola14a}.
For these reasons we come to the conclusion that
the entire C1 class of models (in particular the linear sort $\CC\propto H$) is strongly unfavored and we do not continue its analysis for the rest of this paper.

Overall, we can see that the A1-A2 vacuum models
with the single parameter $\nu$ (or $\nueff$) match quite well the
growth data, that is to say in a way which is
comparable to the $\CC$CDM model (dashed line). We confirm this fact via
a $\chi^{2}_{\rm growth}$ minimization statistical test.
%and a Kolmogorov-Smirnov (KS)
%statistical test respectively.
In particular, for the vacuum
models (including the $\Lambda$CDM) we find that
$\chi^{2}_{\rm growth}/16$ lies in the interval $[0.52-1.25]$\footnote{The growth
sample contains 16 entries \cite{BasilakosNes2013}.}.

On inspecting once more Figs.\,\ref{sigma8TypeA} and \ref{sigma8TypeB}, the data clearly shows that the
growth of structure is hindered near our time, which is evidence of
a positive cosmological constant exerting a negative pressure
against the process of matter collapse. This is well described by
the $\CC$CDM. But it is also comparably well described by the
running vacuum models carrying
an additive constant term in their functional
form [see Eq. (\ref{A1A2B1B2C})]
and a relatively small value of $\nu$ or $\epsilon$ of order $\sim
10^{-3}$. All these features can be seen very clearly in that
figure.

Finally, let us finish with a short discussion concerning the
growth rate index $\gamma$. As we have already mentioned in the introduction,
we can express the linear growth rate of clustering
in terms of $\Omega_{m}(z)$ as follows: $f(z)\simeq \Omega_{m}(z)^{\gamma(z)}$, where
$\gamma$ is the linear growth rate index. For the usual $\Lambda$CDM model, such index is approximated by $\gamma_{\CC} \simeq 6/11\simeq 0.545$. This result is a particular case (for $\omega_D=-1$)  of the theoretical formula $\gamma\simeq 3(\omega_D-1)/(6\omega_D-5)$ corresponding to DE models with a slowly varying  equation of state $\omega_D$\,\cite{WangSteinhardt98}.

%%%%%%%%%%%%%%%%%%%%%%%%%%%%%%%%%%%%%%%%%%%%%%%%%%%%%%%%%%%%%%%%%%%%%%%%%%%%

%%%%%%%%%%%%%%%%%%%%%%%%%%%%%%%%%%%%%%%%%%%%%%%%%%%%%%%%%%%%%%%%%%%%%%%%%%%%%%%%%%%%%

\begin{figure}[!t]
\begin{center}
\includegraphics[scale=0.40]{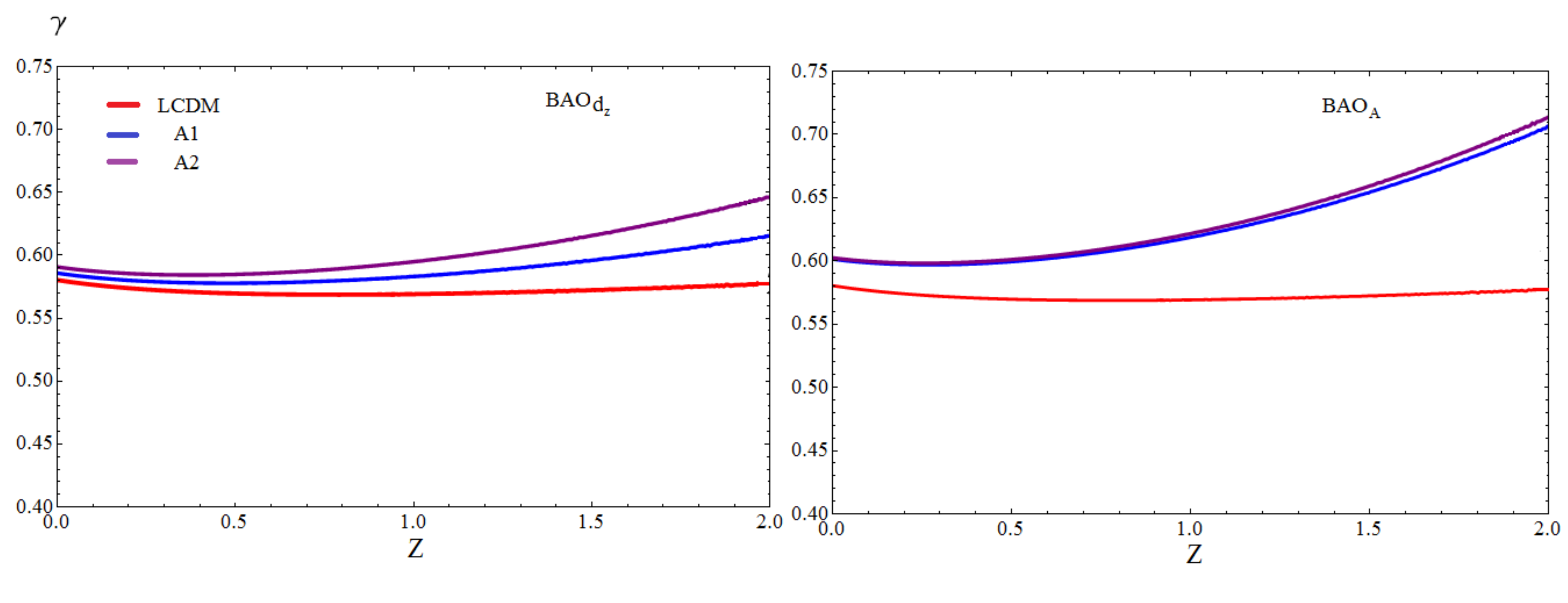}
\caption[]{\footnotesize{The evolution of the growth rate index, Eq.\,(\ref{growingfactor}).
The lines correspond to the
A1 and A2 vacuum models for the best
fit cosmological values discussed in Sec.\,\ref{subsect:combined likelihood}. From top to bottom (in both panels): A2 (purple line), A1 (blue line) and $\CC$CDM (red line), used as reference.
The left panel shows the results based on the
SNIa+CMB+BAO$_{dz}$ fitting while the right panel those of the
SNIa+CMB+BAO$_{A}$ analysis.
\label{GrowthIndexTypeA}}
}
\end{center}
\end{figure}

%%%%%%%%%%%%%%%%%%%%%%%%%%%%%%%%%%%%%%%%%%%%%%%%%%%%%%%%%%%%%%%%%%%%%%%%%%%%%%%%%%%%
To obtain the linear growth index for the dynamical vacuum models studied here we have to use the corresponding linear growth factor, $D(z)$, and
from Eq.(\ref{growingfactor}) we easily
obtain:
\begin{equation}\label{growingfactor2}
\gamma(z)\simeq \frac{\ln\left[-(1+z) \frac{d\ln D}{dz}\right]}
{\ln\Omega_{m}(z)} \;,
\end{equation}
where $D(z)$ for the different vacuum models is given in sections
\ref{sect:perturbationsTypeA} and \ref{sect:perturbationsTypeB}.

\begin{figure}[!t]
\begin{center}
\includegraphics[scale=0.45]{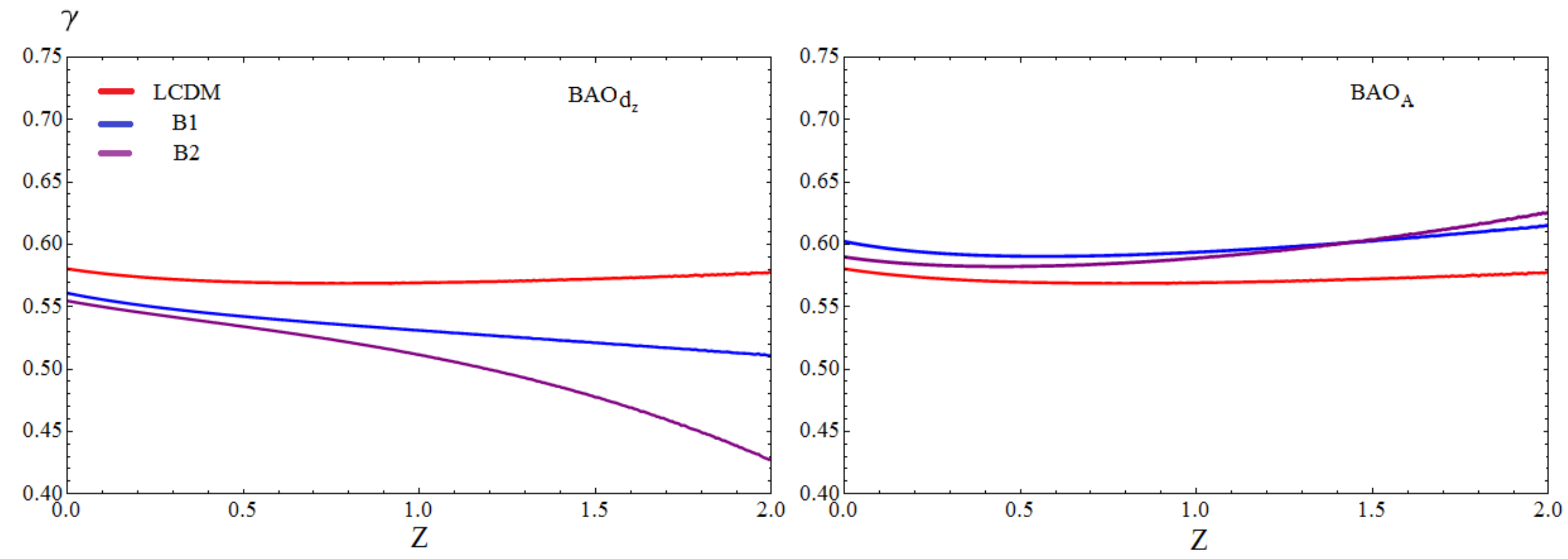}
\caption[]{\footnotesize{The evolution of the growth rate index, similar to the previous figure. In this case the lines correspond to the B1 and B2 vacuum models for the best
fit cosmological values discussed in Sec.\,\ref{subsect:combined likelihood}.
The left panel shows the results based on the
SNIa+CMB+BAO$_{dz}$ fitting while the right panel those of the
SNIa+CMB+BAO$_{A}$ analysis. The uppermost curve (in red) on the left panel corresponds to the $\CC$CDM, the middle one is for B1 (blue line) and the lowest one (in purple) is for B2. On the right panel the $\CC$CDM curve is the lowest one. Near $z=0$ the highest one (in blue) is for B1 and the middle curve (in purple) is for B2.
\label{GrowthIndexTypeB}}
}
\end{center}
\end{figure}

In Figs.\,\ref{GrowthIndexTypeA} and \ref{GrowthIndexTypeB} we present the evolution of the linear growth index
for the A and B type of vacuum models, respectively
(on the left with SNIa+CMB+BAO$_{dz}$ data, and on the right with SNIa+CMB+BAO$_{A}$ data). In the same figures we can also see our determination of $\gamma_{\CC}(z)$ as a function of the redshift, and in particular we find $\gamma_{\CC}(0)\simeq 0.58$.

The comparison shown in the mentioned figures indicates that the growth index of the type-A vacuum models with SNIa+CMB+BAO$_{dz}$ data is well approximated
by the $\Lambda$CDM constant value
for $z \le 1$, while at large redshifts there are deviations. When SNIa+CMB+BAO$_{A}$ data is used, instead, there is a visible deviation from above in all the range, which becomes smaller (at the level $5\%$) for $z \le 1$
%We can appreciate positive deviations of the type-A1 and A2 models with respect to the $\CC$CDM value
%$\gamma(0)_{\CC CDM}\simeq 6/11\simeq 0.545$
%ranging $5-10\%$ at $z=0$
(see Fig. \ref{GrowthIndexTypeA}, right panel).  Let us note that other vacuum models, such as e.g. B1 and B2, depart also from the $\CC$CDM result (in this case from below) when using  SNIa+CMB+BAO$_{dz}$ data (cf. Fig.\,\ref{GrowthIndexTypeB}, left panel). We find that for $z \le 1$ the departure can be of order $5-10\%$. The deviation, on the other hand, is not so pronounced (and of opposite sign) when  SNIa+CMB+BAO$_{A}$ data are used (right panel of the same figure).

It is worth mentioning that the differences we have found with respect to the $\CC$CDM are near the edge of the present experimental limits. For example, in a recent analysis of the clustering properties of Luminous Red Galaxies and the growth rate data provided by the various galaxy surveys it is found that $\gamma=0.56\pm 0.05$ and $\Omo=0.29\pm0.01$\,\cite{Athina2014}. The prediction of $\gamma$ for all our vacuum models lies within $1\sigma$ of that range.

%%%%%%%%%%%%%%%%%%%%%%%%%%%%%%%%%%%%%%%%%%%%%%%%%%%%%%%%%%%%%%%%%%%%%%%%%%%%%%%%%%%%

%%%%%%%%%%%%%%%%%%%%%%%%%%%%%%%%%%%%%%%%%%%%%%%%%%%%%%%%%%%%%%%%%%%%%%%%%%%%%%%%%%%%%%

Since the experimental error on
the $\gamma$-index is of order $10\%$ and some of the vacuum models are bordering these limits, it opens the possibility that the deviations presented by these models might be resolved in the future when more accurate data will be available\,\cite{BGSprogress}. This is quite evident from the results presented in Figs.\,\ref{GrowthIndexTypeA} and \ref{GrowthIndexTypeB} of our analysis. Combining the analysis of the growth rate with the those obtained from the cluster number counts method (studied in the next section), it should be possible to further pin down the nature of these dynamical vacuum models.

\vspace{0.5cm}

\section{Testing the dynamics of vacuum through the cluster number counts method}\label{sec:Number counts}

In the foregoing part of our analysis we have shown that the A and B types of vacuum models can successfully
fit the background cosmological data and the growth of linear
perturbations in a way which in some cases is perfectly comparable to the
$\CC$CDM. This is not so with type-C models, which fail seriously in regard to the expansion data or the structure formation data or both. We have also shown that the A and B vacuum classes have different predictions concerning the linear growth rate index $\gamma$, which in the future may be resolved. In that case we could distinguish between these two sort of vacuum models and also with respect to the $\CC$CDM.

%%%%%%%%%%%%%%%%%%%%%%%%%%%%%%%%%%%%%%%%%%%%%%%%%%%%%%%%%%%%%%%%%%%%%%%%%%%%

\begin{table}[!t]
\tabcolsep 5pt \vspace {0.2cm}
\hspace{3cm}\begin{tabular}{|c|c|ccc|c|c|} \hline \hline
Model    & $\Omega_m^0$& $\nu$ &\phantom{X}& $\epsilon$ &
$\sigma_{8}$
&$\delta_c$ \\
 \hline
$\Lambda$CDM     & 0.292 & 0 & \phantom{X} & 0 & 0.829 & 1.675 \\
\hline
A1 & 0.292 & +0.0013 & \phantom{X} & 0  & 0.813 & 1.666\\
A2  & 0.290 & +0.0024 & \phantom{X} & 0 & 0.797 &1.659
\\ \hline
B1  & 0.297 & 0 & \phantom{X} & -0.014 & 0.859 &1.696\\
B2 & 0.300 & -0.0039 & \phantom{X}
 & -0.0039 & 0.896 & 1.705 \\ \hline
\end{tabular}
\caption[]{Numerical results from fitting SNIa+CMB+BAO$_{dz}$ data (in correspondence with Table \ref{tableFitBAOdz}). The $1^{st}$ column indicates the vacuum
energy model. The $2^{nd}$ shows the central fit value of  $\Omo$. The $3^{rd}$ and $4^{th}$ display the best fit values of the parameters $\nu$ and $\epsilon$, with the understanding that $\nu$ is to be taken $\nueff$ for model A2. Finally, the
$5^{th}$ and $6^{th}$ columns list the computed values of $\sigma_8$ and $\delta_c\equiv\delta_c(z=0)$, respectively. The procedure to compute the
collapse density threshold $\delta_c(z)$ for each model is explained in Appendix ~B.} \label{TableModelsBAOdz}
\end{table}

%%%%%%%%%%%%%%%%%%%%%%%%%%%%%%%%%%%%%%%%%%%%%%%%%%%%%%%%%%%%%%%%%%%%%%%%%%%%

In the meanwhile and in an attempt to define further observational criteria capable of distinguishing the realistic model variants A
and B from the concordance $\Lambda$CDM cosmology, we analyze in this second part of our work
their theoretically predicted cluster-size halo redshift
distributions, i.e. the expected cluster number counts of each model as a function of the redshift. As it turns, this is an efficient method to separate vacuum
models which perform outstanding at the linear perturbation but differ very little in the values of the parameters.

In previous works some of us have described and tested this methodology
for simpler versions of the dynamical vacuum models, see
Refs.\,\cite{BPS09} and \cite{GSBP11}. The method has also been used to place bounds on cosmological parameters and on different types of dark energy models, see e.g.\,\cite{BasilakosPlionisLima2010,Campanelli2011,ChandrachaniDevi}.  The basic tool is the Press-Schechter formalism and its generalization. In the following
we briefly summarize the basics of this method and refer the reader
to the aforesaid references for more details. A crucial
ingredient of the cluster number counts method is the linearly
extrapolated density threshold above which structures collapse,
$\delta_c$. The computation of this model-dependent parameter is a
rather demanding task as it requires to solve the perturbations
equations beyond the linear approximation. In the Appendix B we
compute $\delta_c$ for the models under consideration.

\subsection{Generalized Press-Schechter formalism}\label{sec:PSformalism}

The Press and Schechter  (hereafter PSc) formalism to compute the
fraction of matter in the universe that has formed bounded
structures and its redshift distribution was developed in a pioneering work of these authors $40$ years ago\,\cite{press} and has been generalized and improved since then.
One introduces the so-called halo mass function, $F(M,z)$, representing
the fraction of the universe that has collapsed by the redshift $z$
in halos above some mass $M$, where the primordial density
fluctuation for a given mass $M$ of the dark matter fluid is
described by a random Gaussian field. With this  function and
assuming a mean background mass density $\bar{\rho}$  one may
estimate the (comoving) number density of virialized halos, $n(M,z)$, with
masses within the range $(M, M+\delta M)$: \be n(M,z) dM=
\frac{\partial F(M,z)}{\partial M} \frac{{\bar \rho}}{M} dM\,. \ee
This expression can be rewritten as follows:
\begin{eqnarray}\label{MF}
n(M,z) dM &=& \frac{\bar{\rho}}{M} \frac{d{\rm \ln}\sigma^{-1}}{dM}
f_{\rm PSc}(\sigma) dM,
\end{eqnarray}
where $f_{\rm PSc}(\sigma)=\sqrt{2/\pi} (\delta_c/\sigma)
\exp(-\delta_c^2/2\sigma^2)$. Note that in this approach all the
mass is locked inside halos, according to the normalization
constraint: \be\label{PSnormalization} \int_{-\infty}^{+\infty} f_{\rm PSc}(\sigma) d{\rm
\ln}\sigma^{-1} = 1\;. \ee

%%%%%%%%%%%%%%%%%%%%%%%%%%%%%%%%%%%%%%%%%%%%%%%%%%%%%%%%%%%%%%%%%%%%%%%%%%%%%%%%%%

\begin{table}[!t]
\tabcolsep 5pt \vspace {0.2cm}
\hspace{3cm}
\begin{tabular}{|c|c|ccc|c|c|} \hline \hline
Model   & $\Omega_m^0$& $\nu$ &\phantom{X} & $\epsilon$ &
$\sigma_{8}$
&$\delta_{c}$ \\
 \hline
$\Lambda$CDM    & 0.292 & 0  &\phantom{X} & 0 & 0.829 & 1.675 \\
\hline
A1 & 0.282 & +0.0048 & \phantom{X} & 0  &0.758 & 1.644\\
A2  & 0.283 & +0.0048 &\phantom{X}  & 0 & 0.757 &1.642
\\ \hline

B1  & 0.283 & 0 &\phantom{X}  & +0.005 & 0.820 &1.667\\
B2 & 0.283 & +0.0015 &\phantom{X}
 & +0.0015 & 0.791  & 1.662\\ \hline
\end{tabular}
\caption[]{As in Table \ref{TableModelsBAOdz}, but using the fitting results from SNIa+CMB+BAO$_{A}$ data (in correspondence with Table \ref{tableFitBAOA}).} \label{TableModelsBAOA}
\end{table}

%%%%%%%%%%%%%%%%%%%%%%%%%%%%%%%%%%%%%%%%%%%%%%%%%%%%%%%%%%%%%%%%%%%%%%%%%%%%

The parameter $\delta_{c}$ is the collapse density threshold, i.e. the linearly extrapolated density
threshold above which structures collapse \cite{eke} (see our Appendix B
for more details), while $\sigma^2(M,z)$ is the mass variance of the
smoothed linear density field, which depends on the redshift $z$ at
which the halos are identified. It is given in Fourier space by: \be
\label{sig88} \sigma^2(M,z)=\frac{D^2(z)}{2\pi^2} \int_0^\infty k^2
P(k) W^2(kR) dk \,. \ee In this expression, $D(z)$ is the linear growth
factor of perturbations, which we have computed before for our
models, $P(k)$ is the power-spectrum of the linear density field,
and finally we have the smoothing function $W(kR)=3({\rm
  sin}kR-kR{\rm cos}kR)/(kR)^{3}$, which is the Fourier image of the following geometric top-hat function with spherical symmetry:  $f_{\rm TH}(r)=3/(4\pi R^3)\,\theta(1-r/R)$. Here $\theta(1-r/R)$ is the Heaviside function, which implements the top-hat spherical truncation for $r>R$. Notice that, thanks to it, its Fourier transform also truncates the sum over modes in (\ref{sig88}) since $W(kR)\simeq 1$ for $kR\ll 1$, but $W(kR)\simeq 0$ for $kR\gg 1$. The spherical distribution contains on average a mass $M$ within a radius $R=(3M/ 4\pi
\bar{\rho})^{1/3}$. In this expression $\bar{\rho}$ is the comoving mean mass density at redshift $z$, i.e. the mean mass density of the background divided by $(1+z)^3$. The current background density
is ${\rho_m}=\Omo\,\rco=2.78 \times
10^{11}\Omega_{m}^{0}h^{2}M_{\odot}$Mpc$^{-3}$. We use the CDM power
spectrum:
\begin{equation}\label{Pk}
P(k)=P_{0} k^{n_s} T^{2}(\Omega_{m}^{0},k)\,,
\end{equation}
where $P_0$ is a normalization cconstant (see below), $n_s=0.9603\pm 0.0073$ is the value of the spectral index
measured by Planck+WP\,\cite{PlanckXVI2013}; and $T(\Omega_{m}^{0},k)$
the BBKS transfer function \cite{Bard86,LiddleLyth}:
%%%%%%%%%%%%%%%%%%%%%%%
\begin{eqnarray} \label{jtf}
T(\Omega_{m}^{0},k)&=&\frac{\ln (1+2.34 q)}{2.34 q}\Big[1+3.89 q + (16.1 q)^2 + \, (5.46 q)^3+(6.71 q)^4\Big]^{-1/4}\,,\nonumber\\
q=q(k)&\equiv&\frac{k\cdot {\rm Mpc}}{\Omo h^2}\,
e^{\Omega_b^{0}+\sqrt{2h}\frac{\Omega_b^{0}}{\Omo}}
%=frac{k}{\OMo\,h^2\,  e^{-\,\Omega_b^0\,-\sqrt{2h}\,\left({\Omega_b^0}/{\OMo}\right)}}
\,.
\end{eqnarray}

%%%%%%%%%%%%%%%%%%%%%%%%%%%%%%%%%%%%%%%%%%%%%%%%%%%%%%%%%%%%%%%%%%%%%%%%%%%%
\begin{figure}[!t]
\begin{center}
\includegraphics[scale=0.50]{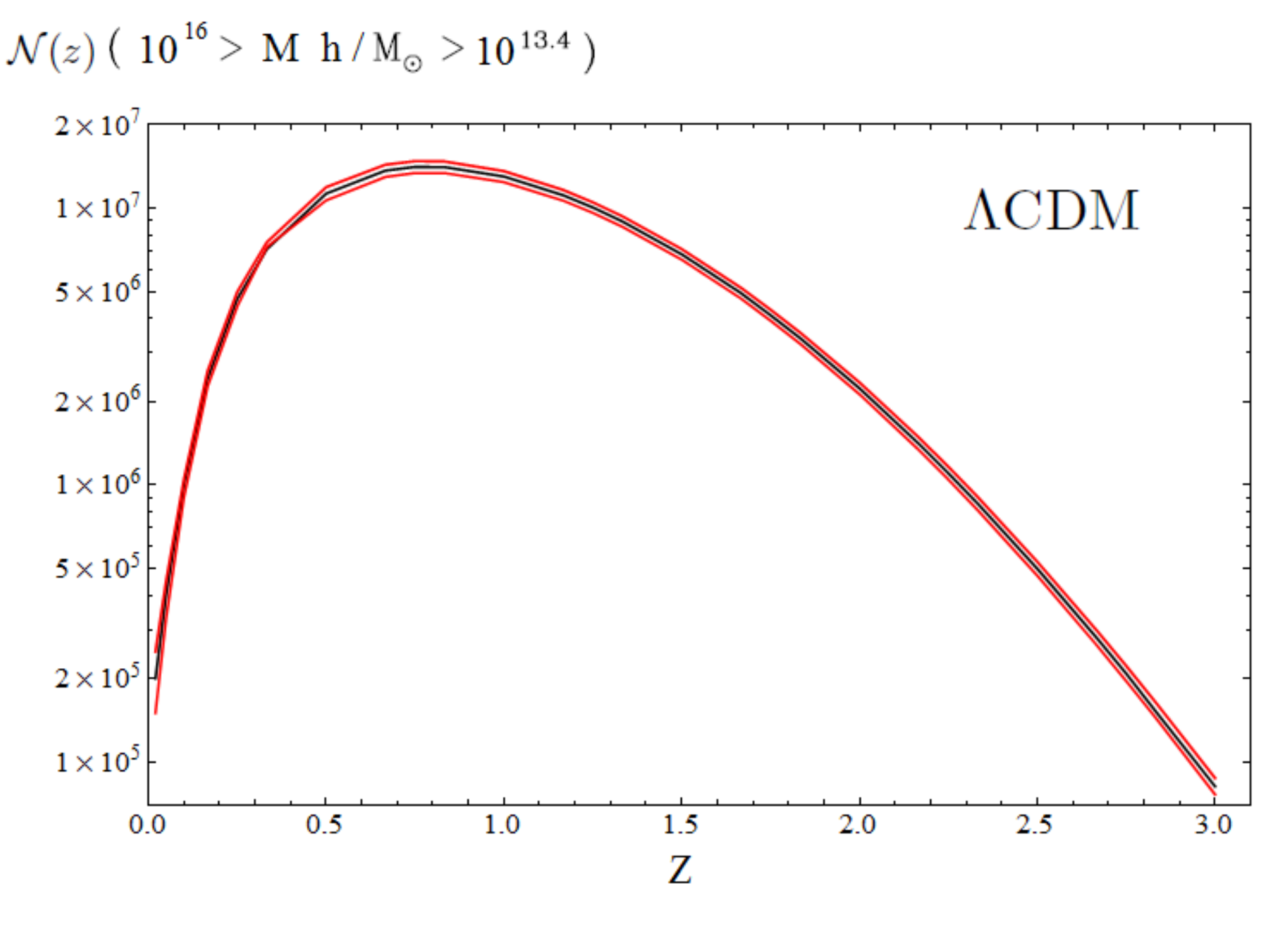}
\caption[]{\footnotesize{The theoretically predicted redshift distribution of the total number of cluster counts, ${\cal N}(z)$, with masses in the range  $10^{13.4}\,h^{-1}\lesssim M/M_{\odot}\lesssim 10^{16}\,h^{-1}$, corresponding to the concordance $\CC$CDM model and using the generalized Press-Schechter function (\ref{PS-Reed}). The curves correspond to the best fit value $\Omo=0.293\pm 0.013$  within $1\sigma$ (cf. Table 1). }}\label{fig:NLCDM}
\end{center}
\end{figure}

%%%%%%%%%%%%%%%%%%%%%%%%%%%%%%%%%%%%%%%%%%%%%%%%%%%%%%%%%%%%%%%%%%%%%%%%

%%%%%%%%%%%%%%%%%%%%%%%
It is traditional to parameterize the mass variance in terms of
$\sigma_8$, the rms mass fluctuation amplitude on scales of $R_{8}=8
\; h^{-1}$ Mpc at redshift $z=0$ [$\sigma_{8} \equiv \sigma_8(0)$]. This allows us to normalize the power spectrum, i.e. to determine $P_0$. Indeed, using equations (\ref{sig88}) and (\ref{Pk})
we have: \be \label{s888} \sigma^2(M,z)=\sigma^2_8(z)
\frac{\int_{0}^{\infty} k^{n_s+2} T^{2}(\Omega_{m}^{0}, k) W^2(kR)
dk} {\int_{0}^{\infty} k^{n_s+2} T^{2}(\Omega_{m}^{0}, k)
W^2(kR_{8}) dk}\,, \ee where \be \label{ss88}
\sigma_8(z)=\sigma_8\frac{D(z)}{D(0)} \;. \ee
Equivalently,
\begin{equation}\label{P0}
P_0=2\pi^2\,\frac{\sigma_8^2}{D^2(0)}\,\left[{\int_{0}^{\infty} k^{n_s+2} T^{2}(\Omega_{m}^{0}, k)
W^2(kR_{8}) dk}\right]^{-1}\,.
\end{equation}
The Planck+WP value of
$\sigma_8$, which we use for our analysis, is $\sigma_8=0.829\pm
0.012$\,\cite{PlanckXVI2013}.

The $\sigma_8$ value for the different dynamical vacuum models can
be estimated by scaling the present time $\CC$CDM value\footnote{In
the following discussion, the quantities referred to the $\CC$CDM
model are distinguished by the subscript `$\CC$' ($\sigma_{8,\CC}$;
$D_\CC$; $\Omega_{m,\Lambda}^{0}$) whereas the corresponding
quantities in the dynamical vacuum models carry no subscript.}
($\sigma_{8, \Lambda}$) using once more equations (\ref{sig88}) and (\ref{Pk}):
%\be
%\sigma_{\rm 8}=\sigma_{8, \Lambda} \frac{D(0)}{D_{\Lambda}(0)}
%\left[\frac{P_{0} \int_{0}^{\infty} k^{n_s+2}
%T^{2}(\Omega_{m}^{0},k) W^2(kR_{8}) dk}
%{P_{\Lambda,0}\int_{0}^{\infty} k^{n_s+2} T^{2}(\Omega_{m,
%\Lambda}^{0},k) W^2(kR_{8}) dk} \right]^{1/2}\,,\label{s88general}
%\ee where
%$P_{0}/P_{\Lambda,0}=(\Omega_{m,\Lambda}^{0}/\Omega_{m}^{0})^{2}$.
%
\be
\sigma_{\rm 8}=\sigma_{8, \Lambda} \frac{D(0)}{D_{\Lambda}(0)}
\left[\frac{\int_{0}^{\infty} k^{n_s+2}
T^{2}(\Omega_{m}^{0},k) W^2(kR_{8}) dk}
{\int_{0}^{\infty} k^{n_s+2} T^{2}(\Omega_{m,
\Lambda}^{0},k) W^2(kR_{8}) dk} \right]^{1/2}\,.\label{s88general}\ee

Overall it follows from the foregoing formulae that the mass variance of the linear density field is determined from
\begin{equation}\label{variance}
\sigma^2(M,z)=\sigma^2_{8,\Lambda}\,\frac{D^2(z)}{D^2_{\Lambda}(0)}
\frac{\int_{0}^{\infty} k^{n_s+2} T^{2}(\Omega_{m}^{0}, k) W^2(kR)
dk} {\int_{0}^{\infty} k^{n_s+2} T^{2}(\Omega_{m,\Lambda}^{0}, k)
W^2(kR_{8}) dk}\,,
\end{equation}
%
%
%%%%%%%%%%%%%%%%%%%%%%%% FIGURE  2 %%%%%%%%%%%%%%%%%%%%%%%%%%%%%%%%%%%%%
%
\begin{figure}[!t]
\begin{center}
{\includegraphics[scale=0.45]{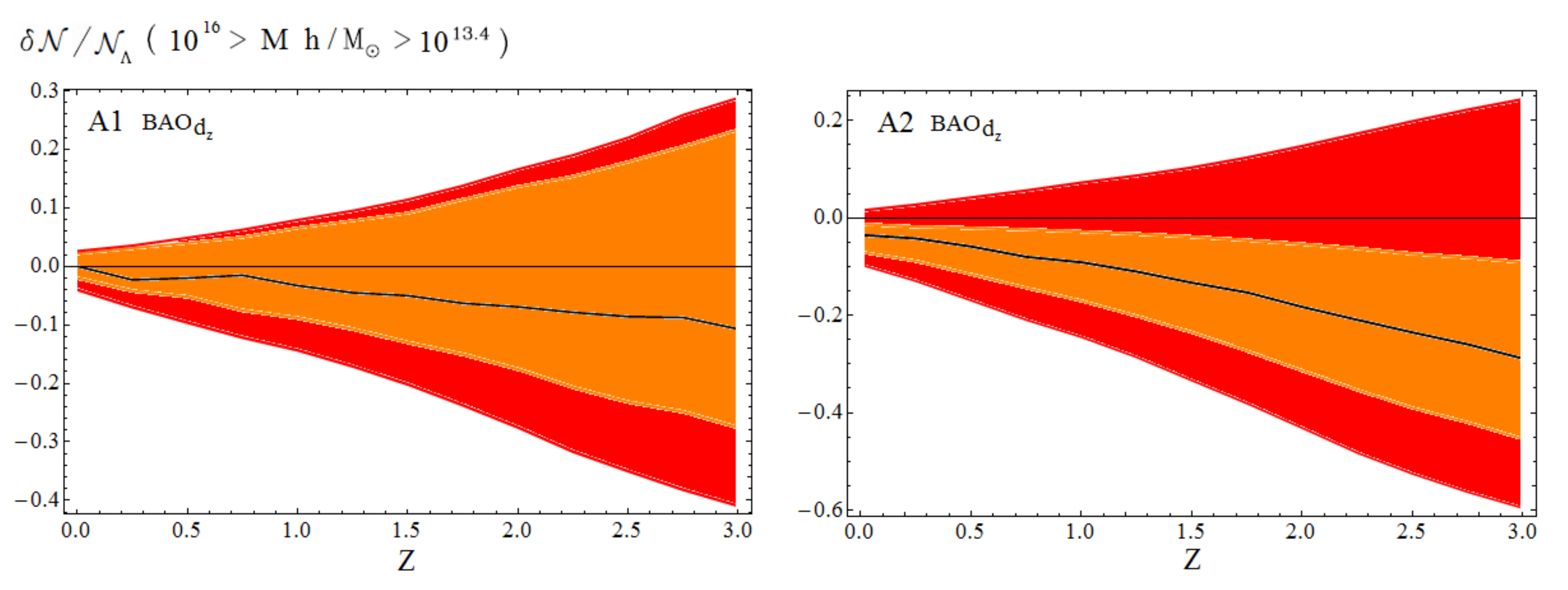}}
\end{center}
%{\includegraphics[scale=0.6]{Figures/fig2a.eps}}\ \ \ \
%{\includegraphics[scale=0.6]{Figures/fig2b.eps}}
\caption[]{\footnotesize{\textbf{Left}: Fractional difference $\delta {\cal N}/{\cal N}$ in the number of counts of clusters between the
vacuum model A1 and the concordance $\CC$CDM model (cf. Fig.\,\ref{fig:NLCDM}) using SNIa+CMB+BAO$_{dz}$ data from Table 1. The continuous solid line represents $\delta {\cal N}/{\cal N}$ for the best fit value from that table,  whereas the innermost (resp. outermost) band comprises the $\delta {\cal N}/{\cal N}$ prediction for the points within the $\pm 1\sigma$ (resp. $\pm 3\sigma$) values around it. \textbf{Right}: As before, but for model A2.}} \label{NC A1A2 BAO_dz}
\end{figure}
%
%%%%%%%%%%%%%%%%%%%%%%%%%%%%%%%%%%%%%%%%%%%%%%%%%%%%%%%%%%%%%%%%%%%%%%%%
%
where  $\sigma_{8,\Lambda}\simeq 0.829$ is the aforementioned $\CC$CDM value extracted from Planck+WP measurements. Furthermore, the numerical value of $\sigma_8$ for the $\CC$CDM and the various vacuum models under consideration has been collected in the last but one column of Tables \ref{TableModelsBAOdz} and \ref{TableModelsBAOA} together with the best fitting values of the parameters according to  each BAO type that we have used (cf. Tables \ref{tableFitBAOdz} and \ref{tableFitBAOA}).

The original Press-Schechter function $f_{\rm PSc}$ was shown to
provide a relatively good first approximation to the halo mass function
obtained by numerical simulations. In Appendix A we use  $f_{\rm
PSc}$ to assess in detail why the number count method is an
efficient one to separate models that may be difficult to
distinguish at the linear perturbation regime. The method, however,
is not tied to the particularly simple form of the original
Press-Schechter function $f_{\rm PSc}$.  More recently  a large
number of works have provided better fitting functions for
$f(\sigma)$. In practice, in our analysis for the various dynamical
vacuum models under consideration we will adopt the generalized one
proposed by  Reed et al. \cite{Reed2007}:
\begin{eqnarray}\label{PS-Reed}
f_R(\sigma, n_{\rm eff}) = A\sqrt{{2b\over\pi}}
\left[1+\left({\sigma^2\over b\delta_c^2}\right)^p + 0.6G_1 + 0.4G_2\right]\,\left({\delta_c\over\sigma}\right)\nonumber\\
\times\exp\left[-{cb\delta_c^2\over2\sigma^2} -{0.03 \over (n_{\rm
eff}+3)^2} \left({\delta_c \over \sigma}
\right)^{0.6} \right]\,,
\end{eqnarray}
where $A = 0.3222,\, p = 0.3,\, b = 0.707,\, c=1.08$, while
$G_1,\,G_2$ and $n_{\rm eff}$, {the slope of the non-linear
power-spectrum at the halo scale}, are given by:
\begin{equation}
G_1 = \exp\left[-{{(\ln\sigma^{-1}-0.4)^2} \over
{2(0.6)^2}}\right],\; G_2 = \exp\left[-{{(\ln\sigma^{-1}-0.75)^2}
\over
    {2(0.2)^2}}\right],\;
n_{\rm eff} = 6 {{\rm d}\ln\sigma^{-1}\over{\rm d}\ln M\phantom{+}}
-3.
\end{equation}

%%%%%%%%%%%%%%%%%%%%%%% FIGURE  2 %%%%%%%%%%%%%%%%%%%%%%%%%%%%%%%%%%%%%

\begin{figure}[!t]
\begin{center}
{\includegraphics[scale=0.45]{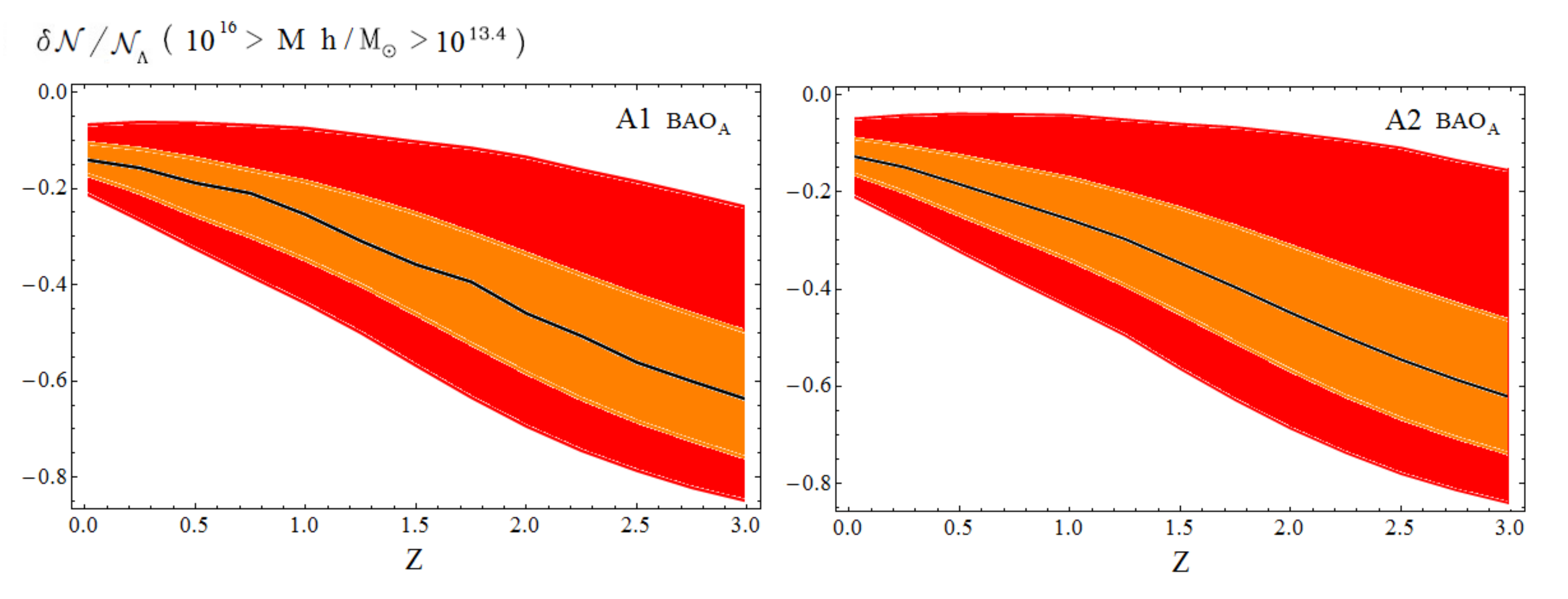}}
\end{center}
%{\includegraphics[scale=0.6]{Figures/fig2a.eps}}\ \ \ \
%{\includegraphics[scale=0.6]{Figures/fig2b.eps}}
\caption[]{\footnotesize{Fractional difference $\delta {\cal N}/{\cal N}$ in the number of counts of clusters between the
vacuum models A1 (left) and A2 (right) and the concordance $\CC$CDM model, but using SNIa+CMB+BAO$_{A}$ data from Table 2. Same notation as in Fig.\,\ref{NC A1A2 BAO_dz}.}} \label{NC:A1A2 BAO_A}
\end{figure}

%%%%%%%%%%%%%%%%%%%%%%%%%%%%%%%%%%%%%%%%%%%%%%%%%%%%%%%%%%%%%%%%%%%%%%%%
%
The previous generalized PS function is a refined variant of an
older function that was used to improve the original PS-formalism by
Sheth and Tormen \,\cite{STfunction1999}:
\begin{equation}\label{PS-ST}
f_{ST}(\sigma) = A'\sqrt{{2b\over\pi}} \left[1+\left({\sigma^2\over
b\delta_c^2}\right)^p \right]\,\left({\delta_c\over\sigma}\right)
\,\exp\left[-{b\delta_c^2\over2\sigma^2} \right]\,,
\end{equation}
where the parameters $b$ and $p$ are the same as in
(\ref{PS-Reed}). Once more $A'$ must be fixed from the normalization (\ref{PSnormalization}). Let us, however, note that the value of the normalization constant cancels in the ratio $\delta\mathcal{N}/\mathcal{N}_{\CC{\rm CDM}}$, where $\delta\mathcal{N}=\mathcal{N}-\mathcal{N}_{\CC{\rm CDM}}$ represents the deviations of the number counts of the given vacuum model with respect to the $\CC$CDM. In fact, the fractional difference $\delta\mathcal{N}/\mathcal{N}_{\CC{\rm CDM}}$ will be the main observable in our test analysis of the number counts for dynamical vacuum models.
%and we find  $A'=0.1611$.
While we have also made use of the parameterization (\ref{PS-ST}) to test the sensibility of our results to the generalized Press-Schechter functions, we will for definiteness only present the final results in terms of the more complete function (\ref{PS-Reed}).

To use that function we need to know the value of the collapse density threshold parameter $\delta_c$. In Appendix B we compute $\delta_c$ by solving the corresponding nonlinear perturbation equations for each vacuum model.  The resulting values  are listed in the last column of Tables \ref{TableModelsBAOdz} and \ref{TableModelsBAOA}, where we have separated them according to the type of BAO used in the best fitting to the SNIa+CMB+BAO cosmological data.

We conclude this section by noticing that the BBKS transfer function (\ref{jtf}), as well as the PS-like functions (\ref{PS-Reed}) and (\ref{PS-ST}) involved in the halo mass function, were all obtained from fits to numerical data assuming strict $\CC$CDM cosmology\,\cite{Bard86,Reed2007,STfunction1999}. We have checked e.g. that the differences in number counts between the mentioned mass functions are within the errors induced in the determination of the model parameters. In this sense we do not consider necessary at this point to further adapt the model dependence of these functions beyond our detailed computation of the $\delta_c$ parameter for each model (cf. Appendix B). The results of our number count analysis (see the next section) within these approximations are already quite suggestive of the rich spectrum of possibilities offered by the dynamical models under study. However, we understand that with the advent of more precision data in the future a more refined treatment might be necessary.

\subsection{Numerical results: number counts of the dynamical vacuum
models}\label{sec:HaloMasFunction}

From the halo mass function (\ref{MF}) we can derive for each vacuum
model the redshift distribution of clusters, ${\cal N}(z)$, within
some determined mass range, say $M_1\le M\le M_2$. This can be
estimated by integrating the expected differential halo mass
function, $n(M,z)$, with respect to mass, namely
\be\label{Nz} {\cal N}(z)=\frac{dV}{dz}\;\int_{M_{1}}^{M_{2}}
n(M,z)dM, \ee where $dV/dz$ is the comoving volume element, which in
a flat universe takes the form:
\be \frac{dV}{dz} =4\pi r^{2}(z)\frac{dr(z)}{dz}, \ee with $r(z)$
denoting the comoving radial distance out to redshift $z$: \be\label{rzdef}
r(z)=\frac{c}{H_{0}} \int_{0}^{z} \frac{dz'}{E(z')}.
%\;\;\;\;\;\frac{dr}{dz}=\frac{c}{H_{0}E(z)}
\ee
It follows that
\begin{equation}\label{Nzb}
\mathcal{N}(z)=4\pi r^2(z)\frac{dr}{dz}\int_{M_1}^{M_2}n(M,z)dM=-\frac{4\pi r^2\,\bar{\rho}(z)}{H_0 E(z)}\,\int_{M_1}^{M_2}\frac{1}{M}\left(\frac{1}{\sigma}\frac{d\sigma}{dM}\right)f(\sigma)dM\,.
\end{equation}
In practice, as we have said, we will use the function (\ref{PS-Reed}) for $f(\sigma)$ in the above expression.

%%%%%%%%%%%%%%%%%%%%%%%% FIGURE  2 %%%%%%%%%%%%%%%%%%%%%%%%%%%%%%%%%%%%%

\begin{figure}[!t]
\begin{center}
{\includegraphics[scale=0.45]{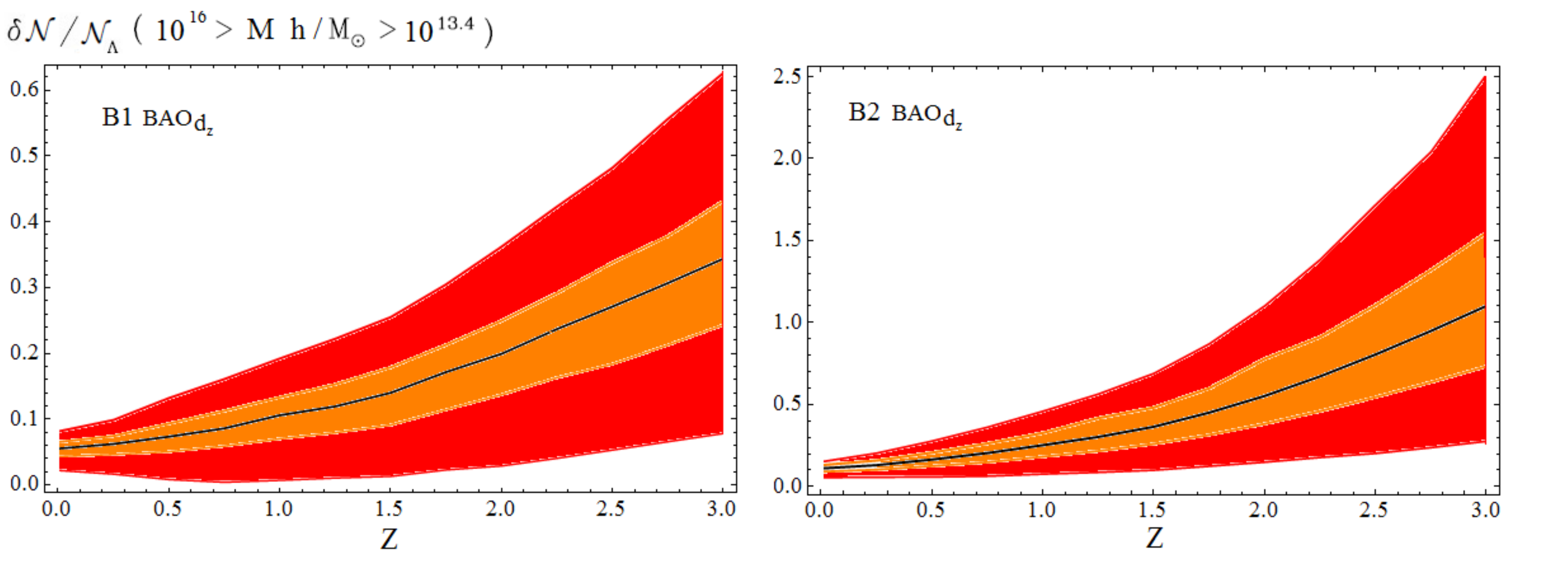}}
\end{center}
%{\includegraphics[scale=0.6]{Figures/fig2a.eps}}\ \ \ \
%{\includegraphics[scale=0.6]{Figures/fig2b.eps}}
\caption[]{\footnotesize{Fractional difference $\delta {\cal N}/{\cal N}$ in the number of counts of clusters between the
vacuum models B1 (left) and B2 (right) and the concordance $\CC$CDM model, using SNIa+CMB+BAO$_{dz}$ data from Table 1. Same notation as in Fig.\,\ref{NC A1A2 BAO_dz}.}} \label{NC:B1B2 BAO_dz}
\end{figure}

%%%%%%%%%%%%%%%%%%%%%%%%%%%%%%%%%%%%%%%%%%%%%%%%%%%%%%%%%%%%%%%%%%%%%%%%

In Fig.~\ref{fig:NLCDM}, we show the theoretically predicted redshift distribution of the total number of cluster counts, ${\cal N}(z)$, with masses in the range  $10^{13.4}\,h^{-1}\lesssim M/M_{\odot}\lesssim 10^{16}\,h^{-1}$, corresponding to the concordance $\CC$CDM model. Notice that there is no significant difference in the best fitted $\CC$CDM value of $\Omo$ when we employ SNIa+CMB+BAO$_{dz}$ or SNIa+CMB+BAO$_{A}$  data, as can be seen in Tables 1 and 2, and therefore the number of counts in Fig.~\ref{fig:NLCDM} does not depend on the BAO data used in the fit. We can see that the total number of counts increases with the redshift up to a maximum point and then decreases steadily, meaning that from that point onwards the larger is the redshift the smaller is the number of counts
of virialized halos with a mass $M$ in the indicated range. The curves shown in that figure (which include the $1\sigma$ error in the fitted value of $\Omo$) define the fiducial $\CC$CDM prediction. We will use it to compare with the corresponding outcome from the dynamical vacuum models under study. Recall that we denote the deviations of the number counts of a given vacuum model with respect to the $\CC$CDM as $\delta\mathcal{N}=\mathcal{N}-\mathcal{N}_{\CC{\rm CDM}}$.

We start our comparison by considering Fig.\,\ref{NC A1A2 BAO_dz}, where we display (on the left plot of it) the fractional difference $\delta {\cal N}/{\cal N}$ in the number of counts of clusters between the
vacuum model A1 and the concordance $\CC$CDM model (cf. Fig.\,\ref{fig:NLCDM}) using SNIa+CMB+BAO$_{dz}$ fitting data from Table 1.
The continuous solid line in the figure represents the predicted deviation $\delta {\cal N}/{\cal N}$ for the best fit value from that table,  whereas the inner and outer bands comprise the $\delta {\cal N}/{\cal N}$ prediction for the points within $\pm 1\sigma$ and  $\pm 3\sigma$ values around it, respectively. The plot on the right of Fig.\,\ref{NC A1A2 BAO_dz} is similar, but for model A2. The corresponding results for models A1 and A2 when SNIa+CMB+BAO$_{A}$ fitting data from Table 2 are used can be seen in the two plots of Fig.\,\ref{NC:A1A2 BAO_A}.

We can summarize the analysis presented in Figs.\,\ref{NC A1A2 BAO_dz} and \ref{NC:A1A2 BAO_A} by saying that $\delta {\cal N}/{\cal N}$  can have both signs in the case of using BAO$_{dz}$ data, provided we consider the points in the $\pm 3\sigma$ band. The narrower $\pm 1\sigma$ band is nevertheless more predominantly bent into the negative sign. As for the  BAO$_{A}$ data, the prediction for  $\delta {\cal N}/{\cal N}$ is negative for all points, even for those in the $\pm 1\sigma$ band. It means that, all in all, models A1 and A2 tend to predict a smaller number of counts as compared to the $\CC$CDM. The fractional decrease can be as significant as $30-60\%$.

%\FIGURE[t]{
%\begin{center}
%\includegraphics[scale=0.6]{NC.A1(BAO_dz)v1.eps}
%\caption{The expected cluster redshift distribution, {over the whole
%sky}, of the two running vacuum models for the case of the two
%future cluster surveys {\tt eROSITA} and SPT (upper panels), and the
%corresponding fractional difference with respect to the reference
%$\Lambda$CDM model (lower panels). The input parameters,
%corresponding to the central fit values, as well as the symbols or
%line-types characterizing the different models are shown in Table
%1.}
%%as in Fig.\,3, for the central fit values of
%%the $\nu$ parameter, see Table 1 for more details.}
%\label{figs:fig3}
%\end{center}
%}

%%%%%%%%%%%%%%%%%%%%%%%% FIGURE  2 %%%%%%%%%%%%%%%%%%%%%%%%%%%%%%%%%%%%%

\begin{figure}[!t]
\begin{center}
{\includegraphics[scale=0.45]{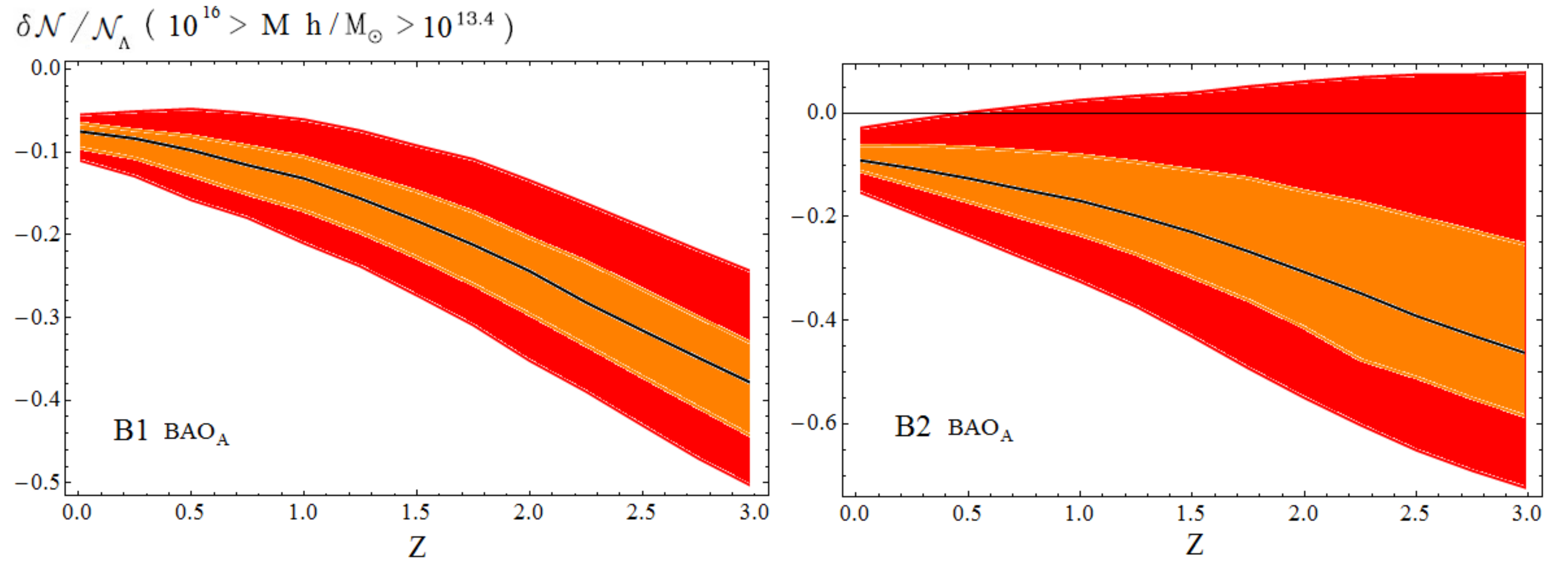}}
\end{center}
%{\includegraphics[scale=0.6]{Figures/fig2a.eps}}\ \ \ \
%{\includegraphics[scale=0.6]{Figures/fig2b.eps}}
\caption[]{\footnotesize{Fractional difference $\delta {\cal N}/{\cal N}$ in the number of counts of clusters between the
vacuum models B1 (left) and B2 (right) and the concordance $\CC$CDM model, using SNIa+CMB+BAO$_{A}$ data from Table 2. Same notation as in Fig.\,\ref{NC A1A2 BAO_dz}.}} \label{NC:B1B2 BAO_A}
\end{figure}

%%%%%%%%%%%%%%%%%%%%%%%%%%%%%%%%%%%%%%%%%%%%%%%%%%%%%%%%%%%%%%%%%%%%%%%%
%

The corresponding deviations in the number of counts for models B1 and B2 are depicted in Figs.\,\ref{NC:B1B2 BAO_dz} and \ref{NC:B1B2 BAO_A}. Here we find a feature that was not present for type-A models, namely we observe from these figures that the deviations with respect to the $\CC$CDM are all positive when the BAO$_{dz}$ data are used (cf. Fig.\,\ref{NC:B1B2 BAO_dz}), whilst they are negative when the BAO$_{A}$ data are utilized (cf. Fig. \,\ref{NC:B1B2 BAO_A}). This may seem surprising, but is related to the sensitivity of the number counts to the best fit value of $\Omo$ employed in the analysis, which is different for each type of BAO. As we have seen from Table 1, the BAO$_{dz}$ fitting data projects a value of $\Omo$ that is closer to the $\CC$CDM value than in the case of BAO$_A$ (cf. Table 2). In the latter, $\Omo$ is significantly smaller than in the $\CC$CDM model. The sign of $\delta{\cal N}/{\cal N}$ is tied to this fact. As it is shown in Appendix A, if a given vacuum model has the same $\Omo$ value (or very similar), the sign of $\delta{\cal N}/{\cal N}$ is opposite to the sign of the vacuum parameter $\nu$ or $\epsilon$ that dominates the model. For models B1 and B2 with BAO$_{dz}$ data, the best fit value of $\Omo$ is indeed very close to the fitted value for the $\CC$CDM.  Thus, since for these models  $\epsilon<0$ we find $\delta{\cal N}/{\cal N}>0$ and moreover this fraction is growing quite fast, up to $50-100\%$ and more (Fig.\,\ref{NC:B1B2 BAO_dz}). At variance with this situation, with BAO$_{A}$ data these models predict a substantial depletion in the number counts as compared to the $\CC$CDM, as shown in Fig.\,\ref{NC:B1B2 BAO_A}, the reason being the smaller preferred value of $\Omo$ as compared to the concordance model.

\begin{figure}[!t]
\begin{center}
{\includegraphics[scale=0.45]{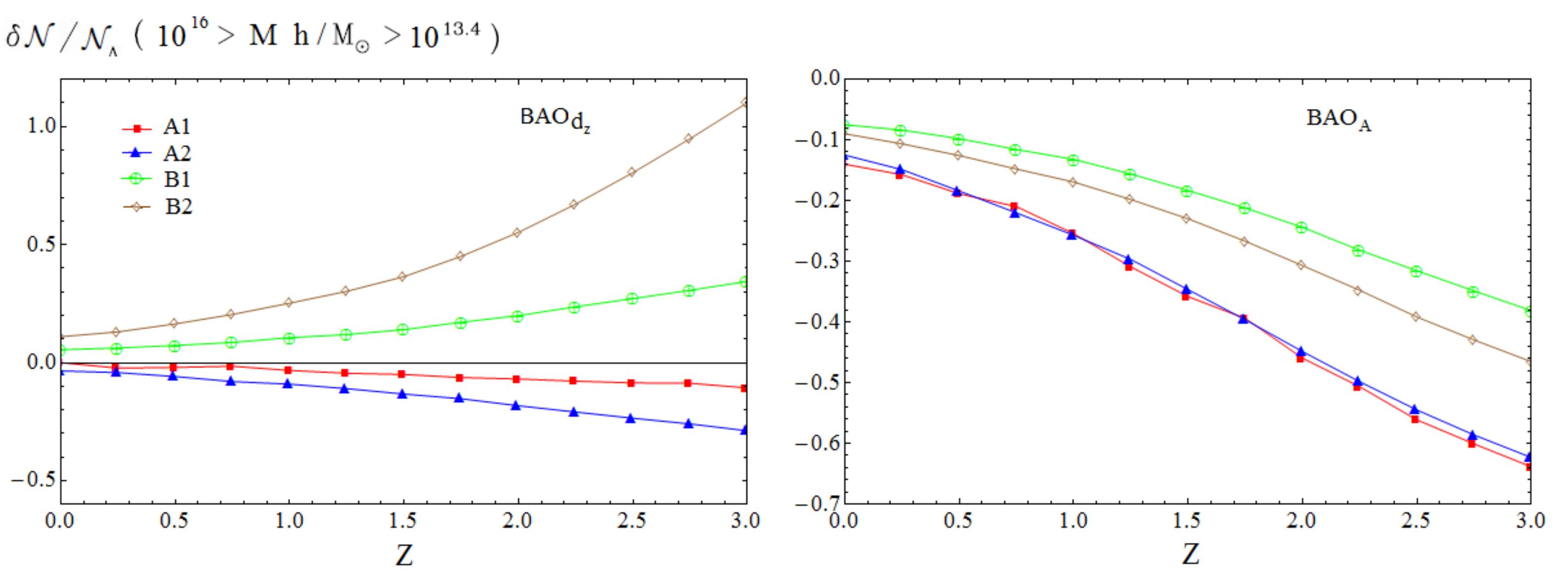}}
\end{center}
%{\includegraphics[scale=0.6]{Figures/fig2a.eps}}\ \ \ \
%{\includegraphics[scale=0.6]{Figures/fig2b.eps}}
\caption[]{\footnotesize{{\bf Left}: Comparison of the fractional difference $\delta {\cal N}/{\cal N}$ in the redshift distribution of cluster number counts of all the
vacuum models with respect to the concordance $\CC$CDM model using the central fit values of the SNIa+CMB+BAO$_{dz}$ data (cf. Table \ref{tableFitBAOdz}); {\bf Right}: As before, but for the SNIa+CMB+BAO$_{A}$ data (cf. Table \ref{tableFitBAOA}).}} \label{NC:CentralFits}
\end{figure}

In Fig.\,\ref{NC:CentralFits} we have put in a nutshell the essential results of our number counts analysis. Namely, we have displayed the fractional differences $\delta{\cal N}/{\cal N}$ with respect to the $\CC$CDM by using only the best fit values of all the vacuum models in the two BAO modalities. Obviously we need an improvement of the two sorts of BAO measurements to see it they can eventually provide a more coincident best fit value of $\Omo$, as this is essential to decide on the sign of $\delta{\cal N}/{\cal N}$. From our point of view perhaps the least model-dependent BAO results are those from BAO$_{A}$, as they are based on low-z data only and therefore are not so tied to the specific behavior of the models around the drag epoch.

%%%%%%%%%%%%%%%%%%%%%%% FIGURE  2 %%%%%%%%%%%%%%%%%%%%%%%%%%%%%%%%%%%%%

%%%%%%%%%%%%%%%%%%%%%%%%%%%%%%%%%%%%%%%%%%%%%%%%%%%%%%%%%%%%%%%%%%%%%%%%

Finally, as a particular case of our general treatment of type-A and type-B models, we briefly mention the situation with the class of C1 models (where e.g.  the linear model $\rL\propto H$ is also included). We have emphasized in Sect.\ref{sect:growthrate} that the C1 models perform a rather bad fit to the linear growth of density perturbations. Recently, however, the number count analysis of the linear model $\rL\propto H$ has been considered in Ref.\,\cite{Carn14}, in which a significant excess in the number of counts is reported as compared to the $\CC$CDM model.  Although it is not part of our main purpose, we have computed in passing the corresponding number of counts for this model. Unfortunately, we do not concur with the results of\,\cite{Carn14}. We do not find an excess in the number of counts as compared to the $\CC$CDM, but a large deficit.

To conclude, in view of the results found in our analysis of the cluster
halo redshift distribution presented in Figures \ref{NC A1A2 BAO_dz}-\ref{NC:CentralFits}, we can assert that it is an efficient method to distinguish the various sorts of dynamical vacuum models with respect to the $\CC$CDM and also among themselves, especially when the different sources of BAO data will become more precise. The sensitivity of the method to the parameters $(\nu,\alpha,\epsilon,...)$ of the vacuum models is large if we take into account that they are relatively small. We have found that the fractional differences $\delta{\cal N}/{\cal N}$ with respect to the $\CC$CDM can typically be as large as $\pm 50\%$ despite the fact that the (absolute) values of those parameters are typically of order $10^{-3}$.

\section{Discussion and conclusions} \label{sect:conclusions}

In this paper, we have analyzed in great detail several classes of dynamical vacuum
models in which the vacuum energy density can be expressed as a power series of
the Hubble function and its cosmic time derivative.
We have singled out model types which are particularly
attractive from the theoretical point of view, namely vacuum models for which the number of time derivatives of the scale
factor is even:
$\rL(t)=c_0+\sum_{k=1} \alpha_{k} H^{2k}(t)+\sum_{k=1}
\beta_{k}\dot{H}^{k}(t)$. These can be well motivated within the
context of quantum field theory (QFT) in curved spacetime since their structure is manifestly compatible with the general covariance
of the effective action and can be linked to the notion of
renormalization group. For the study of the current universe the series naturally terminates at the level of the $H^2$ and $\dot{H}$ terms, but the higher order ones can be very important for a proper description of the early universe and the inflationary phase.

We have stressed the need for the nonvanishing
additive (constant) term, $c_0\neq 0$, in the above class of models. It guarantees a smooth limit converging to the standard $\CC$CDM model when the
coefficients of the dynamical terms go to zero. We have verified
that models with  $c_0=0$ are generally in conflict with
observations, whether with the background data, or with the structure
formation data, or both.

For instance,  we have considered vacuum models of the form
$\CC=a_0+a_1\dot{H}+a_2H^2$, with $c_0\neq 0$ (the class of models that we have called
type-A). They are well-behaved and if the (dimensionless) coefficients $a_1$ and $a_2$ are
sufficiently small, the cosmological term develops just a mild dynamical behavior around
the $\CC$CDM model. Such framework could compete as a good candidate for a consistent description of the Universe in terms of dynamical vacuum energy, an option that should be considered
natural in QFT in curved space-time.
%\,\cite{SS-old1,SS09,MiniReview11}
%-- see e.g.\,\cite{JSP-CCReview13} for a recent and comprehensive review.

In our analysis we have also admitted the possibility that some
terms in the effective structure of $\CC(H,\dot{H})$ could mildly violate
the covariance requirement on phenomenological grounds.
Notwithstanding, we considered this possibility viable only when the expected terms are also present. We do not deem
theoretically sound those vacuum models exclusively constructed from
noncanonical terms (i.e. unexpected terms not satisfying the above mentioned conditions), such as e.g. the model $\CC\propto H$. A model of this
sort has the double inconvenience that $c_0=0$ and that the number
of time derivatives of the scale factor is odd (one derivative in this case). Not surprisingly when
such model is confronted with observations fails on
several accounts. When we add up to it the power $H^2$, we reach
the model $\CC=c_1H+c_2H^2$ (referred to in this work as the type-C1 model). In this extended form the situation of the new model improves at the background level, but is still troublesome at the perturbations level since the model fails to describe the linear growth of structure formation. Similarly, the pure quadratic model
$\CC\propto H^2$ is problematic, but for a different reason. While this
model contains an even power of the Hubble rate, it is actually not sufficient to comply with the phenomenological requirement in the absence of a constant
additive term. The reason is that it does not admit an
inflection point from deceleration to acceleration, and moreover it
does not have a growing mode for structure formation (for reasonable
values of the cosmological parameters). The pure linear and quadratic models, $\CC\propto H$ and $\CC\propto H^2$, are therefore strongly excluded; and their combination, $\CC=c_1H+c_2H^2$, provides a model still considerably crippled to account for the structure formation data. Similar (though not identical) criticisms can be applied to vacuum models of the sort $\CC=c_1\dot{H}+c_2H^2$ (type-C2). Hence, all type-C models are  unfavored, strongly disfavored or simply ruled out.

We have already mentioned that type-A dynamical vacuum models are in very good shape inasmuch as they are perfectly comparable to the concordance $\CC$CDM model when the dynamical components are subdominant in the current universe. Interestingly, another viable variant that we have considered are the type-B models. These are obtained by including an additive term to the type-C1 models, i.e. they have the structure $\CC=b_0+b_1H+b_2H^2$ with $b_0\neq 0$. We have found that they are also, in principle, phenomenologically admissible.
For them the presence of the linear component is not so determinant as in the case of the type-C1, because it can be interpreted as a correction (e.g. a bulk viscosity effect) to the main structure. Most important, the type-B models have a smooth $\CC$CDM limit as in the type-A case, and this fact is again crucial to protect them from departing exceedingly from the concordance $\CC$CDM model near our time.

In the present work we have solved the background and perturbations
cosmology for all these vacuum models and confronted them with
observations.
%thus extending the previous studies of simpler versions considered in the literature\,\cite{BPS09,GSBP11}.
In the light of the most recent observational data on type Ia supernovae,
the Cosmic Microwave Background and the Baryonic Acoustic
Oscillations (BAO), we have obtained a fit to their basic parameters $(\nu, \alpha, \epsilon)$.  From the fitted values we have computed the linear growth factor of structure formation for each model and compared with the observed
linear growth rate of clustering measured from the SDSS galaxies.
Subsequently we have moved to the nonlinear regime and considered
the predicted redshift distribution of cluster-size collapsed
structures as a powerful method to distinguish the models. We have computed the corresponding fractional deviation $\delta {\cal N}/{\cal N}$ in the number of counts of clusters with respect to the $\CC$CDM prediction.

The general conclusion we have reached is that the studied dynamical vacuum models (type-A and
type-B with nonvanishing additive constant term) are able to pass (with some differences) the combined observational tests, including the structure formation data, with a statistical significance that in some cases is comparable or even better than that of the
concordance $\CC$CDM model. The current Universe appears in all these models as FLRW-like, except that the vacuum energy is not a rigid quantity but a mildly evolving one. In fact, the typical values we have obtained for the coefficients $\nu$, $\alpha$ and $\epsilon$ responsible for the
time evolution of $\rL$ in these models lie in the ballpark of $\sim
10^{-3}$. This order of magnitude value is roughly consistent with the theoretical expectations, some of them interpreted in QFT as one-loop $\beta$-functions of the running cosmological constant.

Despite the two types of viable dynamical vacuum models remain close to the $\CC$CDM model, the overall fit from type-B models is not so good as the type-A ones. We have pointed out that this may be due to the fact that the presence of the linear term $\sim \epsilon H$ (characteristic of type-B models, especially the type-B1 ones) is unexpected in the general structure of the effective action in QFT in curved spacetime. This is in contradistinction to the vacuum structure of type-A models, where all included terms are expected. Overall, this feature might be indicative that the A-class of models are both theoretically and phenomenologically preferred to the B-ones. However, it is too early for a final verdict, and more observational work may be necessary to decide. In the meanwhile we have shown that the two types of models could be distinguished from the point of view of the measured redshift distribution of cluster-sized collapsed structures in the Universe. We have found that they can show significant deviations (of order $\pm 50\%$) from the predicted redshift distribution in the concordance $\CC$CDM model. Our expectation is that when the upcoming and present X-ray and Sunyaev-Zeldovich surveys (such as eROSITA and SPT) will have collected enough statistics, it should be possible to decide about the best type of dynamical vacuum model from the phenomenological point of view.

In the course of our analysis we have also briefly pointed out the fact that generally the dynamical models under consideration in this paper, and especially when fitted using the BAO$_A$ observable (which depends on low-redshift data on the acoustic $A(z)$-parameter), tend to provide a value of $\Omo$ significantly smaller than in the $\CC$CDM model. This would seem to be consistent with the possible dynamical character of the dark energy recently claimed in the literature on the basis of model-independent DE diagnostics\,\cite{SahniShafielooStarobinsky2014}.

To summarize, the dynamical vacuum models of the cosmic evolution may
offer an appealing and phenomenologically consistent perspective
for describing dynamical dark energy without introducing
extraneous dark energy fields. In that framework, dark energy is reinforced as
being nothing more, but nothing less, than dynamical $\CC$. This
could help to better understand the origin of the $\CC$-term and the vacuum energy density in the fundamental context of QFT in curved spacetime. Ultimately, it should shed light on the old cosmological constant problem, or at least provide a hint to elucidate the puzzling cosmic coincidence of the current matter and vacuum energy densities.

\vspace{0.5cm}

\acknowledgments The work of AGV has been partially supported by
an APIF predoctoral grant of the Universitat de Barcelona. JS has
been supported in part by FPA2013-46570 (MICINN), Consolider grant
CSD2007-00042 (CPAN) and by DIUE/CUR (Generalitat de Catalunya). SB acknowledges support by the Research Center for
Astronomy of the Academy of Athens
in the context of the program {\it Tracing the Cosmic Acceleration}.

\appendix

\section{Understanding how the cluster number counts method works}\label{sec:understnadingNumberCount}

We have shown that type A and B models of the vacuum energy
successfully fit all known cosmological data, including linear
structure formation, in a way comparable to the $\CC$CDM. However we
would like to find a way to lift their alike performance and be able
to distinguish them in a practical way. The
number counts method is a good method to accomplish this aim. To
understand semianalytically why the method works, it will suffice to
consider the original Press-Schechter function defined in
Sect.\,\ref{sec:PSformalism}. For convenience let us define the
ratios
\begin{equation}
{\cal
T}(M)\equiv\frac{\int_{0}^{\infty}{k^{n+2}T^2(\Omega_m^{(0)},k)W^2(kR)dk}}{\int_{0}^{\infty}{k^{n+2}T^2(\Omega_m^{(0)},k)W^2(kR_8)dk}}\quad
{\rm and}\quad D_N(z)\equiv\frac{D(z)}{D(0)}\,.
\end{equation}
In this way from (\ref{s888}) we have
$\sigma^2(z)=\sigma^2_8\,D_N^2(z)\,{\cal T}(z)$, and we can rewrite
(\ref{Nzb}) as follows:
\begin{equation}\label{N(z)}
\mathcal{N}(z)=-\frac{4\pi r^2\,\bar{\rho}}{H_0 E(z)}\,\int_{M_1}^{M_2}\frac{1}{M}\left(\frac{1}{\sigma}\frac{d\sigma}{dM}\right) f_{PSc}(\sigma)dM
=-4\sqrt{2\pi}\bar{\rho}\frac{c}{H_0}\left(\frac{\delta_c(z)r^2(z)}{E(z)\sigma_8
D_N(z)}\right)\,I^{(\nu)}\,,
\end{equation}
with $r(z)$ given in Eq.(\ref{rzdef}).
In the last step we have used explicitly the original form of the
Press-Schechter function $ f_{PSc}(\sigma)$, and we have defined the
integral
\begin{equation}\label{defIntegral}
I^{(\nu)}\equiv \int_{M_1}^{M_2}\frac{dM}{M}\frac{1}{{\cal
T}}\frac{d\sqrt{{\cal T}}}{dM}e^{-\frac{\delta_c^2}{2\sigma_8^2D_N^2
{\cal T}}}\,.
\end{equation}
Using the generalized forms (\ref{PS-Reed}) or (\ref{PS-ST}) does
not alter the explanation why the method works, and for this reason
we restrict ourselves to the canonical one.

The variations with respect to the $\CC$CDM model  should come from
the variations in the terms in the big parenthesis on the
\textit{r.h.s.} of Eq.\,(\ref{N(z)}), as well as from the
integral (\ref{defIntegral}). The other ingredients of
$\mathcal{N}(z)$ should not depend on the model details in a
significant way. Let us assume that there is only one parameter in
the dynamical vacuum model, say $\nu$. Expanding around $\nu=0$,
i.e. around the $\CC$CDM case, we can get the departure terms:
\begin{equation}
\left(\frac{r^2(z)}{E(z)}\right)^{(\nu)}=\left(\frac{r^2(z)}{E(z)}\right)^{(0)}
+\delta a_1\,;\ \ \ (\sigma_8 D_N(z))^{(\nu)}=(\sigma_8
D_N(z))^{(0)}+\delta a_2\,;\ \ \
\delta_c^{(\nu)}=\delta_c^{(0)}+\delta a_3 \,.
\end{equation}
Notice that all the $\delta a_i$ in the previous expression are
proportional to $\nu$, and therefore very small compared to the
leading terms.  Let us warn the reader that it would be
inappropriate to expand the exponential in the integrand of
(\ref{N(z)}) in the same way, as the linear approximation would
be insufficient for the typical values of $\nu$ found in our
analysis.  The number counts formula (\ref{N(z)}) therefore
yields
\begin{eqnarray}
\mathcal{N}(z)&=&-4\sqrt{2\pi}\bar{\rho}\frac{c}{H_0}\left(\frac{\delta_c(z)r^2(z)}{E(z)\sigma_8 D_N(z)}\right)^{(0)}\nonumber\\
&&\times\left[1+\delta
a_1\left(\frac{E(z)}{r^2(z)}\right)^{(0)}-\frac{\delta
a_2}{(\sigma_8 D_N(z))^{(0)}}+\frac{\delta
a_3}{\delta_c^{(0)}}+\mathcal{O}(\nu^2)\right]\, I^{(\nu)}\,.
\end{eqnarray}
In this way we can compute the  variation in the number of clusters
(at a given redshift) with respect to the $\CC$CDM, i.e.
$\delta\mathcal{N}=\mathcal{N}(\nu)-\mathcal{N}(\nu=0)$. The
corresponding relative variation can be cast as
\begin{equation}\label{N(z)2}
\frac{\delta\mathcal{N}}{\mathcal{N}}=\underbrace{\frac{I^{(\nu)}-I^{(0)}}{I^{(0)}}}_{T0}+\underbrace{\delta
a_1\left(\frac{E}{r^2}\right)^{(0)}\frac{I^{(\nu)}}{I^{(0)}}}_{T1}-\underbrace{\frac{\delta
a_2}{(\sigma_8D_N)^{(0)}}\frac{I^{(\nu)}}{I^{(0)}}}_{T2}+\underbrace{\frac{\delta
a_3}{\delta_c^{(0)}}\frac{I^{(\nu)}}{I^{(0)}}}_{T3},.
\end{equation}
\begin{table}[!t]
\centering
\begin{tabular}{|c|c|c|c|c|c|c|c|c|}
\hline
$\nu$ & $\sigma_8$ & $D_N(z=2)$ & $\delta_c(z=2)$ & $T0$ & $T1$ & $T2$ & $T3$ & $\frac{\delta\mathcal{N}}{\mathcal{N}}$ \\
\hline
-0.0017 & 0.829 & 0.4315 & 1.695 & 0.096 & -0.0034 & -0.015 & 0.0065 & 0.084\\
\hline
0.0017 & 0.794 & 0.4362 & 1.675 & -0.139 & 0.0027 & 0.016 & -0.0051 & -0.125\\
\hline
-0.004 & 0.854 & 0.4282 & 1.710 & 0.276 & -0.0092 & -0.047 & 0.0189 & 0.239\\
\hline
0.004 & 0.770 & 0.4394 & 1.660 & -0.278 & 0.0052 & 0.030 & -0.0107 & -0.254 \\
\hline
\end{tabular}
\caption[]{Numerical evaluation of
${\delta\mathcal{N}}/{\mathcal{N}}$, i.e. the relative variation in
the number of counts as compared to the $\CC$CDM, see Eq.
(\ref{N(z)2}). We consider different values of the $\nu$
parameter at fixed  $z=2$ and provide also the breakdown of the
result in the individual contributions $T0-T3$. To illustrate the
method we have used the consistent set of inputs:
$\Omega_b(z=0)=0.022242h^{-2}$, $\Omega_m(z=0)=0.284$,
$\delta_c^{(0)}=1.675$, $\sigma_8^{(0)}=0.811$ and
$D_N^{(0)}(z=2)=0.4351$. } \label{Appe1}
\end{table}
The numerical evaluation of the various terms of this expression is
displayed in Table \ref{Appe1}. It clearly shows that the dominant
term is $T0$ in (\ref{N(z)2}). The terms $T1-T3$ are all of them
proportional to $\delta a_i$ and hence to $\nu$. Since $\nu={\cal
O}(10^{-3})$ all the terms proportional to it are of the same order
of magnitude. The $T0$-term is not, and it becomes the leading one.
Here is where the main contribution comes from, which is typically
two orders of magnitude larger than $\nu$ and hence it can reach the
order $10\%$ rather than $1$ per mil. This feature is at the root of
the main difference of this method with respect to the linear
perturbations analysis. In the latter  the deviations of the
dynamical vacuum models with respect to the linear growth rate
of the $\CC$CDM are proportional to $\nu$ and therefore cannot be
distinguished. Here, instead, the relative differences become magnified thanks to the nonperturbative effects associated to (\ref{defIntegral}). In addition, we note from Table \ref{Appe1} that
there are significant differences for different values of $\nu$
within the same order of magnitude, which are also sensitive to sign
changes of the parameter. In the present case the sign of $\delta{\cal N}$ is opposite
to the sign of $\nu$, but this is because the value of $\Omo$ for the dynamical vacuum model that we have analyzed is the same as in the $\CC$CDM, but in general there is no such sign correlation. What is important is that using the number count method we expect visible effects that would remain almost invisible in the linear approach owing to the small values of the model parameters. This
is corroborated in the numerical analysis presented in Sect.\,\ref{sec:HaloMasFunction}.

\section{Computing the collapse density threshold $\delta_c$ for the dynamical vacuum models}\label{sec:app}

\setcounter{equation}{0}

In this appendix we present the necessary formulas to compute the
linearly extrapolated density threshold above which structures
collapse, i.e, $\delta_{c}$, for the type-A and type-B dynamical
vacuum models under study. We follow the standard methods available
in the literature -- see e.g. \cite{Pace10} and \cite{Abramo}, and
references therein.  Details of the procedure were also amply
provided in Ref. \cite{GSBP11} where it was applied to other
dynamical vacuum models. We will therefore not repeat these details,
but just the initial setup and the final results for the non-linear
perturbation equations corresponding to the models under
consideration. The linearized part of these equations reduces, of
course, to the perturbations equations that we have derived
previously in Section \ref{sect:perturbations}. Recall that in order
to derive the nonlinear equations it is convenient to start from
the Newtonian formalism for the cosmological fluid, and use  the
continuity, Euler and Poisson equations in the matter dominated
epoch:
\begin{eqnarray}
  \frac{\partial\rho_m}{\partial t}+\nabla_{\vec{r}}\cdot(\rho_m\vec{v})= 0
\label{eqn:cnpert}\;,\\
  \frac{\partial\vec{v}}{\partial
    t}+(\vec{v}\cdot\nabla_{\vec{r}})\,\vec{v}+
  \nabla_{\vec{r}}\,\Phi=0\label{eqn:enpert}\;,\\
  \nabla^2\Phi=4\pi G_{\!N}\sum_i\rho_i(1+3\omega_i)\;,\label{eqn:pnpert}
\end{eqnarray}
where $\vec v$ is the total velocity of the co-moving observer in
three-space, $\Phi$ is the Newtonian gravitational potential,
$\vec{r}$ is the physical coordinate, and  $\omega_i=p_i/\rho_i$ is
the EoS parameter for each component. Introducing comoving
coordinates $\vec{x}=\vec{r}/a$  the perturbations are defined in
the following way:
\begin{eqnarray}
\rho_i(\vec{x},t) & = & \bar{\rho}_i(t)+\delta\rho_i(\vec{x},t)=\bar{\rho}_i(t)(1+\delta_i(\vec{x},t))\;, \label{eqn:rpert} \\
\Phi(\vec{x},t) & = & \Phi_0(\vec{x},t)+\phi(\vec{x},t)\;, \label{eqn:fpert}\\
 \vec{v}(\vec{x},t) & = & a(t)[H(t)\vec{x}+\vec{u}(\vec{x},t)]\;, \label{eqn:vpert}\;.\label{eqn:gpert}
\end{eqnarray}

Here $\vec{u}(\vec{x},t)$ is the comoving peculiar velocity.  Next
we have to insert inserts Eqs.~(\ref{eqn:rpert})--(\ref{eqn:gpert})
into Eqs.~(\ref{eqn:cnpert})--(\ref{eqn:pnpert}) and use the
definition of the gradient with respect to co-moving coordinates.
Notice that $\vec{\nabla}\delta_m=0$, which holds for the spherical
collapse of a top-hat distribution. Following the same systematics
as described in  \cite{GSBP11} we arrive at the nonlinear
perturbations equations for the models under consideration.

%%%%%%%%%%%%%%%%%%%%%%% FIGURE  2 %%%%%%%%%%%%%%%%%%%%%%%%%%%%%%%%%%%%%

\begin{figure}[!t]
\begin{center}
{\includegraphics[scale=0.45]{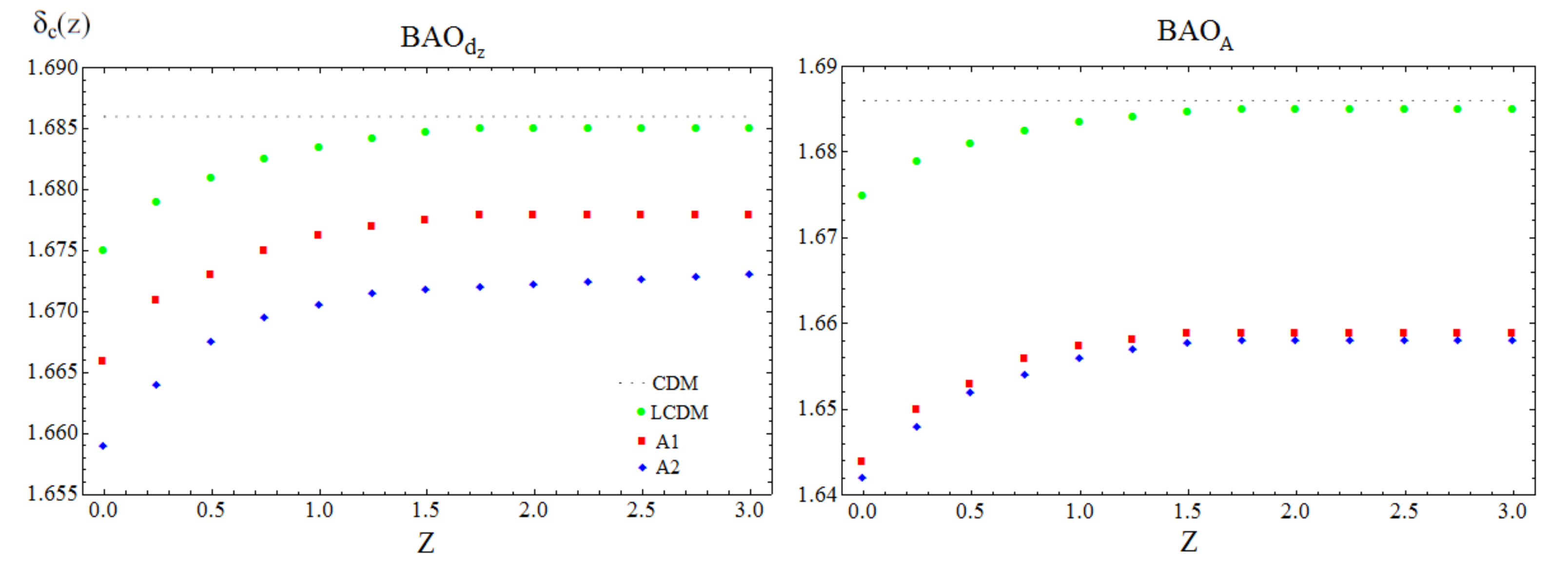}}
\end{center}
%{\includegraphics[scale=0.6]{Figures/fig2a.eps}}\ \ \ \
%{\includegraphics[scale=0.6]{Figures/fig2b.eps}}
\caption[]{\footnotesize{Computation of the collapse density threshold function $\delta_c(z)$ using the best fit values to SNIa+CMB+BAO$_{dz}$ data (left plot) and to SNIa+CMB+BAO$_{A}$ data (right plot). In both plots we include the constant CDM value $\delta_c=\frac{3}{20}(12\pi)^{2/3}\simeq 1.686$ (horizontal dotted line) as well as the $\CC$CDM curve (solid points, in green). The $\delta_c(z)$ curves for the vacuum models A1 and A2 are represented with squares (in red) and with  diamonds (in blue), respectively. The corresponding values at $z=0$ define $\delta_c$ for each model, and are indicated in the last column of Table \ref{TableModelsBAOdz}.} \label{deltac BAOdz}}
\end{figure}

%%%%%%%%%%%%%%%%%%%%%%%%%%%%%%%%%%%%%%%%%%%%%%%%%%%%%%%%%%%%%%%%%%%%%%%%

\subsection{Type-A models}

In this case the corresponding fully non-linear evolution equation
reads as follows:

$$a^2H^2\delta^{\prime\prime}_m+aH\delta_m^\prime\left[3H+Q-\frac{\rho_m}{2H}\right]+\delta_m\left[2HQ+aHQ^\prime-\frac{\rho_m}{2}(1+\delta_m)\right]-$$
\begin{equation}\label{SDE}
-\left[\frac{4a^2H^2\delta_m^{\prime
2}+5aHQ\delta_m\delta_m^\prime+Q^2\delta_m^2}{3(1+\delta_m)}\right]=0
\end{equation}

\noindent where the primes continue denoting derivatives with respect to the
scale factor and
\begin{equation}
Q(a)=3H(a)(1-\xi).
\end{equation}
\noindent The formulas for the non-relativistic matter energy
density $\rmr$ and the Hubble function $H$ can be found in Sect.
\ref{sect:solvingA1A2}.  The numerical solution of the above nonlinear equation is used to compute $\delta_c(z)$ for models A1 and A2 in Fig.\,\ref{deltac BAOdz}, see Sect.\,\ref{sect:numericaldeltac} for details.

\subsection{Type-B models}

For this type of models the nonlinear equation for the perturbations
can be obtained with some extra effort since on this occasion the
calculations cannot be performed analytically in terms of the scale
factor. We write the final result using the variable $y$, which has
been defined in \ref{defx}. We find:
$$\frac{9}{16}H_0^2\mathcal{F}^2(1-y)^2\delta_m^{\prime\prime}+\frac{3}{4}H_0\mathcal{F}\delta_m^\prime(1-y^2)\left[2H+Q-\frac{3}{2}yH_0\mathcal{F}\right]+$$
$$+\left[2HQ+\frac{3}{4}H_0\mathcal{F}(1-y^2)Q^\prime-\frac{\rho_m}{2}(1+\delta_m)\right]\delta_m-\frac{Q^2\delta_m^2}{3(1+\delta_m)}-$$
\begin{equation}\label{SDE2}
-\left[\frac{4\left(\frac{3}{4}H_0\mathcal{F}(1-y^2)\delta_m^\prime\right)^2+
\frac{15}{4}QH_0\mathcal{F}(1-y^2)\delta_m\delta^\prime_m}{3(1+\delta_m)}\right]=0\,,
\end{equation}
where the primes indicate on this occasion derivatives with respect to $y$, defined in
Eq.\,(\ref{defx}) -- the notation should not be confusing with the previous use of primes since we make explicit the argument. The expressions for $Q(y)$, $\rho_m(y)$ and
$H(y)$ for the type-B models can be found in
Sect.\ref{sect:perturbationsTypeB}. The numerical solution of the above nonlinear equation is used to compute $\delta_c(z)$ for models B1 and B2 in Fig.\,\ref{deltac BAOA}, cf. Sect.\,\ref{sect:numericaldeltac}.

The corresponding nonlinear
equation for type-C1 models is a particular case of
Eq.\,(\ref{SDE2}) and is obtained as indicated in
Sect.\,\ref{sect:solvingC1C2}. In particular, for $\epsilon=\OLo $
and $\nu=0$ (hence $\mathcal{F}=\OLo$) we obtain the corresponding
equation for the pure linear model $\rL\propto H$, which we have
ruled out. We shall not consider the computation of the number counts for these models here.

\subsection{Numerical procedure to determine $\delta_c$}
\label{sect:numericaldeltac}

%%%%%%%%%%%%%%%%%%%%%%%% FIGURE  2 %%%%%%%%%%%%%%%%%%%%%%%%%%%%%%%%%%%%%

\begin{figure}[!t]
\begin{center}
{\includegraphics[scale=0.45]{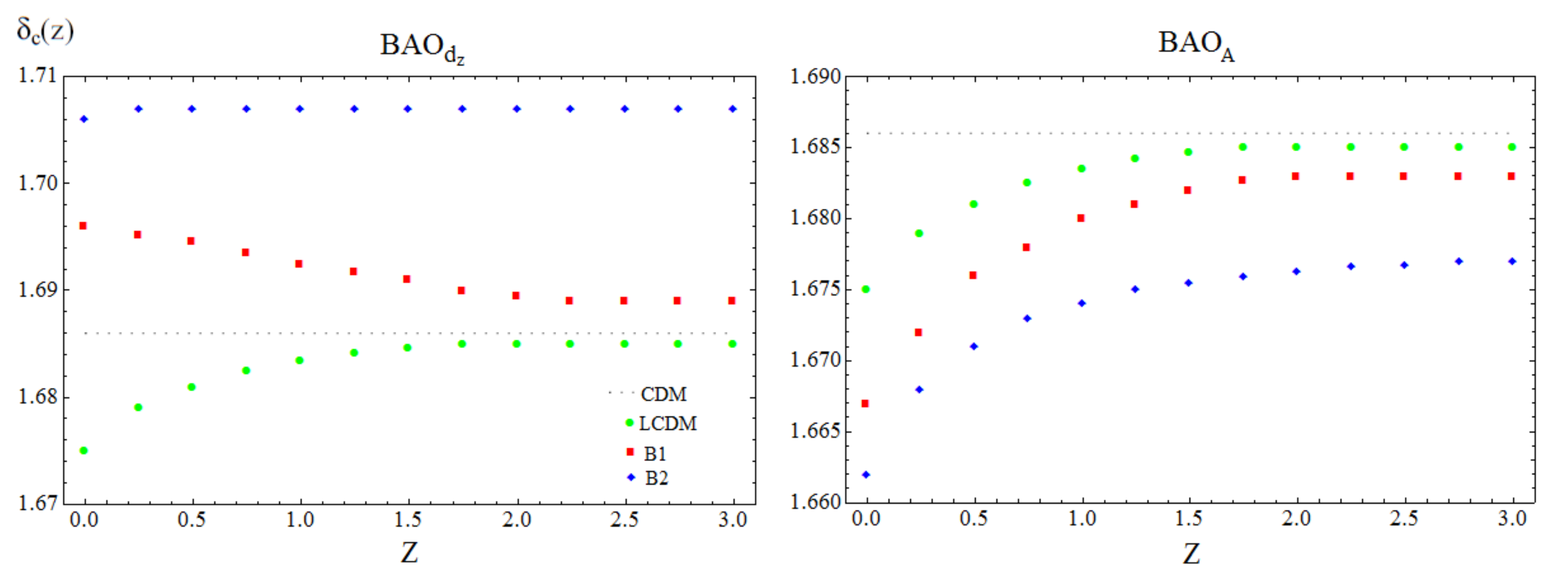}}
\end{center}
%{\includegraphics[scale=0.6]{Figures/fig2a.eps}}\ \ \ \
%{\includegraphics[scale=0.6]{Figures/fig2b.eps}}
\caption[]{\footnotesize{Computation of the collapse density threshold function $\delta_c(z)$} for the B1 and B2 vacuum models. The rest of the notation is as in Fig.\,\ref{deltac BAOdz}. The values of $\delta_c(z)$ at $z=0$ are indicated in the last column of Table \ref{TableModelsBAOA}. \label{deltac BAOA}}
\end{figure}

%%%%%%%%%%%%%%%%%%%%%%%%%%%%%%%%%%%%%%%%%%%%%%%%%%%%%%%%%%%%%%%%%%%%%%%%

Next we follow the prescriptions of \cite{Pace10}, which was also
described in detail (and applied to previous vacuum models) in
\cite{GSBP11}. We compute $\delta_c(z_f)$  by numerically
integrating the above nonlinear equations between $z_i$ and $z_f$
(where the initial redshift $z_i$ is sufficiently large, for
instance $10^{6}$). The aim is to find the initial value
$\delta_m(z_i)$ for which the collapse takes place at $z=z_f$, i.e.
such that $\delta_m(z_f)$ is very large, say $10^5$ or $10^9$ (the
result does not change significantly). Second, we use the previously
determined value of $\delta_m(z_i)$ together with a small value of  $\delta'_m(z_i)$. In fact, we know it is zero for a sphere, so we may take $\delta'_m(z_i)\sim 10^{-6}-10^{-4}$\,\cite{Pace10}. These are then used as the initial conditions for solving the corresponding linear perturbations equations. The value
of $\delta_m(z_f)$ obtained in the second step of this procedure
defines $\delta_c(z_f)$, and the value of this quantity at $z_f=0$
defines $\delta_c\equiv\delta_c(z_f=0)$.  Notice that the linear
equations are simply obtained from  (\ref{SDE}) and (\ref{SDE2}) upon neglecting all terms of ${\cal O}(\delta_m^2)$ and ${\cal
O}(\delta_m'^2)$. The equations obtained in this way are, of course,
the ones already presented in Sect.\,\ref{sect:perturbations} for
both types of models A and B.  The values of $\delta_c$ obtained by
this method for each model are displayed in the last column of Tables \ref{TableModelsBAOdz} and \ref{TableModelsBAOA}
(cf. Sect. \ref{sec:HaloMasFunction}).  The numerical solutions $\delta_c(z)$ for each model are displayed in Figs.\,\ref{deltac BAOdz} and \ref{deltac BAOA}.

%%%%%%%%%%%%%%%%%%%%%%%%%%%%%%%%%%%%%%%%%%%%%%%%%%%%%%%%%%%%%%%%%%%%%%
%%%%
%%%%%%%%%%%%%%%%%%%%%%%%%%%%%%%%%%%%%%%%%%%%%%%%%%%%%%%%%%%%%%%%%%%%%%
%\newcommand{\JHEP}[3]{{ J. of High Energy Physics } {JHEP} {#1} (#2)  {#3}}
\newcommand{\JHEP}[3]{ {JHEP} {#1} (#2)  {#3}}
\newcommand{\NPB}[3]{{ Nucl. Phys. } {\bf B#1} (#2)  {#3}}
\newcommand{\NPPS}[3]{{ Nucl. Phys. Proc. Supp. } {\bf #1} (#2)  {#3}}
\newcommand{\PRD}[3]{{ Phys. Rev. } {\bf D#1} (#2)   {#3}}
\newcommand{\PLB}[3]{{ Phys. Lett. } {\bf B#1} (#2)  {#3}}
\newcommand{\EPJ}[3]{{ Eur. Phys. J } {\bf C#1} (#2)  {#3}}
\newcommand{\PR}[3]{{ Phys. Rep. } {\bf #1} (#2)  {#3}}
\newcommand{\RMP}[3]{{ Rev. Mod. Phys. } {\bf #1} (#2)  {#3}}
\newcommand{\IJMP}[3]{{ Int. J. of Mod. Phys. } {\bf #1} (#2)  {#3}}
\newcommand{\PRL}[3]{{ Phys. Rev. Lett. } {\bf #1} (#2) {#3}}
\newcommand{\ZFP}[3]{{ Zeitsch. f. Physik } {\bf C#1} (#2)  {#3}}
\newcommand{\MPLA}[3]{{ Mod. Phys. Lett. } {\bf A#1} (#2) {#3}}
%%%%%%%%%%%%%%%%%%%%%%%%%%%%%%%%%%%%%%%%%%%%%%%%%%%%%%%%%%%%%%%%%%%%%%%%%
%%%%%%%%%%%%%%%%%%%%%%%%%%%%%%%%%%%%%%%%%%%%%%%%%%%%%%%%%%%%%%%%%%%%%%%%%
\newcommand{\CQG}[3]{{ Class. Quant. Grav. } {\bf #1} (#2) {#3}}
\newcommand{\JCAP}[3]{{ JCAP} {\bf#1} (#2)  {#3}}
\newcommand{\APJ}[3]{{ Astrophys. J. } {\bf #1} (#2)  {#3}}
\newcommand{\AMJ}[3]{{ Astronom. J. } {\bf #1} (#2)  {#3}}
\newcommand{\APP}[3]{{ Astropart. Phys. } {\bf #1} (#2)  {#3}}
\newcommand{\AAP}[3]{{ Astron. Astrophys. } {\bf #1} (#2)  {#3}}
\newcommand{\MNRAS}[3]{{ Mon. Not. Roy. Astron. Soc.} {\bf #1} (#2)  {#3}}
\newcommand{\JPA}[3]{{ J. Phys. A: Math. Theor.} {\bf #1} (#2)  {#3}}
\newcommand{\ProgS}[3]{{ Prog. Theor. Phys. Supp.} {\bf #1} (#2)  {#3}}
\newcommand{\APJS}[3]{{ Astrophys. J. Supl.} {\bf #1} (#2)  {#3}}
%%%%%%%%%%%%%%%%%%%%%%%%%%%%%%%%%%%%%%%%%%%%%%%%%%%%%%%%%%%%%%%%%%%%%%%%%

\newcommand{\Prog}[3]{{ Prog. Theor. Phys.} {\bf #1}  (#2) {#3}}
\newcommand{\IJMPA}[3]{{ Int. J. of Mod. Phys. A} {\bf #1}  {(#2)} {#3}}
\newcommand{\IJMPD}[3]{{ Int. J. of Mod. Phys. D} {\bf #1}  {(#2)} {#3}}
\newcommand{\GRG}[3]{{ Gen. Rel. Grav.} {\bf #1}  {(#2)} {#3}}

%  Example:  \NPB {\bf 20} {1992}  {200}

%%%%%%%%%%%%%%%%%%%%%%%%%%%%%%%%%%%%%%%%%%%%%%%%%%%%%%%%%%%%%%%%%%%%%%%%%
%\newpage
%%%%%%%%%%%%%%%%%%%%%%%%%%%%%%%%%%%%%%%%%%%%%%%%%%%%%%

\end{document}